\documentclass[a4paper,11pt]{article}
\PassOptionsToPackage{comma,square}{natbib}
\usepackage{jheppub}
\pdfoutput=1

\usepackage{xspace, amsthm, relsize, bbm, enumitem}
\usepackage[bbgreekl]{mathbbol}
\usepackage{amsfonts}
\usepackage[T1]{fontenc}
\usepackage{mathtools}
\usepackage{comment}

\DeclareSymbolFontAlphabet{\mathbb}{AMSb}
\DeclareSymbolFontAlphabet{\mathbbl}{bbold}

\newcommand{\Lg}{{\mathfrak{g}}}

\theoremstyle{definition}
\newtheorem{theorem}{Theorem}
\newtheorem{lemma}{Lemma}
\newtheorem{corollary}[theorem]{Corollary}
\newtheorem{proposition}{Proposition}
\newtheorem{remark}{Remark}

\newtheorem{definition}{Definition}

\newtheorem{convention}{Convention}

\DeclareFontFamily{OT1}{pzc}{}
\DeclareFontShape{OT1}{pzc}{m}{it}{<-> s * [1.200] pzcmi7t}{}
\DeclareMathAlphabet{\mathpzc}{OT1}{pzc}{m}{it}

\DeclareFontFamily{U}{bbold}{}
\DeclareFontShape{U}{bbold}{m}{n}
 {  <-5.5> s*[1.05] bbold5
    <5.5-6.5> s*[1.05] bbold6
    <6.5-7.5> s*[1.05] bbold7
    <7.5-8.5> s*[1.05] bbold8
    <8.5-9.5> s*[1.05] bbold9
    <9.5-11.5> s*[1.05] bbold10
    <11.5-16> s*[1.05] bbold12
    <16-> s*[1.05] bbold17
 }{}

\renewcommand{\Im}{\mathbbl{m}}

\newcommand{\beq}{\begin{equation}}
\newcommand{\eeq}{\end{equation}}
\newcommand{\beqnn}{\begin{equation*}}
\newcommand{\eeqnn}{\end{equation*}}
\newcommand{\bea}{\begin{eqnarray}}
\newcommand{\eea}{\end{eqnarray}}
\newcommand{\bean}{\begin{eqnarray*}}
\newcommand{\eean}{\end{eqnarray*}}

\newcommand{\place}[3]{\vbox to0pt{\kern-\parskip\kern-7pt\kern-#2truein\hbox{\kern#1truein #3}\vss}\nointerlineskip}

\DeclareFontFamily{U}{wncy}{}
\DeclareFontShape{U}{wncy}{m}{n}{<->wncyr10}{}
\DeclareSymbolFont{mcy}{U}{wncy}{m}{n}
\DeclareMathSymbol{\sha}{\mathord}{mcy}{"58}

\newcommand{\del}{{\partial}}
\newcommand{\delb}{{\bar\partial}}

\newcommand{\dd}{{\text{d}}}

\renewcommand{\Im}{\text{Im~}}

\newcommand{\tr}{\text{tr}\hskip2pt}

\makeatletter
\g@addto@macro\bfseries{\boldmath}
\makeatother

\numberwithin{equation}{section}
\title{On a Dual Heterotic Kodaira--Spencer Theory}

\author[a,b]{Siddhartha Mangamuri}
\author[a]{Eirik Eik Svanes}
\author[c]{Anuthi Tiwary}

\affiliation[a]{Department of Mathematics and Physics,
Faculty of Science and Technology, University of Stavanger,
N-4036 Stavanger, Norway}

\affiliation[b]{Department of Physics,
Indian Institute of Science Education and Research,
Dr. Homi Bhabha Road, Pune 411008, India}

\affiliation[c]{School of Physics, Indian Institute of Science Education and Research, Thiruvananthapuram, Vithura, Kerala - 695551, India}

\emailAdd{siddhartha.mangamuri03@gmail.com}
\emailAdd{eirik.e.svanes@uis.no}
\emailAdd{anuthi21@alumni.iisertvm.ac.in}
\abstract{
We consider a dual version of heterotic Kodaira--Spencer theory on a Calabi--Yau three-fold $X$ with background fluxes, related to the type I BCOV theory of Costello--Li. We implement the Green--Schwarz mechanism of Costello--Li, where the Beltrami differential plays the role of the Green--Schwarz field. On a compact Calabi--Yau three-fold with background gauge fluxes $F$, where $\tr(F^2)$ is trivial in cohomology, retaining gauge anomalies only and leaving gravitational anomalies to future work, we incorporate the Costello--Li Green--Schwarz mechanism into the classical deformation complex. This gives a theory governed by a nilpotent operator $\bar D$ on $Q=T^{1,0}X\oplus\operatorname{End}(V)\oplus T^{*1,0}X$, whose extension maps carry holomorphic derivatives, so that $(Q,\bar D)$ is not a holomorphic bundle. Using diagonal $\delb^\dagger$-gauge, the one-loop gauge anomaly cancels, assuming that the extension maps of $\bar D$ do not modify it, which we argue is plausible. Introducing an auxiliary $(0,2)$-form, we construct a cubic interacting theory whose gauge symmetries are governed by an $L_3$ algebra. On compact Calabi--Yau three-folds, and for sufficiently small fields, we show that the equation of motion is equivalent to the Maurer--Cartan equation, together with a divergence-free constraint on the Beltrami differential. Finally we trivialise $\bar D$ locally to the Dolbeault operator. The (non-tensorial) transition maps glue into a quasi-holomorphic extension sheaf $\widetilde{\mathcal Q}$, for which we prove a Dolbeault theorem, $H^{0,q}_{\bar D}(Q)\cong\check H^q(\widetilde{\mathcal Q})$.
}

\begin{document}
\maketitle
\flushbottom

\section{Introduction}
Heterotic string compactifications provide a particularly rich framework for studying the interplay between string theory, gauge theory, and complex geometry. Among the consistent ten-dimensional superstring theories, the $E_8\times E_8$ heterotic string is of particular interest from a phenomenological perspective, since the gauge group $E_8$ contains the Standard Model gauge group $SU(3)\times SU(2)\times U(1)$ as a subgroup. At the same time, heterotic compactifications exhibit a mathematical structure which, though perhaps richer, is less well understood than that of their type II counterparts. In particular, the non-Kähler geometries that arise naturally in heterotic compactifications lie beyond the framework in which many of the powerful tools of algebraic geometry, such as mirror symmetry and Gromov--Witten theory, have been developed. A systematic mathematical understanding of these compactifications is therefore of interest both for string phenomenology and for the study of geometric structures arising in quantum field theory.

Supersymmetric heterotic compactifications on a six-dimensional manifold are governed by the Hull--Strominger system \cite{Hull:1986kz, Strominger:1986uh}, whose solutions include Calabi--Yau compactifications as a special case \cite{Candelas:1985en}, but more generally allow for non-Kähler geometries with $SU(3)$-structure \cite{Becker:2006et, fu_theory_2008, Fernandez:2008wa, Fei:2014aca, Otal:2016bgn, Garcia-Fernandez:2016azr, Fei:2017ctw, Garcia-Fernandez:2018qcl, Fino:2019mvp}. The deformation theory of these compactifications is correspondingly more intricate than that of ordinary holomorphic bundles on Kähler manifolds, since complex-structure, gauge-bundle, and Hermitian deformations are coupled by curvature terms. 

The infinitesimal deformations of the Hull--Strominger system are described by the first cohomology of a differential complex associated to the bundle
\begin{equation}
\label{eq:Qintro}
Q=T^{1,0}X\oplus\operatorname{End}(V)\oplus T^{*1,0}X\: ,
\end{equation}
where the holomorphic tangent bundle encodes complex structure deformations, ${\rm End}(V)$ encodes deformations of the gauge connection, and the holomorphic cotangent bundle encodes (complexified) Hermitian deformations. The bundle $Q$ is equipped with a differential $\bar D$ whose off-diagonal entries couple the three sectors. This description, and the identification of the infinitesimal moduli with $H^{0,1}_{\bar D}(Q)$, was developed in \cite{Anderson:2014xha, delaOssa:2014cia, Garcia-Fernandez:2015hja} and has since been studied from a number of directions, including the moduli-space metric and its universal geometry \cite{Candelas:2016usb, Candelas:2018lib}, the heterotic superpotential \cite{LopesCardoso:2003dvb, Gurrieri:2004dt, Ashmore:2018ybe, McOrist:2026pbw}, and holomorphic string algebroids \cite{Garcia-Fernandez:2018emx, Garcia-Fernandez:2020awc}, see \cite{Ibarra:2024ntg} for a review. In particular, a classical six-dimensional heterotic Kodaira--Spencer theory was constructed in \cite{Ashmore:2018ybe}, by deforming the heterotic superpotential around (flux) backgrounds. This theory (and its ten-dimensional counterpart) has since been studied in the context of generalised geometry \cite{Kupka:2026usx, Jonker:2026cok}, and its one-loop partition function and anomalies have been analysed \cite{Ashmore:2023vji, Ashmore:2025fxr}. Recent work in the context of heterotic $G_2$ compactifications is also of note \cite{delaOssa:2017pqy, deIaOssa:2019cci, McOrist:2024ivz, Silva:2024fvl, ossa2026quantum, Tringas:2025bwe, Aikot:2026tzn, Brongers:2026uxb, Brongers:2026pyc, Brongers:2026gbo, Ashmore:2026vgf, Aikot:2026btp}.

In this paper we define and study a dual theory to the one studied in \cite{Ashmore:2018ybe}. The word \emph{dual} in the title refers to the arrangement of \eqref{eq:Qintro}. In the formulations cited above the extension maps, given by background fluxes, run from the complex-structure deformation through $\operatorname{End}(V)$ and into the Hermitian sector. Ultimately, this is a consequence of the usual Green--Schwarz mechanism \cite{Green:1984sg} of the heterotic string and heterotic supergravity \cite{Bergshoeff:1989de}, applied to the heterotic moduli problem. Another viewpoint has recently developed in the context of type I BCOV theory \cite{Costello:2015xsa, Costello:2019jsy, Li:2026eyh}, where the Beltrami differential, or complex structure deformation, is used as the Green--Schwarz field instead. This has since been utilized to a great extent in other related contexts, such as twistor theory and celestial holography, see e.g. \cite{Williams:2018ows, Costello:2021bah, Costello:2021kiv, Costello:2022wso, Bittleston:2022nfr, Bittleston:2024efo, Bittleston:2026pai, Sharma:2025ntb}. 

As we shall see, employing this dual Green--Schwarz mechanism leads to the flux-induced extension maps running the other way. The differential \eqref{eq:barD} is upper triangular with the cotangent field mapping to the gauge field and thence to the complex-structure slot. As a consequence, the Beltrami field $\mu$ enters the bracket only linearly, the self-interaction being carried instead by the divergence-free vector $(\del x)$ built from the cotangent field. This exchange is responsible for most of what distinguishes the results below from their counterparts in the standard arrangement. 

Furthermore, in the formulation adopted here with the same field content as the standard heterotic Kodaira--Spencer theory \cite{Ashmore:2018ybe}, $\bar D$ contains derivative-dependent extension maps, and therefore does not define a conventional holomorphic vector bundle. Heuristically, $\bar D$ takes the form
\begin{equation}
    \bar D =
    \begin{pmatrix}
        \bar\partial & F\!\cdot\!\nabla & T \\
        0 & \bar\partial & F\!\cdot\!\nabla \\
        0 & 0 & \bar\partial
    \end{pmatrix}\:,
\end{equation}
where the extension maps $F\!\cdot\!\nabla$ and $T$ are defined below, $T$ being a second order map in holomorphic derivatives. This presents an obstruction to applying standard sheaf-theoretic methods directly to the corresponding dual heterotic deformation complex, and we analyse some of the consequences of this in Section \ref{sec:local-triv}.

A few assumptions should be stated at the outset. We work throughout on a compact Calabi--Yau three-fold, so that the geometry is Kähler and the $\del\delb$-lemma is available. The quantum theory possesses both gravitational and gauge anomalies. For simplicity and clarity, we will only focus on gauge anomalies in the current paper. The anomaly cancellation, with local counter terms, then forces $\tr(F^2)$ to be trivial in cohomology. The Hermitian Yang--Mills condition is nowhere imposed, the connection on $V$ being only the Chern connection of a Hermitian metric. The construction of Section~\ref{sec:local-triv} does not itself use the $\del\delb$-lemma or Hodge theory, and extends to non-K\"ahler backgrounds provided these carry a holomorphic volume form and a global $(2,1)$-form
$\omega^{2,1}$ with $\delb\omega^{2,1}=\tr(F^2)$. The cubic theory of Section~\ref{sec:cubic} needs in addition $\del\omega^{2,1}=0$. On K\"ahler backgrounds the $\del\delb$-lemma supplies such an $\omega^{2,1}$. In general
its existence is an additional assumption. It should be noted that torsional backgrounds satisfying the Hull--Strominger system are not of this kind, since there $\tr(F^2)$ is balanced against $\tr(R^2)$ in the Bianchi identity. Including such backgrounds requires the gravitational anomaly, and is left for future work.

A central result of the paper is a geometric interpretation of the apparently non-holomorphic complex $(\Omega^{0,p},\bar D)$. We show that $\bar D$ can be locally trivialised to the ordinary $\bar\partial$ operator. The nontrivial information is consequently transferred to the transition maps between local patches, which are now no longer tensorial. These transition maps define a globally consistent \emph{quasi-holomorphic extension} \cite{MurgasIbarra:2025ylk}, represented by a sheaf $\widetilde{\mathcal Q}$, which is canonically the sheaf of $\bar D$-closed smooth sections of $Q$. We obtain a modified Dolbeault theorem relating the $\bar D$ cohomology to the sheaf cohomology of $\widetilde{\mathcal Q}$. Since the transition operators themselves contain holomorphic derivatives they are not $\mathcal{O}_X$-linear, so $\widetilde{\mathcal Q}$ is a sheaf of $\mathbb{C}$-vector spaces rather than a coherent analytic sheaf. What the theorem supplies is \v{C}ech cohomology, a fine resolution, and the comparison with $H^{0,q}_{\bar D}(Q)$. This is already enough to bring the heterotic deformation complex within reach of global, sheaf-theoretic arguments and more formal algebraic geometry.

The anomaly analysis also leads to an important subtlety. As mentioned, we ignore gravitational anomalies in the current paper, focusing only on strict gauge transformations of the gauge bundle. Furthermore, since $\bar D$ is not the $(0,1)$-part of an ordinary connection, the standard formulation of the one-loop anomaly in terms of a gauge-fixed kinetic operator cannot be applied immediately. The study of anomalies for such kinetic operators is left to future works. Here, we instead show that the usual diagonal $\bar\partial^\dagger$ gauge condition, supplemented by a condition on the cotangent field, selects exactly one representative of each gauge-equivalence class, and take the one-loop anomaly to be that of the flux-less quadratic theory, where the kinetic operator is $\delb$. 

The Green--Schwarz counterterm cancels the ${\rm ch}_4$ anomaly of the quadratic theory, and its variation under the additional transformation generated by $\bar D$ is cancelled by a further counterterm involving the Chern--Simons form. This assumes that the extension maps of $\bar D$, which are not $(0,1)$-parts of a connection, do not modify the anomaly. As argued below, this is to be expected, though the full analysis is left to future work. 

We go on to study the interacting theory and construct a cubic term, using a bracket compatible with the corrected differential and the natural pairing. The resulting bracket satisfies the graded Leibniz rule and cyclicity properties required for the cubic theory, but does not obey the Jacobi identity strictly. Its Jacobiator is, however, $\partial$-exact and is controlled by a Yukawa cubic obtained from the cotangent field. This structure points naturally toward an $L_\infty$ description. We show by a degree count that no obvious local $\ell_3$ exists on $Q$ itself, and that enlarging the complex and extending the action by a scalar $(0,\bullet)$-form, as was also done in \cite{Ashmore:2018ybe}, yields an extended $L_3$ algebra whose products are completely determined, with $\ell_k=0$ for $k\geq4$ as a consequence of the Jacobiator identity in complex dimension three. The enlarged action is invariant under the $L_3$ gauge transformations with divergence-free parameters, the same restriction as in \cite{Ashmore:2018ybe}. There, however, invariance under the vector-type parameters holds only once the multiplier has been integrated out, that is, on the constraint surface. Here it holds with the multiplier present: exactly on the constraint surface (Theorem~\ref{thm:constraint}), and off it up to an on-shell-trivial term in the multiplier's transformation. On the constraint surface the extended action reduces to the cubic action, and the theory to a differential graded Lie algebra with an invariant pairing, as in \cite{Ashmore:2018ybe}. The Maurer--Cartan equation is covariant under the full gauge algebra. Furthermore, on a compact Calabi--Yau three-fold and for small fields, the equation of motion implies the Maurer--Cartan equation of the extended $L_3$ algebra.

The paper is organised as follows. In Section~\ref{sec:quad}, we review the quadratic heterotic Kodaira--Spencer theory, analyse its gauge anomaly, and construct the Green--Schwarz-corrected differential $\bar D$. We prove its nilpotency, determine its cohomology on Kähler backgrounds, and discuss gauge fixing and anomaly cancellation. In Section~\ref{sec:cubic}, we construct the cubic bracket, establish its graded anti-symmetry, compatibility with $\bar D$, and cyclicity with respect to the pairing. Using this, we construct a cubic action respecting the gauge symmetries. We then analyse the resulting Maurer--Cartan equation and compute the Jacobiator, identifying the obstruction with the Yukawa cubic, and construct the extended $L_3$ algebra which resolves it, together with an $L_3$-compatible cubic action. Finally, in Section~\ref{sec:local-triv}, we discuss the local and sheaf-theoretic interpretation of the derivative-dependent extension maps and their relation to quasi-holomorphic extensions. We conclude in section \ref{sec:concl} with an outlook on future directions. Some lengthy computations and proofs are left to appendices. 

\section{Anomalies and the quadratic theory}\label{sec:quad}
We start by considering a heterotic Kodaira--Spencer theory on a Kähler Calabi--Yau manifold $X$ of complex dimension $3$, where the classical field content is
\begin{align*}
    x&\in\Omega^{0,1}(T^{*1,0}X)\\
    \alpha&\in\Omega^{0,1}({\mathfrak g})\\
    \mu&\in\Omega^{0,1}(T^{1,0}X)\:,
\end{align*}
where $\mu$ is also taken to be divergence-free, in the sense that
\begin{equation}
    \label{eq:divfree}
    \del\chi=\del\left(\tfrac{1}{2}\mu^{a}\Omega_{abc}\dd z^{bc}\right)=0\:,
\end{equation}
where $\Omega\in\Omega^{3,0}(X)$ is the holomorphic top-form of the Calabi--Yau, and 
\begin{equation}
    \chi=\tfrac{1}{2}\mu^{a}\Omega_{abc}\dd z^{bc}\:.
\end{equation}
Here $\mathfrak g$ is the Lie algebra of a holomorphic vector bundle $V\rightarrow X$. Throughout, the connection on $V$ is the Chern connection of a Hermitian metric, so that $F^{0,2}=F^{2,0}=0$. In terms of moduli fields, we may think of $x$ as complexified Hermitian degrees of freedom, $\alpha$ as the gauge degree of freedom, and $\chi$ as the complex structure deformation, corresponding to the Beltrami differential $\mu$. This is the same moduli field content as that of the heterotic Kodaira--Spencer theory described in \cite{Ashmore:2018ybe}.

The quadratic part of the action without fluxes then reads
\begin{equation}
\label{eq:ClassActionQuad}
    S=\int_X \left(x_a\delb\mu^a-\tfrac12\tr(\alpha\delb\alpha)\right)\wedge\Omega\:, 
\end{equation}
where we have picked holomorphic gauge on the background gauge bundle where $(\dd_A)^{0,1}=\delb$. As shown in \cite{Ashmore:2025fxr, Kupka:2026usx, Jonker:2026cok}, the theory \eqref{eq:ClassActionQuad} can be related to the more standard formulation of type I BCOV \cite{Costello:2015xsa}, via a non-local field redefinition.

\subsection{The gauge anomaly}
It is well known that the theory \eqref{eq:ClassActionQuad} suffers from both gravitational and gauge anomalies. We consider only the ${\rm ch}_4({\rm End}\,V)$ part of the gauge anomaly. The mixed term $\tfrac1{12}c_2(X)\,{\rm ch}_2({\rm End}\,V)$ in the index density is gauge dependent but has a gravitational coefficient, and we treat it together with the gravitational anomalies, which we leave to later work. Anomalies are associated to an anomalous phase transformation of the partition function,
\begin{equation}
    Z=\int{\cal D}\phi\, e^{iS(\phi)}\:,
\end{equation}
where the path integral is over a Lagrangian sub-manifold in the space of (off-shell) fields, collectively denoted $\phi$. Here $S(\phi)$ is the BV action, and $\phi$ also includes ghosts and anti-fields, as in the BV formalism. The Lagrangian sub-manifold is determined by a gauge fixing constraint, where Lorenz gauge is a common choice. To impose Lorenz gauge, one picks a metric on the space of fields and constructs the corresponding adjoint of the kinetic operator. The path integral is then over fields in the kernel of the adjoint, which for the gauge sector reads
\begin{equation}
    \label{eq:adj}
    \delb^\dagger=-*\del_{A_0}*\:,
\end{equation}
where $A_0$ is the (Hermitian) background connection of type $(1,0)$. 

The point is that even though the (off-shell) fluctuations $\phi$ are gauge fixed in the path integral under $\delb$-exact transformations, the theory $S(\phi)$ is still invariant under background gauge rotations of the background gauge bundle (and other geometric/gravitational rotations). This leads to an anomalous transformation of the path integral measure, and in turn an anomalous transformation of the phase of the partition function. For the gauge bundle sector, the gauge anomaly takes the form\footnote{As the main objective here is to demonstrate that the anomaly may be cancelled with appropriate counter-terms, we do not worry about the numerical factors in front of these terms.}

\begin{equation}
    \label{eq:gaugeAnomaly}
    \delta\log(Z)\propto i\hbar\int_X{\rm Tr}\left(F^3\epsilon\right)\:,
\end{equation}
where $\epsilon\in\Omega^0({\mathfrak g})$ is the infinitesimal gauge transformation of the background gauge field, which also leads to a classical gauge transformation of the classical gauge field $\alpha$ as $\delta\alpha=\delb\epsilon$ in the classical action \eqref{eq:ClassActionQuad}.\footnote{A small gauge transformation of the classical background connection $A=A_0$ reads 
\begin{equation}
    \delta A=\dd_{A}\epsilon=\del_{A_0}\epsilon+\delb\epsilon\:.
\end{equation}
The first term corrects $\del_{A_0}$ in \eqref{eq:adj}, contributing to the anomalous phase along with the background gauge rotations of the ${\mathfrak g}$-valued off-shell modes. The second term may be viewed as a gauge transformation of the classical gauge field deformation $\alpha$ in the classical action.} We have also included a factor of $\hbar$ in \eqref{eq:gaugeAnomaly} to indicate the quantum one-loop nature of the anomaly. 
Here, $\rm Tr(\bullet)$ denotes the trace in the adjoint representation, while $\tr$ denotes the fundamental trace. To cancel the gauge anomaly, it is then common to assume a gauge group satisfying trace identities of the form
\begin{equation}
    \label{eq:Factor}
    {\rm Tr}(T^4)\propto\tr(T^2)^2\:,
\end{equation}
where $T$ is a generator of the given Lie algebra, and $\tr(\bullet)$ is the trace in the fundamental representation. For example, for a gauge group of $SU(3)$, we have
\begin{equation}
    {\rm Tr}(T^4)=9\,\tr(T^2)^2\:.
\end{equation}
Assuming the factorization condition \eqref{eq:Factor}, the anomaly becomes
\begin{equation}
    \delta\log(Z)\propto i\hbar\int_X\tr(F^2)\tr(F\epsilon)\:.
\end{equation}
A common way to cancel such anomalies is then to introduce a counter term of the form
\begin{equation}
    S_{\rm counter}(X)\propto\int_X\tr(F^2)\wedge x\:,
\end{equation}
and introduce the Green--Schwarz type transformation of the Hermitian degree of freedom \cite{Green:1984sg}
\begin{equation}
    \delta x=\hbar\,\tr(F\epsilon)\:.
\end{equation}
This approach was followed in e.g. \cite{Ashmore:2023vji, Ashmore:2025fxr}.

For the current paper, we wish to take a slightly different approach, following the Green--Schwarz type mechanism of Costello--Li \cite{Costello:2015xsa, Costello:2019jsy}, which has been employed extensively in recent works, see e.g. \cite{Williams:2018ows, Costello:2021kiv, Costello:2021bah, Bittleston:2022nfr, Costello:2022wso, Bittleston:2024efo}. To use this approach with globally well-defined counter terms, we must make the assumption that $\tr(F^2)$ vanishes in cohomology,\footnote{\label{foot:YMflat} Without including the gravitational anomaly, this condition excludes background solutions such as the Hull--Strominger system \cite{Hull:1986kz, Strominger:1986uh}, where the background connection is a Yang--Mills instanton. Indeed, for such solutions on compact geometries, this condition forces the bundle to be flat \cite{Garcia-Fernandez:2023vah}.} that is
\begin{equation}
    \tr(F^2)=\del\omega^{1,2}
\end{equation}
for some global $(1,2)$-type Chern--Simons form $\omega^{1,2}$, which we may also take to be $\delb$-exact by the $\del\delb$-lemma. The anomaly then becomes
\begin{equation}
    \delta\log(Z)\propto i\hbar\int_X\omega^{1,2}\wedge\tr(F\del_A\epsilon)\:.
\end{equation}
The anomaly is then cancelled by adding a counter term
\begin{equation}
    \label{eq:counterChi}
    S_{\rm counter}(\chi)\propto\int_X\omega^{1,2}\wedge\chi
\end{equation}
and introducing a Green--Schwarz transformation for $\chi$ as
\begin{equation}
    \label{eq:GSchi}
    \delta\chi=\hbar\,\tr(F\del_A\epsilon)\:.
\end{equation}
The counter-term \eqref{eq:counterChi} is invariant under the classical $\delb$-exact transformation of $\chi$, and at one loop the transformation \eqref{eq:GSchi} leaves the classical action \eqref{eq:ClassActionQuad} unchanged, since the remaining fields may be taken on-shell modulo higher-loop corrections.

\subsection{The one-loop effective action}
One main objective of this paper, however, is to incorporate the transformation \eqref{eq:GSchi} into the gauge algebra of the classical theory \eqref{eq:ClassActionQuad}, to construct an exactly gauge invariant classical action. This amounts to promoting the classical $\delb$ operator to a new extension operator $\bar D$. Collecting the fields as $y=(\chi,\alpha,x)$, we can write the corrected quadratic classical action as
\begin{equation}
    \label{eq:corrAction}
    S=\tfrac{1}{2}\int_X\langle y,\bar D y\rangle\:,
\end{equation}
where the pairing of $y_1$ and $y_2$ is
\begin{equation}
    \langle y_1,y_2\rangle=\chi_1\wedge x_2+(-1)^{y_2}x_1\wedge\chi_2-\tr(\alpha_1\wedge\alpha_2)\wedge\Omega\:,
\end{equation}
where $y_2$ in the exponent denotes the anti-holomorphic degree of $y_2$. This pairing is appropriately symmetric.
\begin{lemma}
\label{lem:sym}
    The pairing $\langle\,,\,\rangle$ satisfies
    \begin{equation}
        \langle y_1,y_2\rangle=(-1)^{y_1y_2}\langle y_2,y_1\rangle\:.
    \end{equation}
\end{lemma}
\begin{proof}
    This follows from a short computation
    \begin{align*}
        \langle y_1,y_2\rangle&=\chi_1\wedge x_2+(-1)^{y_2}x_1\wedge\chi_2-\tr(\alpha_1\wedge\alpha_2)\wedge\Omega\\
        &= (-1)^{y_1(y_2+1)}x_2\wedge\chi_1+(-1)^{y_2+(y_1+1)y_2}\chi_2\wedge x_1-(-1)^{y_1y_2}\tr(\alpha_2\wedge\alpha_1)\wedge\Omega\\
        &=(-1)^{y_1y_2}\left(\chi_2\wedge x_1+(-1)^{y_1}x_2\wedge\chi_1-\tr(\alpha_2\wedge\alpha_1)\wedge\Omega\right)=(-1)^{y_1y_2}\langle y_2,y_1\rangle\:.
    \end{align*}
\end{proof}
A nilpotent operator which also integrates by parts in the action (Lemma \ref{lem:IBP} below), and 
which incorporates the gauge transformation \eqref{eq:GSchi}, can be schematically written as
\begin{equation}
    \label{eq:barD}
    \bar D =
    \begin{pmatrix}
        \bar\partial & F\!\cdot\!\nabla & T \\
        0 & \bar\partial & F\!\cdot\!\nabla \\
        0 & 0 & \bar\partial
    \end{pmatrix}\:,
\end{equation}
acting on $(0,p)$-forms valued in
\begin{equation}
    Q=\Lambda^2(T^{*1,0}X)\oplus{\mathfrak g}\oplus T^{*1,0}X\:.
\end{equation}
The three slots carry $\chi$, $\alpha$ and $x$, respectively. Note that on the Calabi--Yau three-fold, contraction with $\Omega$ identifies $\Lambda^2T^{*1,0}X\cong T^{1,0}X$, matching the conventions of Sections \ref{sec:cubic}--\ref{sec:local-triv}. The extension maps act as follows. For $x\in\Omega^{0,q}(T^{*1,0}X)$ we have
\begin{equation}
    F\!\cdot\!\nabla x=F_a\,\partial_b x_c\,\bar\Omega^{abc}\:,
\end{equation}
where $\bar\Omega^{abc}$ is the holomorphic trivector inverse to $\Omega$, normalised as in \eqref{eq:Omnorm}. For $\alpha\in\Omega^{0,q}({\mathfrak g})$, we have
\begin{equation}
    F\!\cdot\!\nabla\alpha=\tr(F\wedge\del_A\alpha)\:.
\end{equation}
This is the map which incorporates the transformation \eqref{eq:GSchi}, and we have set $\hbar=1$. Note that 
\begin{equation}
    F\!\cdot\!\nabla\alpha=\del\,\tr(F\wedge\alpha)\:,
\end{equation}
so that the divergence-free condition \eqref{eq:divfree} is preserved. Finally, the map $T$ acts as
\begin{equation}
    \label{eq:Ttrans}
    T(x)= -\tfrac{1}{2}\,\partial\left(\omega^{2,1}_a\,
      \partial_b x_c\,\bar\Omega^{abc}\right)\:,
\end{equation}
again preserving the divergence-free condition. Here $\omega^{2,1}$ is a
global Chern--Simons form of pure type $(2,1)$,
\begin{equation}\label{eq:dCS}
  \dd\omega^{2,1}=\tr(F\wedge F)\:.
\end{equation}
Since $\tr(F\wedge F)$ has type $(2,2)$, \eqref{eq:dCS} is equivalent to
\begin{equation}\label{eq:dCScomp}
  \delb\omega^{2,1}=\tr(F\wedge F)\:,\qquad \del\omega^{2,1}=0\:.
\end{equation}
Such a form exists whenever $\tr(F\wedge F)$ is exact. Indeed, by the
$\del\delb$-lemma there is a $(1,1)$-form $\beta$ with
\[
  \tr(F\wedge F)=\del\delb\beta\:,\qquad \omega^{2,1}\coloneqq-\del\beta\:,
\]
and this $\omega^{2,1}$ satisfies \eqref{eq:dCScomp}. Only the first
condition in \eqref{eq:dCScomp} enters $\bar D^2=0$; the second is used in
the Leibniz rule of Proposition~\ref{prop:leibniz}. We write
$\omega^{2,1}_a=\iota_{\del_a}\omega^{2,1}$ for the components along the
holomorphic coordinate vector fields; the combination entering
\eqref{eq:Ttrans} is the global form $\iota_{\del x}\omega^{2,1}$. The
components satisfy
\begin{equation}
    \label{eq:CW}
    \delb\omega_a^{2,1}=-2\,\tr(F\wedge F_a)\:,
\end{equation}
where $F_a=F_{a\bar b}\dd z^{\bar b}=\iota_{\del_a}F$, so that
$F=\dd z^a\wedge F_a$. The identity \eqref{eq:CW} follows by contracting
$\delb\omega^{2,1}=\tr(F\wedge F)$ with $\iota_{\del_a}$: the contraction
anticommutes with $\delb$, since $\{\delb,\iota_{\del_a}\}$ is the
$(-1,1)$-component of the Lie derivative $\mathcal{L}_{\del_a}$, which
preserves bidegree and hence vanishes; and it acts as an antiderivation
on the two identical curvature slots, giving
$\iota_{\del_a}\tr(F\wedge F)=2\,\tr(F\wedge F_a)$ by cyclicity of the
trace. The transformation \eqref{eq:Ttrans} also preserves the divergence-free condition \eqref{eq:divfree}.

We refer to the three off-diagonal entries as the \emph{metric-to-gauge}, \emph{gauge-to-complex} and \emph{metric-to-complex} entries of $\bar D$, after the fields they connect. In the vector description of Section~\ref{sec:cubic} they map the cotangent slot to the gauge slot, the gauge slot to the tangent slot, and the cotangent slot to the tangent slot respectively.

We then have the following proposition:

\begin{proposition}
\label{prop:nilp}
    The operator $\bar D$ as constructed above is nilpotent. That is
    \begin{equation}
        \bar D^2=0\:.
    \end{equation}
\end{proposition}
\begin{proof}
Since $\bar D$ is block upper-triangular with $\delb$ on the diagonal,
$\bar D^2$ has entries only in the $(1,2)$, $(2,3)$ and $(1,3)$ slots.
Throughout we abbreviate $(\del x)^a\coloneqq \del_b x_c\,\bar\Omega^{abc}$. 

The $(2,3)$ entry is $\delb\bigl(F_a(\del x)^a\bigr)
+F_a(\del\,\delb x)^a$, which vanishes because $\delb F_a=0$: this is
the $(1,2)$-component of the Bianchi identity $\delb_A F=0$ together
with $F^{0,2}=0$, and $\delb$ commutes with the holomorphic derivatives
in $(\del x)^a$.

The $(1,2)$ entry is
\[
\delb\bigl(\tr(F\wedge\del_A\alpha)\bigr)
+\tr\bigl(F\wedge\del_A(\delb_A\alpha)\bigr)
=\tr\bigl(F\wedge\{\delb_A,\del_A\}\alpha\bigr)
=\tr\bigl(F\wedge[F,\alpha]\bigr)=0\:,
\]
where the first equality uses the Leibniz rule with $\delb_A F=0$, the
second the standard identity $\{\delb_A,\del_A\}=[F,\,\cdot\,]$, and the
last the cyclicity of the trace.

For the $(1,3)$ entry we must show
\begin{equation}
\label{eq:nilpx}
\delb\bigl(T(x)\bigr)+T(\delb x)
+\tr\Bigl(F\wedge\del_A\bigl(F_a(\del x)^a\bigr)\Bigr)=0\:.
\end{equation}
Using the Leibniz rule and $\del\delb=-\delb\del$, the first two terms
combine into a single total derivative,
\[
\delb\bigl(T(x)\bigr)+T(\delb x)
=\tfrac12\,\del\bigl(\delb\omega^{2,1}_a\wedge(\del x)^a\bigr)
\overset{\eqref{eq:CW}}{=}
-\,\del\bigl(\tr(F\wedge F_a)\wedge(\del x)^a\bigr)\:,
\]
where the Chern--Weil identity \eqref{eq:CW} has been used. Expanding
the remaining $\del$ by the Leibniz rule, and using
$\del\,\tr(F\wedge F_a)=\tr\bigl(F\wedge\del_A F_a\bigr)$ (the
$(2,1)$-Bianchi identity $\del_A F=0$ with cyclicity of the trace),
\[
\delb\bigl(T(x)\bigr)+T(\delb x)
= -\,\tr\bigl(F\wedge\del_A F_a\bigr)\wedge(\del x)^a
  +\tr(F\wedge F_a)\wedge\del(\del x)^a\:.
\]
The third term of \eqref{eq:nilpx}, expanded once by the Leibniz rule,
is
\[
\tr\Bigl(F\wedge\del_A\bigl(F_a(\del x)^a\bigr)\Bigr)
= \tr\bigl(F\wedge\del_A F_a\bigr)\wedge(\del x)^a
  -\tr(F\wedge F_a)\wedge\del(\del x)^a\:,
\]
and the two displays cancel identically. Note that the relative
normalisation is fixed: writing $T(x)=c\,\del(\omega^{2,1}_a(\del x)^a)$
and $\delb\omega^{2,1}_a=\kappa\,\tr(F\wedge F_a)$, the cancellation of
the $\del(\del x)^a$-terms requires $c\,\kappa=1$, which the values
$c=-\tfrac12$ and $\kappa=-2$ satisfy.
\end{proof}
Furthermore, the operator $\bar D$ satisfies integration by parts. That is, we have the following lemma.
\begin{lemma}
\label{lem:IBP}
    The operator $\bar D$ satisfies integration by parts, in the sense that
    \begin{equation}
        \int_X\langle\bar D y_1,y_2\rangle=(-1)^{y_1+1}\int_X\langle y_1,\bar D y_2\rangle\:,
    \end{equation}
    where $y_1$ and $y_2$ have the correct total degree so that $y_1+y_2+1=3$.    
\end{lemma}
\begin{proof}
Write $y_i=(\chi_i,\alpha_i,x_i)$. Since $\bar D$ is upper-triangular
and the pairing is anti-diagonal in the $(\chi,x)$ slots, the claim
splits into three independent statements, one for the diagonal and one
for each off-diagonal coupling.

\medskip\noindent\textbf{Diagonal.} As $\Omega$ is $\delb$-closed, $\del$-closed and $X$ is
compact, $\int_X\delb\langle y_1,y_2\rangle=0$; expanding by the Leibniz
rule and using Lemma~\ref{lem:sym} gives the stated sign
$(-1)^{y_1+1}$ for the $\delb$-diagonal.

\medskip\noindent\textbf{The coupling $T$.} In the pairing, $T$ is paired against itself:
the relevant terms are $\int_X T(x_1)\wedge x_2$ and
$\int_X x_1\wedge T(x_2)$. Since $T(x)=-\tfrac12\del(\omega^{2,1}_a
(\del x)^a)$ is a total $\del$-derivative and $\del\Omega=0$, the
$\del$ may be integrated by parts onto $x_2$. A short computation with
$\bar\Omega^{abc}$ shows that $\del$ acting on $x_2$ reassembles into
the same contraction $(\del x_2)^e$, so both terms reduce to the common
bilinear expression
\[
\int_X K_{ae}\wedge(\del x_1)^a\wedge(\del x_2)^e\:,
\qquad
K_{ae}\coloneqq \omega^{2,1}_a\wedge\iota_{\del_e}\Omega\:.
\]
Because $\omega^{2,1}$ is a form, its holomorphic components are
antisymmetric, $K_{ae}=-K_{ea}$; combined with the graded
anti-symmetry of the two $(\del x)$-factors this reproduces exactly the
sign $(-1)^{y_1+1}$ of the diagonal blocks.

\medskip\noindent\textbf{The couplings $F\!\cdot\!\nabla$.} These exchange the
$\mathbf{g}$-slot with the $\chi$- and $x$-slots, so the two off-diagonal
entries are paired against each other rather than against themselves.
Consider the gauge-to-complex entry acting on $\alpha_1$, paired against $x_2$.
Since $\del_AF=0$, this entry is a total $\del$-derivative, and integrating by
parts gives
\begin{equation}
    \label{eq:IBPFa}
    \int_X\tr(F\wedge\del_A\alpha_1)\wedge x_2
    =\int_X\del\tr(F\wedge\alpha_1)\wedge x_2
    =(-1)^{y_1+1}\int_X\tr(F\wedge\alpha_1)\wedge\del x_2\:,
\end{equation}
where the sign is the Koszul sign of moving $\del$ past
$\tr(F\wedge\alpha_1)$, which has total degree $y_1+2$. By \eqref{eq:OO}, the
form $\del x_2$ is expressed through the vector $(\del x_2)^e$ as
\begin{equation}
    \label{eq:IBPFb}
    \del x_2=(-1)^{y_2}\,(\del x_2)^e\,\iota_{\del_e}\Omega\:,
\end{equation}
the sign arising from moving $\dd z^b$ past the component $(0,y_2)$-form.
Inserting \eqref{eq:IBPFb} into \eqref{eq:IBPFa}, and using the contraction
identity $F\wedge\iota_{\del_e}\Omega=\Omega\wedge F_e$, the cyclicity of the
trace and $y_1+y_2=2$, we find
\begin{equation}
    \label{eq:IBPFc}
    \int_X\tr(F\wedge\del_A\alpha_1)\wedge x_2
    =(-1)^{y_1+1}\Bigl(-\int_X\tr\bigl(\alpha_1\wedge F_a(\del x_2)^a\bigr)\wedge\Omega\Bigr)\:.
\end{equation}
The bracket on the right is precisely the pairing of $\alpha_1$ against the
metric-to-gauge entry acting on $x_2$, so this coupling again comes with the
sign $(-1)^{y_1+1}$ of the diagonal blocks.

Collecting the three blocks gives the stated identity.
\end{proof}
It follows from Lemma \ref{lem:sym} and Lemma \ref{lem:IBP} that, for variations with $\del\delta\chi=0$, the equations of motion of the corrected classical theory \eqref{eq:corrAction} are
\begin{equation}
    \label{eq:EOMconstr}
    (\bar Dy)_\chi=0\:,\qquad(\bar Dy)_{\mathfrak g}=0\:,\qquad
    (\bar Dy)_x=\delb x\in\del\,\Omega^{0,2}(X)\:,
\end{equation}
the last because, on compact K\"ahler $X$, $\int_X\delta\chi\wedge\delb x$ vanishes for all $\del$-closed $\delta\chi$ precisely when $\delb x$ is $\del$-exact. This is the linearisation of \eqref{eq:MC}, whose $\del$-exact remainder is identified in Section~\ref{sec:MCext}. Besides the $\bar D$-exact transformations, the action is invariant under $x\mapsto x+\del\beta$, since $(\del\,\del\beta)^a=0$ and $(\bar Dy)_\chi$ is $\del$-closed. Solutions modulo gauge are therefore classes in $H^1(K)$, where
\begin{equation}
    \label{eq:Kcomplex}
    K^p=\bigl\{y\in\Omega^{0,p}(Q):\del\chi=0\bigr\}\big/\bigl\{(0,0,\del\beta):\beta\in\Omega^{0,p}(X)\bigr\}\:,
\end{equation}
on which $\bar D$ is well defined.
\begin{lemma}
\label{lem:constr}
    On a compact K\"ahler three-fold $X$, $H_{\bar D}^p(K)\cong H^{0,p}_{\bar D}(Q)$ for all $p$.
\end{lemma}
\begin{proof}
    Let $C\subset\Omega^{0,\bullet}(Q)$ be the subspace of fields $y=(\chi,\alpha,x)$ with $\del\chi=0$. It is preserved by $\bar D$: the operator $\delb$ anticommutes with $\del$, and the two extension maps into the $\chi$-slot,
    \begin{equation}
        F\!\cdot\!\nabla\alpha=\del\tr(F\wedge\alpha)\:,\qquad
        T(x)=-\tfrac12\,\del\bigl(\omega^{2,1}_a(\del x)^a\bigr)\:,
    \end{equation}
    are $\del$-exact. Next, the fields $(0,0,\del\beta)$ with $\beta\in\Omega^{0,\bullet}(X)$ form a subcomplex of $C$. Indeed, $(\del\,\del\beta)^a=0$ by the anti-symmetry of $\bar\Omega^{abc}$, as in \eqref{eq:autodiv}, so both extension maps out of the cotangent slot annihilate $\del\beta$, and
    \begin{equation}
        \bar D(0,0,\del\beta)=(0,0,-\del\delb\beta)\:,
    \end{equation}
    which is again of this form. By \eqref{eq:Kcomplex}, $K$ is the quotient of $C$ by this subcomplex. We therefore have two short exact sequences of complexes,
    \begin{gather}
        0\rightarrow C\rightarrow\Omega^{0,\bullet}(Q)\xrightarrow{\;y\,\mapsto\,\del\chi\;}\del\Omega^{2,\bullet}(X)\rightarrow 0\:,
        \label{eq:sesC}\\
        0\rightarrow\del\Omega^{0,\bullet}(X)\rightarrow C\rightarrow K\rightarrow 0\:,
        \label{eq:sesK}
    \end{gather}
    where the outer complexes carry the differential $\delb$. (In \eqref{eq:sesC} the map $y\mapsto\del\chi$ intertwines $\bar D$ with $-\delb$, since $\del\delb=-\delb\del$; this sign does not affect cohomology.) Both outer complexes are acyclic by the $\del\delb$-lemma: a $\delb$-closed $\del$-exact form can be written as
    \begin{equation}
        \del\delb\eta=\delb(-\del\eta)\:,
    \end{equation}
    and is therefore $\delb$-exact within the complex of $\del$-exact forms. The long exact sequences in cohomology associated with \eqref{eq:sesC} and \eqref{eq:sesK} then give
    \begin{equation}
        H^{0,p}_{\bar D}(Q)\cong H^p_{\bar D}(C)\cong H^p_{\bar D}(K)\:.
    \end{equation}
\end{proof}
It follows from Lemma \ref{lem:constr} that solutions of the equations of motion modulo gauge transformations are counted by
\begin{equation}
    [y]\in H^{0,1}_{\bar D}(Q)\:.
\end{equation}

\subsection{Comments on $\bar D$}
The kinetic operator $\bar D$ has several interesting features that we will comment on. Firstly, note that the off-diagonal extension maps involve holomorphic derivatives, and so they are not the usual tensorial maps as a holomorphic extension bundle would require. Rather, $(Q,\bar D)$ defines a kind of holomorphic sheaf, the properties of which we explore further in section \ref{sec:local-triv}. Such sheaves appear to be ubiquitous in higher-derivative heterotic supergravity \cite{McOrist:2021dnd, McOrist:2024zdz, deLazari:2024zkg, Chisamanga:2024xbm, Ashmore:2025fxr}, in addition to appearing elsewhere \cite{Kjelsnes:2026oox}.  

In spite of the non-tensorial structure of the extension maps, one can still parametrise the cohomology $H^{0,1}_{\bar D}(Q)$ in terms of long exact extension sequences and homological algebra. However, for Kähler backgrounds, the story simplifies. 

\begin{proposition}\label{prop:kahler}
    For Kähler backgrounds $X$, we have
    \begin{equation}
        H^{0,p}_{\bar D}(Q)\cong H^{2,p}_{\delb}(X)\oplus H_{\delb}^{0,p}({\mathfrak g})\oplus H_{\delb}^{1,p}(X)\:.
    \end{equation}
\end{proposition}
\begin{proof}
    Let us first write $\bar D$ in a form that makes the double extension clear
    \begin{equation}
        \bar D =
    \begin{pmatrix}
        \bar\partial & F\!\cdot\!\nabla & T \\
        0 & \bar\partial & F\!\cdot\!\nabla \\
        0 & 0 & \bar\partial
    \end{pmatrix} =
    \begin{pmatrix}
        \bar\partial & {\cal H} \\
        0 & \bar D_1
    \end{pmatrix}\:.
    \end{equation}
    The double extension then reads
    \begin{align}
        &0\rightarrow\Omega^{0,p}_{\delb}({\mathfrak g})\rightarrow\Omega^{0,p}_{\bar D_1}(Q_1)\rightarrow\Omega^{0,p}_{\delb}(T^{*1,0}X)\rightarrow 0\\
        &0\rightarrow\Omega^{2,p}_{\delb}(X)\rightarrow\Omega^{0,p}_{\bar D}(Q)\rightarrow\Omega^{0,p}_{\bar D_1}(Q_1)\rightarrow 0\:,
    \end{align}
    leading to long exact sequences in cohomology. The usual sequence chasing gives the cohomology
    \begin{equation}
        H^{0,p}_{\bar D}(Q)\cong \frac{H^{2,p}_{\delb}(X)}{{\rm Im}({\cal H})}\oplus\ker{{\cal H}}\:,
    \end{equation}
    where 
    \begin{equation}
        \label{eq:Hcoh}
        {\cal H}\::\;\;\;H^{0,p}_{\bar D_1}(Q_1)\rightarrow H^{2,p+1}_{\delb}(X)\:,
    \end{equation}
    where we use the same symbol for the map between cohomologies by slight abuse of notation. By construction, the map $\cal H$ maps a $\bar D_1$-closed form into $\del$-exact forms, which should also represent a class in $H^{2,p+1}_{\delb}(X)$. By the $\del\delb$-lemma, this class has to vanish, and so the cohomology map \eqref{eq:Hcoh} is trivial. It follows that
    \begin{equation}
        H^{0,p}_{\bar D}(Q)\cong H^{2,p}_{\delb}(X)\oplus H^{0,p}_{{\bar D}_1}(Q_1)\:,
    \end{equation}
    where
    \begin{equation}
        H^{0,p}_{{\bar D}_1}(Q_1)\cong\frac{H^{0,p}_{\delb}({\mathfrak g})}{{\Im}(F\!\cdot\!\nabla)}\oplus\ker(F\!\cdot\!\nabla)\:,
    \end{equation}
    for the cohomology map
    \begin{equation}
        F\!\cdot\!\nabla\::\;\;\;H^{1,p}_{\delb}(X)\rightarrow H^{0,p+1}_{\delb}({\mathfrak g})\:.    
    \end{equation}
    Next, note that the map $F\!\cdot\!\nabla$ involves the holomorphic derivative $\del x$ on a form $x\in\Omega^{0,p}(T^{*1,0}X)$. If $x$ is $\delb$-closed it again follows by the $\del\delb$-lemma and Bianchi identity for $F$ that $F\!\cdot\!\nabla x$ is trivial in cohomology. Hence
    \begin{equation}
        H^{0,p}_{{\bar D}_1}(Q_1)\cong H^{0,p}_{\delb}({\mathfrak g})\oplus H_{\delb}^{1,p}(X)\:.
    \end{equation}
    This completes the proof. 
\end{proof}
Although the operator $\bar D$ produces the standard cohomology on Kähler backgrounds, the operator still possesses intriguing geometric properties worth studying further. We elaborate on these properties in Section \ref{sec:local-triv}. In particular, $(Q,\bar D)$ provides an example of a ``quasi-holomorphic extension'' \cite{MurgasIbarra:2025ylk}, which as mentioned appears to be ubiquitous in higher-order supergravity, particularly for heterotic supergravity.  

\subsection{Anomalies in the corrected theory}
The anomalies of the corrected theory \eqref{eq:corrAction} are somewhat subtle. As mentioned, in order to do the path integral over off-shell modes, one has to pick a Lagrangian sub-manifold in the space of fields. This boils down to a gauge-fixing constraint on the fields. A common choice is Lorenz gauge, where we pick a metric on $Q$, and construct a corresponding adjoint $\bar D^\dagger$, and then only integrate over the kernel of $\bar D^\dagger$. However, as mentioned, the operator $\bar D$ is not a usual $(0,1)$-part of a connection on $Q$, due to the non-tensorial extension terms. The usual anomaly formula, deriving from transforming the measure in the path integral, hence does not apply directly for such operators. It is beyond the scope of this paper to explore how the anomaly formulas are corrected for such operators, but we will explore this in future works.

Instead, here we shall use the standard diagonal $\delb^\dagger$-closed gauge when performing the path integral. 
We argue below that, with this choice, the anomaly is expected to remain the same as for the classical action \eqref{eq:ClassActionQuad}. Still, we need to make sure that this actually fixes the gauge, or is elliptic in the appropriate sense. Recall from the discussion before \eqref{eq:Kcomplex} that the gauge transformations of the quadratic theory are
\begin{equation}
    \label{eq:fullgauge}
    y\;\mapsto\;y+\bar Dz+(0,0,\del\beta)\:,
\end{equation}
that is, besides the $\bar D$-exact transformations, the cotangent field may be shifted by a $\del$-exact form. The condition $\delb^\dagger y=0$ does not fix the second kind of transformation. Indeed, by the K\"ahler identity $\delb^\dagger\del=-\del\,\delb^\dagger$, for any $\gamma\in\Omega^{0,p+1}(X)$ the shift
\begin{equation}
    x\;\mapsto\;x+\del\,\delb^\dagger\gamma\:,
\end{equation}
preserves $\delb^\dagger y=0$, and is generically non-zero. We therefore supplement the $\delb^\dagger$-closed gauge by the extra constraint $\del^\dagger x=0$ on the cotangent field.

We now first show that $\delb^\dagger y=0$ leaves no residual $\bar D$-exact transformations, and then that the two conditions together fix the gauge completely (Proposition \ref{prop:gaugeFix}).

\begin{lemma}
    \label{lem:gaugeFix}
    If $y\in\Omega^{0,p}(Q)$ is both $\bar D$-exact and $\delb^\dagger$-closed, then $y=0$.
\end{lemma}
\begin{proof}
    Let us assume $y=\bar D z$ for some $z=(\chi,\alpha,x)\in\Omega^{0,p-1}(Q)$. We compute
    \begin{equation}
        \label{eq:gaugefix}
        \delb^\dagger\bar Dz=\begin{pmatrix}
        \delb^\dagger\bar\partial\chi +\delb^\dagger\left(F\!\cdot\!\nabla\alpha + T(x)\right) \\
        \delb^\dagger\bar\partial\alpha +\delb^\dagger\left(F\!\cdot\!\nabla x\right) \\
        \delb^\dagger\bar\partial x
    \end{pmatrix}=0\:.
    \end{equation}
    As $y=(..,..,\delb x)$, we may without loss of generality assume that $x$ is $\delb^\dagger$-closed. Or we can shift $z$ by $\bar D(0,0,b)=(..,..,\delb b)$ to ensure that $x$ in $\delb x$ is $\delb^\dagger$-closed. For compact $X$, the bottom line then ensures that $x$ is harmonic. As the extension maps acting on $x$ both involve $\del x$, it follows that
    \begin{equation}
        T(x)=F\!\cdot\!\nabla x=0\:.
    \end{equation}
    Hence,
    \begin{equation}
        y=\bar D(\chi,\alpha,x)=(\delb\chi+F\!\cdot\!\nabla\alpha,\delb\alpha,0)\:.
    \end{equation}
    The second line of \eqref{eq:gaugefix} then implies that $\delb\alpha=0$. As
    \begin{equation}
        F\!\cdot\!\nabla\alpha=\del\,\tr(F\wedge\alpha)\:,
    \end{equation}
    and $\tr(F\wedge\alpha)$ is $\delb$-closed, the $\del\delb$-lemma
    implies that $F\!\cdot\!\nabla\alpha$ is $\delb$-exact. The first line of \eqref{eq:gaugefix} then implies that 
    \begin{equation}
        \delb\chi+F\!\cdot\!\nabla\alpha=0\:.
    \end{equation}
    It follows that $y=\bar D z=0$.
\end{proof}

\begin{proposition}
\label{prop:gaugeFix}
Let $X$ be compact K\"ahler, and let $C^p=\{y\in\Omega^{0,p}(Q):\del\chi=0\}$ be the field space, with gauge transformations \eqref{eq:fullgauge} for $z\in C^{p-1}$ and $\beta\in\Omega^{0,p}(X)$. Then every $y\in C^p$ is gauge equivalent to exactly one field satisfying
\begin{equation}
    \label{eq:fullgaugecond}
    \delb^\dagger y=0\:,\qquad \del^\dagger x=0\:.
\end{equation}
\end{proposition}
\begin{proof}
We first show existence, beginning with $\delb^\dagger y=0$. That is, we wish to find a $\bar D z$, so that
\begin{equation}
    \label{eq:barDexist}
    \delb^\dagger(y+\bar Dz)=0\:.
\end{equation}
Since $\bar D$ is upper-triangular, we can solve \eqref{eq:barDexist} slot by slot from the bottom. The $x$-component of \eqref{eq:barDexist} reads
\begin{equation}
    \delb^\dagger y_x+\delb^\dagger\delb z_x=0\:.
\end{equation}
Assuming $z_x$ to be $\delb^\dagger$-exact without loss of generality, we find that this equation is solved by
\begin{equation}
    z_x=-\Delta^{-1}\delb^\dagger y_x\:,
\end{equation}
where $\Delta^{-1}$ is the Green's function of $\Delta_{\delb}=\Delta_{\del}$ on differential forms on the Calabi--Yau $X$. Next, the $\alpha$-component of \eqref{eq:barDexist} reads
\begin{equation}
    \delb^\dagger y_\alpha+\delb^\dagger\delb z_\alpha +\delb^\dagger\left( F\!\cdot\!\nabla z_x\right)=0\:.
\end{equation}
Again, assuming $z_\alpha$ to be $\delb^\dagger$-exact without loss of generality, we find 
\begin{equation}
    z_\alpha=-\Delta_{A}^{-1}\delb^\dagger\bigl(y_\alpha+F\!\cdot\!\nabla z_x\bigr)\:,
\end{equation}
where $\Delta_{A}^{-1}$ is the Green's function of the $\delb_A$-Laplacian on gauge-valued forms. Finally, a similar reasoning holds for the $\chi$-slot. Taking
\begin{equation}
    \label{eq:zsol}
    \begin{gathered}
    z_\chi=-\Delta^{-1}\delb^\dagger\bigl(y_\chi+F\!\cdot\!\nabla z_\alpha+T(z_x)\bigr)\:,
    \end{gathered}
\end{equation}
together with $z_x$ and $z_\alpha$ above then solves \eqref{eq:barDexist}. Note also that $z_\chi$ is $\del$-closed, since $\del y_\chi=0$ and the extension terms are $\del$-exact, and $\del$ (anti-)commutes appropriately with the Green's function and $\delb^\dagger$ by Kähler identities. 
We then reach $\del^\dagger x=0$ with the transformation $\beta=-\Delta^{-1}\del^\dagger\,x$, that is
\begin{equation}
    x\;\mapsto\;x'=x-\del\,\Delta^{-1}\del^\dagger\,x\:.
\end{equation}
It follows that $\del^\dagger x'=\del^\dagger x-\del^\dagger x=0$ and $\delb^\dagger x'=-\del\,\Delta^{-1}\del^\dagger\,\delb^\dagger x=0$ by Kähler identities, so the first condition is preserved, and the other slots are unchanged.

For uniqueness, let $y$ and $y'$ both satisfy \eqref{eq:fullgaugecond}, with $d\coloneqq y'-y=\bar Dz+(0,0,\del\beta)$. The cotangent slot $u\coloneqq\delb z_x+\del\beta$ of $d$ is both $\delb^\dagger$-closed and $\del^\dagger$-closed, so
\begin{equation}
    \|u\|^2=\langle\delb^\dagger u,z_x\rangle+\langle\del^\dagger u,\beta\rangle=0\:,
\end{equation}
where $\langle\,,\,\rangle$ is the standard inner-product on differential forms of $X$, given by the Calabi--Yau metric. Hence $\del\beta=-\delb z_x$. By the $\del\delb$-lemma, $\del\beta=\delb\del\eta$ for some $\eta$. It follows that
\begin{equation}
    d=y'-y=\bar Dz+\bar D(0,0,\del\eta)\:,
\end{equation}
where $\bar D(0,0,\del\eta)=(0,0,\delb\del\eta)=(0,0,\del\beta)$, since the extension maps act on the cotangent slot only through $(\del z_x)^a$, and $(\del\,\del\eta)^a=0$. Hence $d$ is $\bar D$-exact and $\delb^\dagger$-closed, and Lemma~\ref{lem:gaugeFix} gives $d=0$.
\end{proof}

Lemma~\ref{lem:gaugeFix} and Proposition~\ref{prop:gaugeFix} now show that the usual Lorenz gauge condition of being $\delb^\dagger$-closed, supplemented by $\del^\dagger x=0$, completely fixes the gauge.
It should follow that the usual anomaly formulas apply, though an in-depth analysis of this is left to future work.\footnote{Note that the additional condition $\del^\dagger x=0$ only involves the scalar parameter $\beta$, which is uncharged under the gauge bundle, and so should not affect the gauge-bundle part of the anomaly.} Briefly, the $(1,0)$-parts of connections used to construct the curvatures in the full gauge anomaly polynomial derive from the adjoint through the Lorenz gauge fixing condition. In the case at hand, if the adjoint is given by the diagonal
\begin{equation}
    \delb^\dagger=-*D*\:,
\end{equation}
where 
\begin{equation}
    D=\begin{pmatrix}
        \partial_\nabla & 0 & 0 \\
        0 & \partial_A & 0 \\
        0 & 0 & \partial_\nabla
    \end{pmatrix}\:,
\end{equation}
and where $\del_\nabla$ is the usual Chern-connection. The curvature
\begin{equation}
    F_{\bar D}=D\bar D+\bar D D
\end{equation}
is hence upper-triangular, and a trace of any polynomial in $F_{\bar D}$, such as anomaly polynomials, will pick out only the diagonal part. As mentioned, here we are only interested in the strict gauge-bundle part of the anomaly, with curvature $F=\delb\del_A+\del_A\delb = \delb A$. In particular, equation \eqref{eq:gaugeAnomaly}, describing the anomaly of an infinitesimal gauge transformation of the gauge bundle, is still expected to hold.

As discussed above, the Green--Schwarz type counter term \eqref{eq:counterChi} cancels the gauge anomaly when the Green--Schwarz transformation \eqref{eq:GSchi} is implemented. However, under a $\bar D$-exact gauge transformation, $\chi$ also gets transformed as
\begin{equation}
    \delta\chi=T(b)=-\tfrac{1}{2}\,\partial\left(\omega^{2,1}_a\,
      \partial_b b_c\,\bar\Omega^{abc}\right)\:,
\end{equation}
where $b\in\Omega^{1,0}(X)$. The counter-term \eqref{eq:counterChi} is not invariant under such a transformation. Note, however, that we can write the extra transformation of the counter term as
\begin{align}
    \int_X\omega^{1,2}\wedge\delta\chi=-\tfrac{1}{2}\int_X\tr(F^2)\wedge\omega^{2,1}_a\,
      \partial_b b_c\,\bar\Omega^{abc}\:.
\end{align}
A short computation shows that
\begin{equation}
    \delb\left(\omega^{2,1}\wedge\omega^{2,1}_a\right)=2\,\tr(F^2)\wedge\omega^{2,1}_a\:.
\end{equation}
The transformation of the counter term may then be written as
\begin{equation}
    \int_X\omega^{1,2}\wedge\delta\chi=-\tfrac14\int_X\omega^{2,1}\wedge\omega^{2,1}_a\partial_b \delb b_c\,\bar\Omega^{abc}\:.
\end{equation}
As $x$ transforms as $\delta x=\delb b$, this may then be cancelled by adding an additional counter term of the form
\begin{equation}
    S_{\rm counter}(x)\propto\int_X\omega^{2,1}\wedge\omega^{2,1}_a\partial_b x_c\,\bar\Omega^{abc}\:,
\end{equation}
ensuring, under the assumption that the extension maps of $\bar D$ do not modify the anomaly as argued above, that the corrected theory is anomaly free, at least at the one-loop level.\footnote{We again stress that the gravitational part of the anomaly has been ignored, which we will return to in future works.}

\subsection{The one-loop partition function}
Before we move to discuss the cubic extension of the quadratic theory, let us comment briefly on the one-loop partition function itself. The above discussion concerns anomalies related to the phase of the partition function. Indeed, the partition function should be viewed as a section of a (determinant) line bundle over the space of background geometric structures. If this line bundle has vanishing curvature (which follows from $[\tr(F\wedge F)]=0$ when gravitational anomalies are ignored), anomalies may be cancelled, and the partition function may be given a well-defined phase.\footnote{Vanishing curvature of the determinant line bundle cancels the local, or perturbative, anomalies. A flat line bundle can still have non-trivial holonomy around non-contractible loops in the space of background structures, which would signal a global anomaly \cite{Witten:1985xe}. We do not address global anomalies here.} 

As for the absolute value of the partition function, we first consider the one-loop partition function of ordinary holomorphic Chern--Simons theory, where the quadratic action reads
\begin{equation}
    S_2(\alpha)=\tfrac{1}{2}\int_X\tr(\alpha\delb\alpha)\wedge\Omega\:.
\end{equation}
The usual BV quantisation procedure then promotes $\alpha$ to a poly-field $\alpha\in\Omega^{0,\bullet}(\Lg)$, with bosonic fields of odd degree and fermionic fields of even degree. The one-loop partition function is then defined as
\begin{equation}
    Z=\int{\cal D\alpha}\,e^{iS_2(\alpha)}\:,
\end{equation}
where the integral is over Lorenz-constrained fields $\alpha\in\ker(\delb^\dagger)$. This produces the absolute value of the partition function, ignoring harmonic forms\footnote{Including harmonic volumes extends the holomorphic Ray--Singer torsion to the Quillen measure on the space of fields.}
\begin{equation}
    \vert Z\vert=\left(\prod_{p=0}^{3}\vert\Delta_{\delb}^{0,p}\vert^{(-1)^{p+1}p}\right)^{\tfrac{1}{4}}=I_{\delb}(\Lg)^{\tfrac{1}{2}}\:,
\end{equation}
where $\Delta^{0,p}_{\delb}$ are the Laplacians of $\delb$ on the corresponding gauge-valued $(0,p)$-forms, and $I_{\delb}(\Lg)$ is the holomorphic Ray--Singer torsion. 

Turning to the theory at hand \eqref{eq:corrAction}, one might expect a similar form of the absolute value of the partition function as $\vert Z\vert=I(Q)^{\tfrac{1}{2}}$. This however comes with the subtlety that the tangent-valued fields are divergence-free, together with the extra gauge transformation of the cotangent-valued fields by $\del$-exact terms. Following the computation given at the start of Section 3 of \cite{Ashmore:2025fxr}, these correct the one-loop partition function to become
\begin{equation}
    \vert Z\vert=\left(\prod_{p=0}^{3}\vert\Delta_{\bar D}^{0,p}\vert^{(-1)^{p+1}p}\right)^{\tfrac{1}{4}}\left(\prod_{p=0}^{3}\vert\Delta_{\delb}^{0,p}\vert^{(-1)^{p+1}p}\right)^{-\tfrac{1}{2}}=I_{\bar D}(Q)^{\tfrac{1}{2}}\left(I_{\delb}^0\right)^{-1}\:,
\end{equation}
where $I_{\delb}^0$ is the holomorphic Ray--Singer torsion of $(0,p)$-forms. 

Choosing the diagonal gauge above, where we impose $\bar D^\dagger=\delb^\dagger$, the determinants of Laplacians reduce to the determinants of Laplacians of the individual bundles, and the holomorphic Ray--Singer torsion and partition function split as
\begin{equation}
    \label{eq:DiagPartFunc}
    \vert Z\vert^2=\frac{I_{\delb}(T^{*1,0}X)I_{\delb}(\Lg)I_{\delb}(T^{1,0}X)}{\left(I_{\delb}^0\right)^2}\:.
\end{equation}
The metric dependence and corresponding holomorphic anomaly equations may be extracted from the results of Bismut et al. \cite{bismut1988analytic1, bismut1988analytic2, bismut1988analytic3}, and were studied explicitly in \cite{Ashmore:2023vji} for the combination of Ray--Singer torsions given by \eqref{eq:DiagPartFunc}. 

Perhaps more naturally, or even more rigorously, if one instead picks a metric on $Q$ and defines the corresponding adjoint $\bar D^\dagger$ with respect to this metric, the expressions for the corresponding Laplacians and determinants become more involved. Again, as $\bar D$ is not the $(0,1)$-part of a connection, the geometric dependence of the corresponding holomorphic Ray--Singer torsion is not covered by the theorems of Bismut et al. It is beyond the current paper to investigate this dependence, but we will return to it in future works.

\section{The cubic theory}
\label{sec:cubic}

The quadratic theory \eqref{eq:corrAction} of Section~\ref{sec:quad} is governed by the nilpotent operator $\bar D$ of \eqref{eq:barD}. The main goal of this section is to extend the theory to a cubic interacting theory, similar to the cubic heterotic Kodaira--Spencer theory of \cite{Ashmore:2018ybe}, but with the present operator $\bar D$ -- a dual heterotic Kodaira--Spencer theory. To do so, we require a bracket on $Q$ compatible with $\bar D$ and with the pairing $\langle\,,\,\rangle$. In this section we construct that bracket, establish its graded anti-symmetry, prove that $\bar D$ acts on it as a graded derivation, and show that the pairing is cyclic. Together these are the conditions needed for the cubic action
\eqref{eq:cubicAction} below to be well defined and gauge covariant.

Throughout this section we write the field triple as $y=(\mu,\alpha,x)$,
where $\mu\in\Omega^{0,p}(T^{1,0}X)$ is the Beltrami-type vector of
Section~\ref{sec:quad}, related to the two-form field via
\begin{equation}
    \chi=\iota_\mu\Omega=\tfrac12\,\mu^a\Omega_{abc}\,\dd z^{bc}\:,
\end{equation}
through the isomorphism $\Lambda^2T^{*1,0}X\cong T^{1,0}X$ given by $\Omega$. In the frames of Convention~\ref{conv:frame} below the divergence-free condition \eqref{eq:divfree} reads \begin{equation}\label{eq:constraint}   \partial_a\mu^a=0\:, \end{equation} and we refer to the locus of fields satisfying \eqref{eq:constraint} as the constraint surface, 
and the pairing of \eqref{eq:corrAction} is rewritten in this description in Definition~\ref{def:pairing-mu}.

\subsection{Normalisation conventions}
\label{sec:normalisation}
Before we begin, we note that the following convention is in force throughout this section and is used retroactively in parts of Section~\ref{sec:quad}. 

\begin{convention}[Koszul signs]\label{conv:koszul}
All the fields are $(0,p)$-form valued, and Koszul signs are retained
throughout: whenever two factors of odd total parity are transposed
relative to the printed order, the corresponding sign $(-1)^{pq}$ is
displayed explicitly. Component formulas below are to be read in this
graded sense.
\end{convention}

\begin{convention}[Frame for $\Omega$]\label{conv:frame}
The component computations are performed in local holomorphic frames in
which the coefficients $\Omega_{abc}$ are constant, so that
$\partial_d$ passes freely through $\Omega_{abc}$ and
$\bar\Omega^{abc}$. Equivalently, holomorphic derivatives on tensor
indices may be read as the Chern connection of the Ricci-flat metric,
with respect to which $\Omega$ is covariantly constant. In a general
coordinate frame $\bar\Omega^{abc}=f^{-1}\epsilon^{abc}$ with $f$
holomorphic and non-constant, and $\partial_a\bar\Omega^{abc}\neq0$.
\end{convention}

The bracket below involves both $\Omega_{abc}$ and $\bar\Omega^{abc}$,
and we fix their relative normalisation by the convention
\begin{equation}
    \label{eq:Omnorm}
    \bar\Omega^{abc}\,\Omega_{dbc}=2\,\delta^a_d\:,
\end{equation}
and write throughout
\begin{equation}
    \label{eq:dxdef}
    (\partial x)^a\coloneqq \bar\Omega^{abc}\,\partial_b x_c\:,
\end{equation}
with no additional factor. Note that \eqref{eq:dxdef} is automatically
divergence free, as
\begin{equation}
    \label{eq:autodiv}
    \partial_a(\partial x)^a
    =\bar\Omega^{abc}\,\partial_a\partial_b x_c=0\:,
\end{equation}
by anti-symmetry of $\bar\Omega^{abc}$.

We also note two consequences of the normalisation \eqref{eq:Omnorm} in complex dimension three, which will be used repeatedly:
\begin{align}
    \bar\Omega^{cba}\,\Omega_{cde}
      &=\delta^b_d\,\delta^a_e-\delta^b_e\,\delta^a_d\:,
      \label{eq:OO}\\
    \Omega_{cba}\,(\partial x)^c
      &=\partial_b x_a-\partial_a x_b\:,
      \label{eq:inv}
\end{align}
the second of which inverts the map \eqref{eq:dxdef}. We also record the dimension-three Schouten identity, the vanishing of the anti-symmetrisation of four indices in three dimensions,
\begin{equation}
    \label{eq:schouten}
    \delta^d_a\Omega_{bce}-\delta^d_b\Omega_{ace}
    +\delta^d_c\Omega_{abe}-\delta^d_e\Omega_{abc}=0\:.
\end{equation}
We shall use repeatedly the contraction identity
\begin{equation}
    \label{eq:contrOm}
    \tfrac12\,\Omega_{abc}\,\dd z^{bc}\wedge\dd z^d
    =\delta^d_a\,\Omega\:,
\end{equation}
which holds precisely when the components of $\Omega$ are normalised as
\begin{equation}
    \Omega=\tfrac16\Omega_{abc}\,\dd z^{abc}\:.
\end{equation}

We manipulate vector fields of the form \eqref{eq:dxdef} using the Cartan
calculus, for which we record the following. For a $(0,1)$-form valued (graded) vector
field $V$ the contraction $\iota_V$ has even total parity, so $[\partial,\iota_V]=L_V$ defines the graded Lie derivative on holomorphic indices, and $[L_{V_i},\iota_{V_j}]=-\iota_{[V_i,V_j]}$ for the graded Lie bracket
\begin{equation}
[V_i,V_j]^a=V_i^b\partial_bV_j^a+V_j^b\partial_bV_i^a\:.    
\end{equation}
Note that in this language, \eqref{eq:autodiv} is the statement
\begin{equation}
    \label{eq:LVOmega}
    L_{V_i}\Omega=\partial\,\iota_{V_i}\Omega
    =-\bigl(\partial_aV_i^a\bigr)\,\Omega=0\:,
\end{equation}
i.e. $\Omega$ is preserved by divergence-free (graded) diffeomorphisms.

\subsection{The pairing in the vector description}
\label{sec:pairing-mu}

The pairing $\langle\,,\,\rangle$ of Section~\ref{sec:quad} was written
in terms of the $(2,1)$-form field $\chi$. For convenience, in the current section we work with the vector $\mu$ instead, and we record the pairing here in that description. Throughout, component $(0,p)$-forms are written to the left of the holomorphic basis forms, so that $x=x_a\,\dd z^a$ and $\chi=\tfrac12\mu^a\Omega_{abc}\,\dd z^{bc}$. The one exception is the curvature, for which we keep the convention of Section~\ref{sec:quad}, \begin{equation}   F=\dd z^a\wedge F_a\:,\qquad F_a\coloneqq\iota_{\del_a}F\:; \end{equation} since $F_a$ is a one-form, writing $F=F_a\wedge\dd z^a$ instead would flip the sign of $F_a$. 

\begin{definition}[Pairing in the vector description]
\label{def:pairing-mu}
For $y=(\mu,\alpha,x)\in\Omega^{0,p}(Q)$ and
$y'=(\mu',\alpha',x')\in\Omega^{0,q}(Q)$,
\begin{equation}
    \label{eq:pairing-mu}
    \langle y,y'\rangle
    =\bigl(\mu^a\,x'_a+x_a\,\mu'^a
      -\tr(\alpha\wedge\alpha')\bigr)\wedge\Omega\:.
\end{equation}
\end{definition}

This is the pairing of Section~\ref{sec:quad}, not a new one. Indeed,
by \eqref{eq:contrOm} the first term of that expression is
\[
\chi\wedge x'=\tfrac12\,\mu^a\Omega_{abc}\,x'_d\,
\dd z^{bc}\wedge\dd z^d=\mu^a x'_a\wedge\Omega\:,
\]
with no sign, the component $x'_d$ crossing the even form
$\dd z^{bc}$; while
\[
x\wedge\chi'=(-1)^q\,x_a\,\mu'^a\wedge\Omega\:,
\]
the sign arising as $\mu'^b$ crosses the odd form $\dd z^a$. This is
exactly cancelled by the explicit prefactor $(-1)^{y'}$ of
Section~\ref{sec:quad}, so that the only explicit sign left in
\eqref{eq:pairing-mu} is the one carried by the gauge term. In this
form Lemma~\ref{lem:sym} becomes immediate: transposing the
two component form factors in each of the three terms of
\eqref{eq:pairing-mu} produces the single factor $(-1)^{pq}$, using
$\tr(\alpha'\wedge\alpha)=(-1)^{pq}\tr(\alpha\wedge\alpha')$ in the
gauge term.

As a check on the normalisations \eqref{eq:Omnorm},
\eqref{eq:contrOm} and \eqref{eq:pairing-mu} taken together, the
quadratic action \eqref{eq:corrAction} reads, ignoring the
extension maps,
\[
\tfrac12\int_X\langle y,\bar Dy\rangle
=\tfrac12\int_X\bigl(\mu^a\delb x_a+x_a\delb\mu^a
 -\tr(\alpha\,\delb\alpha)\bigr)\wedge\Omega
=\int_X\bigl(x_a\delb\mu^a
 -\tfrac12\tr(\alpha\,\delb\alpha)\bigr)\wedge\Omega\:,
\]
where we have integrated by parts in the second equality. Note that this is exactly \eqref{eq:ClassActionQuad}.

\subsection{The bracket}
\label{sec:bracket}
We now come to define the bracket on $\Omega^{0,p}(Q)$. With some trial and error, we find a bracket that satisfies the Leibniz rule with respect to $\bar D$, 
\begin{equation}
    [\:,\:]\::\;\;\;\Omega^{0,p}(Q)\times\Omega^{0,q}(Q)\rightarrow\Omega^{0,p+q}(Q)\:,
\end{equation}
to be defined as follows.
\begin{definition}
\label{def:bracket}
For $y=(\mu,\alpha,x)\in\Omega^{0,p}(Q)$ and $y'=(\mu',\alpha',x')\in\Omega^{0,q}(Q)$, the bracket
$[\,y,y'\,]$ has components
\begin{align}
    [y,y']^a&= -\Bigl(\mu^b\wedge\partial_b(\partial x')^a
      - \partial_b\mu^a\wedge(\partial x')^b
      + (\partial x)^b\wedge\partial_b\mu'^a
      - \partial_b(\partial x)^a\wedge\mu'^b\Bigr)\notag\\
    &-\tfrac{1}{2}\Bigl(\nabla_c\tr(\alpha\nabla_b\alpha')
      -(-1)^{pq}\,\nabla_c\tr(\alpha'\nabla_b\alpha)\Bigr)
      \bar\Omega^{cba}\:,
\label{eq:bracket_tangent}
      \\
    [y,y']_{\mathfrak g}
    &= [\alpha,\alpha']_{\mathfrak g}
      - (\partial x)^a\nabla_a\alpha' + (-1)^{pq}\,(\partial x')^a\nabla_a\alpha\:,
\label{eq:bracket_gauge}
    \\
    [y,y']_a&= \Omega_{abc}\,(\partial x)^b\wedge(\partial x')^c\:.
\label{eq:bracket_cotangent}
\end{align}
\end{definition}
It is a straightforward computation to check that the tangent-valued slot of the bracket remains divergence free on the constraint surface \eqref{eq:constraint}. Note also that the bracket is of second order in holomorphic derivatives, in contrast to type I BCOV theory \cite{Costello:2015xsa, Costello:2021kiv}, where $(\del x)^a$ is promoted to an independent divergence-free field.\footnote{Similar higher-derivative brackets have appeared elsewhere in the literature, see e.g. \cite{Hohm:2014eba, Hohm:2026sff}.}

Note also that the three components of the bracket are not independent. The cotangent component \eqref{eq:bracket_cotangent} is characterised by the requirement that it reproduce the Lie bracket of the associated holomorphic vector fields $V^a=(\partial x)^a$ via
\begin{equation}
    \label{eq:cot_char}
    \bigl(\partial\,[y,y']_\bullet\bigr)^a
    = -\,[V,V']^a
    = -\Bigl(V^b\partial_bV'^a-(-1)^{pq}\,V'^b\partial_bV^a\Bigr)\:,
\end{equation}
where $[\,,\,]$ is the graded Lie bracket on vector-valued forms.
This follows from \eqref{eq:dxdef}, \eqref{eq:OO} and the divergence-freeness \eqref{eq:autodiv} of $V$ and $V'$, and it is the property from which \eqref{eq:bracket_cotangent} was obtained. Note that since \eqref{eq:dxdef} has a kernel, \eqref{eq:cot_char} fixes $[y,y']_a$ only up to elements annihilated by $\partial$. This ambiguity is the origin of the higher $L_3$-bracket of Section~\ref{sec:linfty}.

The next comment concerning the switching of the roles of $x$ and $\mu$ in the dual theory, as compared with the theory of \cite{Ashmore:2018ybe}, deserves its own remark:

\begin{remark}
\label{rem:nomumu}
Note that the bracket contains no $[\mu,\mu']$ term. That is, $\mu$ enters linearly in every component in which it is present. Comparing with \cite{Ashmore:2018ybe}, the corresponding self-interaction of ``complex-structure deformations'' is carried entirely by $V^a=(\partial x)^a$. This is the exchange of the roles of $\mu$ and $x$ relative to type I BCOV and heterotic Kodaira--Spencer theory \cite{Costello:2015xsa, Ashmore:2018ybe, Costello:2021kiv}. Correspondingly, the Yukawa cubic $W=\Omega_{abc}V_1^aV_2^bV_3^c$ appears both as the on-shell $(\mu,x,x)$ vertex \eqref{eq:yukawa} and as the potential of the Jacobiator in Proposition~\ref{prop:jacobi}.
\end{remark}

We now show that the bracket has the appropriate symmetry properties, and satisfies the graded Leibniz rule.
\begin{lemma}
\label{lem:brasym}
The bracket of Definition~\ref{def:bracket} is graded antisymmetric,
\begin{equation}
    \label{eq:brasym}
    [y',y]=-(-1)^{pq}\,[y,y']\:.
\end{equation}
\end{lemma}
\begin{proof}
We check each component separately. In \eqref{eq:bracket_tangent} the four terms of the first line are exchanged in pairs under $y\leftrightarrow y'$, the first with the fourth and the second with the third, each exchange producing the Koszul factor $(-1)^{pq}$ from commuting the $(0,p)$- and
$(0,q)$-forms past one another. In the second line the two summands
are interchanged at the cost of the explicit prefactor $-(-1)^{pq}$,
which is the required sign. In \eqref{eq:bracket_gauge} the term
$[\alpha,\alpha']_{\mathfrak g}$ is graded antisymmetric because
the pointwise Lie bracket is, and the remaining two terms are
exchanged into one another. In \eqref{eq:bracket_cotangent} the
exchange costs $(-1)^{pq}$ from the two form factors and a further
$-1$ from the anti-symmetry of $\Omega_{abc}$ in $b$ and $c$.
\end{proof}

The bracket has been constructed to satisfy the graded Leibniz rule, i.e.\ $\bar D$ is a graded derivation on the bracket.

\begin{proposition}
\label{prop:leibniz}
The operator $\bar D$ acts as a graded derivation of the bracket,
\begin{equation}
    \label{eq:leibniz}
    \bar D[y,y']
    = [\bar Dy,y']+(-1)^{p}\,[y,\bar Dy']\:.
\end{equation}
\end{proposition}
\begin{proof}
See Appendix~\ref{app:leibniz} for the full proof. We give a brief sketch of the proof here:

Since $\bar D$ is block upper-triangular, \eqref{eq:leibniz} splits into one identity per component of the bracket, and within each the terms group by which of $\mu,\alpha,x$ occur. The cotangent component is the Leibniz rule for $\bar\partial$, using that $\Omega_{abc}$ is holomorphic. The gauge component uses the commutator $[\bar\partial,\nabla_a]\alpha=-[F_a,\alpha]$ and the Bianchi identity $\nabla_aF_b=\nabla_bF_a$, the latter supplying the anti-symmetry which cancels the bilinear terms in $x$. The tangent component uses the same commutator for the bilinear terms in $\alpha$, while the terms bilinear in $x$ are linear in $\omega^{2,1}$ and do not involve its $\delb$-derivative; they cancel by the second condition in \eqref{eq:dCScomp}, and not for a $(2,1)$-form satisfying only the first. It is this computation which fixes the relative coefficients in Definition~\ref{def:bracket}.
\end{proof}

Next we show that the pairing is appropriately cyclic, in the sense that we have the following lemma.

\begin{lemma}
\label{lem:cyclic}
The pairing is cyclic with respect to the bracket: for
$y_i\in\Omega^{0,p_i}(Q)$ with $p_1+p_2+p_3=3$,
\begin{equation}
    \label{eq:cyclic}
    \int_X\langle y_1,[y_2,y_3]\rangle
    =(-1)^{p_3(p_1+p_2)}\int_X\langle y_3,[y_1,y_2]\rangle\:,
\end{equation}
which at degrees $(1,1,1)$ carries no sign. The families (i) and (ii) of the
proof hold off the constraint surface. Family (iii) uses $\del_a\mu^a=0$
once, and the term it produces off the constraint surface is recorded in
\eqref{eq:cyclicdefect} below.
\end{lemma}
\begin{proof}
See Appendix~\ref{app:cyclic} for the full proof. We give a brief sketch of the proof here:

As both sides of \eqref{eq:cyclic} are trilinear, it suffices to check the identity on arguments each of which lies in a single summand of $Q$. Matching the field content of the bracket against that of the pairing leaves
exactly three families of unordered content, $(\alpha,\alpha,\alpha)$,
$(x,\alpha,\alpha)$ and $(\mu,x,x)$, which we denote by (i), (ii), and (iii) respectively. The first is the cyclicity of the trace. The second is one holomorphic integration by parts, and is what fixes the coefficient $-\tfrac12$ of the gauge bilinear in \eqref{eq:bracket_tangent}, consistently with Proposition~\ref{prop:leibniz}. The third uses the inversion identity
\eqref{eq:inv} together with $\del_a\mu^a=0$, and fixes the relative
normalisation between \eqref{eq:bracket_cotangent} and the first line of
\eqref{eq:bracket_tangent}.
\end{proof}

Note that the third family evaluates the $(\mu,x,x)$ Yukawa vertex, which is worth emphasizing for later. Using \eqref{eq:contrOm}, it reads
\begin{equation}
\label{eq:yukawa}
\int_X\langle\mu_1,[x_2,x_3]\rangle
=\int_X\Omega_{aef}\,\mu_1^a\,V_2^e\,V_3^f\wedge\Omega\:,
\end{equation}
where we will often write $V_i^a=(\partial x_i)^a$.

\subsection{The cubic action}
\label{sec:cubicAction}
With Proposition~\ref{prop:leibniz} and Lemma~\ref{lem:cyclic} in place
we may extend the quadratic action \eqref{eq:corrAction} to a cubic action. Notably, a {\it dual Kodaira--Spencer} action is of the form
\begin{equation}
    \label{eq:cubicAction}
    S=\int_X\Bigl(\tfrac{1}{2}\langle y,\bar Dy\rangle
      +\tfrac{1}{6}\langle y,[y,y]\rangle\Bigr)\:.
\end{equation}
This action should be compared with the action of equation $(6.4)$ of \cite{Ashmore:2018ybe}, possibly with different sign conventions. Note that cyclicity guarantees that the cubic term is well defined, that is,
independent of which of the three arguments is placed in the first slot
of the pairing. This is a useful relation for when we come to perform gauge transformations of the action.

The infinitesimal gauge transformation leaving the cubic action \eqref{eq:cubicAction} invariant is
\begin{equation}
    \label{eq:gaugeMC}
    \delta y=\bar D\Lambda+[y,\Lambda]+\del\beta\:,
    \quad \Lambda=(\xi_\mu,\xi_\alpha,\xi_x)\in\Omega^{0,0}(X,Q)\:,\quad \del_a\xi_\mu^a=0\:,\quad \beta\in\Omega^{0,1}(X)\:,
\end{equation}
where $\del\beta$ is in the cotangent slot, and is an allowed gauge transformation due to the divergence-free constraint on $\mu$, and the properties of $\bar D$ and the bracket. The restriction to divergence-free $\xi_\mu$ is the condition for \eqref{eq:gaugeMC} to preserve the constraint surface. 
Under the gauge transformation \eqref{eq:gaugeMC}, using Lemmas~\ref{lem:IBP} and~\ref{lem:cyclic}, the action transforms as
\begin{equation}
    \delta S=\tfrac{1}{6}\int_X\langle\Lambda,J(y,y,y)\rangle
    =\tfrac{1}{6}\int_X(\del_a\xi_\mu^a)\,W\wedge\Omega\:,
\end{equation}
which vanishes by the restriction $\del_a\xi_\mu^a=0$ imposed in \eqref{eq:gaugeMC}. Here $J$ is the Jacobiator \eqref{eq:Sdef} defined below, and the second equality uses Proposition~\ref{prop:jacobi}, by which $J$ is $\del$-exact, together with holomorphic integration by parts. Note that if the restriction were dropped, it is the first equality that would acquire further terms proportional to $\del_a\xi_\mu^a$, since Lemma~\ref{lem:cyclic} was proved on the constraint surface. We investigate this in Section \ref{sec:MCext}, where we construct an extended action along the lines of equation $(4.23)$ of \cite{Ashmore:2018ybe}, which was derived from deforming the heterotic superpotential.

Varying \eqref{eq:cubicAction} and using Lemmas~\ref{lem:IBP} and~\ref{lem:cyclic} gives the equation of motion
\begin{equation}
    \label{eq:MC}
    \bar Dy+\tfrac{1}{2}[y,y]=\del\text{-exact}\:,
\end{equation}
where the right hand side $\del$-exact part concerns the cotangent slot of $Q$, and appears due to the divergence-free constraint on $\mu$. Throughout the present section that constraint is imposed on the off-shell field space. In Section~\ref{sec:MCext} it is relaxed, and the $\del$-exact remainder on the right hand side of \eqref{eq:MC} is identified.\footnote{\label{foot:quotient} One could instead consider a field space where we quotient the cotangent slot by $\del$-exact forms, i.e. the space of fields is given by \eqref{eq:Kcomplex}. Here the Jacobiator of Proposition~\ref{prop:jacobi}, being $\del$-exact, vanishes identically and \eqref{eq:MC} becomes the Maurer--Cartan equation of a differential graded Lie algebra, or $L_2$-algebra. We do not take this route, since Section~\ref{sec:extended} instead gives the $\del$-exact remainder a name, and the full structure is retained when extended to an $L_3$-algebra.} 

A gauge transformation of the left-hand side of the equation of motion gives, using Proposition~\ref{prop:leibniz},
\begin{equation}
    \label{eq:MCvar}
    \delta\Bigl(\bar Dy+\tfrac12[y,y]\Bigr)
    =\bigl[\bar Dy+\tfrac12[y,y],\;\Lambda\bigr]
    +\delb\del\beta-\tfrac12\,J(\Lambda,y,y)\:,
\end{equation}
with $J$ the Jacobiator \eqref{eq:Sdef} below, so that the Maurer--Cartan equation is covariant up to the Jacobiator. By Proposition~\ref{prop:jacobi}, the obstruction $J(\Lambda,y,y)$ is valued purely in the cotangent slot, where it is $\del$-exact, and depends only on the cotangent components of $\Lambda$ and $y$. It vanishes identically when $\Lambda$ has no cotangent component, and is otherwise absorbed by the $\ell_3$-corrected gauge transformations of the extended structure of Section~\ref{sec:extended}.

\subsection{Failure of the Jacobi identity}
\label{sec:linfty}

As mentioned, the bracket of Definition~\ref{def:bracket} does not satisfy the graded Jacobi identity. However, the failure is very mild. We now compute this failure:

\begin{proposition}
\label{prop:jacobi}
Write
\begin{equation}
    \label{eq:Sdef}
    J(y_1,y_2,y_3)\coloneqq [[y_1,y_2],y_3]
    +(-1)^{(p_1+p_2)p_3}[[y_3,y_1],y_2]
    +(-1)^{(p_2+p_3)p_1}[[y_2,y_3],y_1]\:.
\end{equation}
Then $J$ vanishes identically in the tangent and gauge components, and
in the cotangent component is the total derivative
\begin{equation}
    \label{eq:Sresult}
    J(y_1,y_2,y_3)_a=-\,\partial_a W\:,
    \qquad
    W\coloneqq \Omega_{bcd}\,(\partial x_1)^b\wedge(\partial x_2)^c
       \wedge(\partial x_3)^d\:.
\end{equation}
In particular, the obstruction depends only on the cotangent components
of its arguments.
\end{proposition}

\begin{proof}
The complete proof is given in Appendix~\ref{app:jacobi}. Here we give a brief sketch of the proof. 

The tangent and gauge components vanish identically. By field content, each reduces to a graded Jacobi identity, for the Lie bracket on vector-valued forms and for the bracket on $\mathfrak g$ respectively, the gauge component using $F^{2,0}=0$ once, where the nested transport terms produce a curvature commutator. The cotangent component is Cartan calculus $W=\iota_{V_1}\iota_{V_2}\iota_{V_3}\Omega$, and applying $\partial$ with $\partial\Omega=0$ produces exactly the three nested brackets of \eqref{eq:Sdef}. The only divergence-freeness used is that of the vector fields $V_i=(\del x_i)$, which holds identically by \eqref{eq:autodiv}. It enters through \eqref{eq:LVOmega}, removing the terms in which a Lie derivative reaches $\Omega$; for general vector fields these terms would survive and $J$ would not be a gradient. No condition on $\mu$ is needed.
\end{proof}

Two features of \eqref{eq:Sresult} are worth recording. First, the
obstruction is annihilated by the map \eqref{eq:dxdef}: applying
$\bar\Omega^{ajk}\partial_j$ to \eqref{eq:Sresult} gives zero, by the
same anti-symmetry as \eqref{eq:autodiv}. Equivalently, the vector
fields $V^a=(\partial x)^a$ satisfy the Jacobi identity exactly, and
the failure of Jacobi for the bracket lies entirely in the kernel of
$x\mapsto(\partial x)^a$. This is precisely the ambiguity left open by
the characterisation \eqref{eq:cot_char}. Second, $W$ may be thought of as the Yukawa cubic of a Kodaira--Spencer theory, where $(\del x)^a$ plays the role of the Beltrami differential. 

It is natural to wonder if the bracket is the binary operation of an $L_\infty$-structure, with $\ell_1=\bar D$, $\ell_2=[\,,\,]$, and a $3$-bracket $\ell_3$ witnessing \eqref{eq:Sresult} through the generalised Jacobi identity\footnote{The signs are those for an $\ell_3$ which lowers the degree of the complex by one while preserving the anti-holomorphic form degree of its arguments, as the $\ell_3$ constructed in Section~\ref{sec:extended} does. The relative minus sign on the insertion terms and the factor $(-1)^{p_1+p_2+p_3}$ are the Koszul signs
of that shift, and they are verified component by component in Proposition~\ref{prop:relations}. After the substitution $\ell_3\mapsto(-1)^{p_1+p_2+p_3}\ell_3$ the relation takes the standard form \cite{Lada:1994mn} in which all terms enter with a plus sign.}
\begin{multline}
    \label{eq:Linf3}
    \ell_1\bigl(\ell_3(y_1,y_2,y_3)\bigr)
    -\sum_{i}(-1)^{p_1+\dots+p_{i-1}}
      \ell_3(y_1,\dots,\ell_1 y_i,\dots,y_3)\\
      +(-1)^{p_1+p_2+p_3}\,J(y_1,y_2,y_3)=0\:.
\end{multline}
However, we find no obvious candidate for $\ell_3$ which works. We have the following remark:

\begin{remark}
\label{prop:noQl3}
No such obvious (local) $\ell_3$ exists on $Q$ itself. It would have to map
\begin{equation}
    \Omega^{0,p_1}(Q)\otimes\Omega^{0,p_2}(Q)\otimes\Omega^{0,p_3}(Q)\rightarrow\Omega^{0,p_1+p_2+p_3-1}(Q)\:.
\end{equation}
Both $\bar D$ and the bracket are built from
wedge products of the component $(0,p)$-forms, holomorphic derivatives, and
the tensors $\Omega$, $\bar\Omega$, $F$ and $\omega^{2,1}$, and any trilinear
operation of this kind takes arguments of anti-holomorphic degrees
$p_1,p_2,p_3$ to a $(0,p_1+p_2+p_3)$-form, one degree too high. Lowering the
form degree would require contracting an anti-holomorphic index, using a
Hermitian metric, which is foreign to the holomorphic setting, or resorting to non-local operators such as $\delb^{-1}$. 
\end{remark}

The obstruction is one of degree rather than of coefficient, and it is removed
by enlarging the complex by a slot whose anti-holomorphic form degree exceeds
its degree in the original complex by one, so that a $(0,p_1+p_2+p_3)$-form can sit in
degree $p_1+p_2+p_3-1$.
This mimics the $L_3$-algebra of \cite{Ashmore:2018ybe}, and is what we do next.

\subsection{The extended complex, and an \texorpdfstring{$L_3$}{L3}-algebra}
\label{sec:extended}

Note that a scalar $(0,\bullet)$-form is already implicit in three places in the foregoing: as the unnamed $\del$-exact term on the right of the equation of motion \eqref{eq:MC}, as the parameter $\beta$ of the extra gauge transformation in \eqref{eq:gaugeMC}, and as the quotient by $\del$-exact forms noted in footnote \ref{foot:quotient}. These are the same object seen three times. Naming it removes the quotient, makes \eqref{eq:gaugeMC} structural, and gives $\ell_3$ somewhere to live.

\begin{definition}
\label{def:extended}
Set
\begin{equation}
    \label{eq:Ydef}
    \mathcal{Y}_n\coloneqq \Omega^{0,n}(Q)\oplus\Omega^{0,n+1}(X),
    \qquad Y=(y,b),
\end{equation}
so that the physical fields sit at $n=1$, with $y\in\Omega^{0,1}(Q)$ and
$b\in\Omega^{0,2}(X)$, and the gauge parameters at $n=0$. We write
$(y,y')$ for the pairing \eqref{eq:pairing-mu} with the factor $\wedge\,\Omega$
stripped, so that $\langle y,y'\rangle=(y,y')\wedge\Omega$ and
$(y,\del b)=\mu^a\del_ab$.
\end{definition}

Note that no divergence-free condition is imposed on $\mathcal{Y}$: the $\mu$-component of $y$ is now an arbitrary section of $\Omega^{0,n}(T^{1,0}X)$. The $L_\infty$ relations established below use the divergence-freeness only of the vector fields $(\del x)^a$, which holds identically by \eqref{eq:autodiv}, and never that of $\mu$. The constraint \eqref{eq:constraint} enters only through the cyclicity of the pairing, Lemma~\ref{lem:cyclic}, and hence only through the action. In Section~\ref{sec:MCext} it is imposed by the equation of motion of the auxiliary field $b$, as in \cite{Ashmore:2018ybe}.  Note that the auxiliary slot carries form degree one above the
$\mathcal{Y}$-degree, so that $\del b$ lands in the cotangent slot of $Q$ one
degree up. This is what allows it to absorb the obstruction of
Remark~\ref{prop:noQl3}. The image $\del b$ of $b$ in $Q$ lies in the kernel of $x\mapsto(\del x)^a$, which is precisely where the failure of Jacobi was located above.

\begin{definition}
\label{def:products}
On $\mathcal{Y}$ define
\begin{align}
    \ell_1(Y)&=\bigl(\bar Dy+\del b,\ \delb b\bigr),
    \label{eq:l1def}\\
    \ell_2(Y,Y')&=\bigl([y,y'],\ 0\bigr),
    \label{eq:l2def}\\
    \ell_3(Y,Y',Y'')&=\bigl(0,\ \Omega_{abc}\,V^aV'^bV''^c\bigr)
      =\bigl(0,\ W\bigr),
    \label{eq:l3def}\\
    \ell_{k\geq4}&=0,
    \label{eq:l4def}
\end{align}
where $V^a=(\del x)^a$ as in \eqref{eq:dxdef}.
\end{definition}

We now show that Definition \ref{def:products} indeed gives an $L_3$-algebra on $\mathcal{Y}$. This is a direct consequence of Proposition \ref{prop:l1sq}, Lemma \ref{lem:central}, and Propositions \ref{prop:relations}--\ref{prop:l4} below. 

\begin{proposition}
\label{prop:l1sq}
In the conventions of Section~\ref{sec:normalisation}, with
component forms written to the left of the holomorphic basis forms, $\ell_1^2=0$.
\end{proposition}
\begin{proof}
Write $\xi=\del b$, a purely cotangent element. Then
$(\del\xi)^a=\bar\Omega^{abc}\del_b\del_cb=0$ by anti-symmetry of $\bar\Omega^{abc}$, as in \eqref{eq:autodiv}. Both extensions map out of the cotangent slot, and $F\!\cdot\!\nabla$ and $T$ act only through $(\del x)^a$, so both annihilate
$\xi$ and $\bar D(0,0,\xi)=(0,0,\delb\xi)$. With $\bar D^2=0$ from Proposition~\ref{prop:nilp}, we find
\begin{equation}
    \ell_1^2(Y)=\left(\delb\del b+\del\delb b,\delb^2b\right)=0\:.    
\end{equation}
\end{proof}

\begin{lemma}
\label{lem:central}
The image of the auxiliary slot is central for the bracket:
$[\,(0,0,\del b),\,y\,]=0$ for every $y$.
\end{lemma}
\begin{proof}
Every occurrence of the cotangent slot in Definition~\ref{def:bracket} is
through the map $x\mapsto(\del x)^a$.
For $x=\del b$ that map vanishes, as in the proof
of Proposition~\ref{prop:l1sq}. The remaining terms of
\eqref{eq:bracket_tangent} and \eqref{eq:bracket_gauge} each require a $\mu$ or
an $\alpha$ in the same argument, and $(0,0,\del b)$ has neither.
\end{proof}

\begin{proposition}
\label{prop:relations}
The products of Definition~\ref{def:products} satisfy the $L_\infty$ relations.

\end{proposition}
\begin{proof}
At $n=2$ the $Q$ component reads
\begin{equation}
    \bar D[y_1,y_2]
    =[\bar Dy_1+\del b_1,y_2]+(-1)^{p_1}[y_1,\bar Dy_2+\del b_2]\:.
\end{equation}
The $\bar D$ terms cancel by Proposition~\ref{prop:leibniz} and the $\del b$ terms
by Lemma~\ref{lem:central}.

At $n=3$ both components of \eqref{eq:Linf3} are non-trivial. In the
auxiliary component $\ell_1\ell_3$ contributes $\delb W$, the insertion terms
contribute $W(\delb V_1,V_2,V_3)$ and its two partners, since the cotangent
component of $\ell_1Y_i$ is $\delb x_i+\del b_i$ and $(\del\del b_i)^a=0$, and
$J$ contributes nothing. The relation is then the Leibniz rule
\[
\delb W=W(\delb V_1,V_2,V_3)+(-1)^{p_1}W(V_1,\delb V_2,V_3)
+(-1)^{p_1+p_2}W(V_1,V_2,\delb V_3)\:,
\]
which is where the minus sign on the insertion
terms in \eqref{eq:Linf3} comes from. 

In the $Q$ component the insertion terms contribute nothing, since $\ell_3$
has no $Q$ component. To keep track of normalisations, we allow general
coefficients $\varepsilon$ and $w$ in \eqref{eq:l1def} and \eqref{eq:l3def},
\begin{equation}
    \ell_1(Y)=\bigl(\bar Dy+\varepsilon\,\del b,\ \delb b\bigr)\:,\qquad
    \ell_3(Y_1,Y_2,Y_3)=\bigl(0,\ w\,W\bigr)\:.
\end{equation}
Then $\ell_1\ell_3$ contributes $\varepsilon w\,\del W$, whose cotangent
component carries the sign $(-1)^{p_1+p_2+p_3}$ of
Convention~\ref{conv:koszul}. Together with $J_a=-\del_aW$ from
Proposition~\ref{prop:jacobi}, which holds at all degrees, the $Q$ component
of \eqref{eq:Linf3} reads
\begin{equation}
    (-1)^{p_1+p_2+p_3}\,\bigl(\varepsilon w-1\bigr)\,\del_aW=0\:.
\end{equation}
This is where the factor $(-1)^{p_1+p_2+p_3}$ on the last term of
\eqref{eq:Linf3} comes from, and the relation holds precisely when
$\varepsilon w=1$. The auxiliary component above holds for any $w$, and
$\ell_1^2=0$ holds for any $\varepsilon$ by the proof of
Proposition~\ref{prop:l1sq}, so the two coefficients are constrained only
through their product. The residual freedom is the rescaling of $b$, which is
a choice of units for an auxiliary field. We have used it to set
$\varepsilon=w=1$ in Definition~\ref{def:products}. The convention of
\cite{Ashmore:2018ybe} corresponds instead to $\varepsilon=-\tfrac12$ and
$w=-2$.

The relation at $n=4$ is Proposition~\ref{prop:l4} below.
\end{proof}

\begin{proposition}
\label{prop:l4}
Set $\ell_k=0$ for $k\geq4$. Then the products of Definition~\ref{def:products} satisfy the $L_\infty$ relations for all $n\geq4$. The relations with $n\geq5$ hold term by term, and the relation with $n=4$ is equivalent to the identity
\begin{equation}
    \label{eq:n4}
    \sum_{\sigma\in\mathrm{Sh}(2,2)}\chi(\sigma)\,
    W\bigl([V_{\sigma1},V_{\sigma2}],V_{\sigma3},V_{\sigma4}\bigr)=0\:,
\end{equation}
where $W(U_1,U_2,U_3)\coloneqq\Omega_{abc}\,U_1^aU_2^bU_3^c$ and $V_i=(\del x_i)$. The identity \eqref{eq:n4} holds identically on $X$.
\end{proposition}
\begin{proof}[Sketch of proof]

The full proof is given in Appendix~\ref{app:l4}; here we outline its
structure. By Definition~\ref{def:products}, $\ell_2$ depends on its
arguments only through their $Q$-components and $\ell_3$ only through their
cotangent components, while $\ell_3$ takes values in the auxiliary slot. Hence
\begin{equation}
    \label{eq:l3compose}
    \ell_2\bigl(\ell_3(Y_1,Y_2,Y_3),Y_4\bigr)=0\:,\qquad
    \ell_3\bigl(\ell_3(Y_1,Y_2,Y_3),Y_4,Y_5\bigr)=0\:.
\end{equation}
The $n$-th $L_\infty$ relation is a sum of compositions $\ell_j\circ\ell_i$
with $i+j=n+1$. Since $\ell_k=0$ for $k\geq4$, by \eqref{eq:l3compose} every
such composition vanishes for $n\geq5$, so these relations hold term by term.
For $n=4$ only the compositions $\ell_3(\ell_2(\cdot,\cdot),\cdot,\cdot)$
survive. They are summed over the six ways $\mathrm{Sh}(2,2)$ of splitting the
four arguments into two pairs, each with its Koszul sign $\chi(\sigma)$, and
by \eqref{eq:cot_char} this sum is \eqref{eq:n4}.

The identity \eqref{eq:n4} itself uses three inputs:
Proposition~\ref{prop:jacobi}, contracted with a fourth vector; the
divergence-freeness \eqref{eq:autodiv} of the vectors $(\del x_i)$; and the
dimension of $X$, through \eqref{eq:schouten}. After reducing to vector fields
$v_i=(\del x_i)$ of degree zero, one computes the derivative of the function
\begin{equation}
    W_{234}\coloneqq W(v_2,v_3,v_4)
\end{equation}
along $v_1$ in two ways. The Schouten identity, together with
$\del_av_1^a=0$, expresses $v_1^a\del_aW_{234}$ as the three terms of
\eqref{eq:n4} in which $v_1$ sits inside the bracket.
Proposition~\ref{prop:jacobi}, contracted with $v_1$, expresses it as minus
the three terms in which $v_1$ sits outside the bracket. The six terms
therefore sum to zero. No condition on $\mu$ enters.
\end{proof}

It should be noted that the proof of the $L_3$-algebra identities, and Proposition \ref{prop:l4} in particular, relies on the fact that $X$ is a complex three-fold. We have the following remark: 

\begin{remark}
\label{cor:dim3}
The bracket, $W$ and $\bar\Omega^{abc}$ are defined here only on a three-fold,
and we make no claim about other dimensions. For a structure of the same type
on a Calabi--Yau $n$-fold with $n>3$, in which the cotangent field is
\begin{equation}
  x\in\Omega^{0,p}\bigl(\Lambda^{n-2}T^{*1,0}X\bigr)\:,
\end{equation}
of holomorphic degree $n-2$ 
so that $V^a=(\del x)^a$ is again a vector, $W(V_2,V_3,V_4)$ is an $(n-3)$-form rather than a function, and the trace identity \eqref{eq:trace}, which requires all slots of $\Omega$ to be filled, fails. The step \eqref{eq:n4first} in the proof of Proposition~\ref{prop:l4}, Appendix~\ref{app:l4}, then acquires additional terms, and \eqref{eq:n4} no longer follows from the Jacobiator identity. The truncation at $\ell_3$ is in
this sense a three-dimensional statement.
\end{remark}

\subsection{The extended action and the equation of motion}
\label{sec:MCext}
We now wish to construct an extended action, including an auxiliary field. To do so, one might try to construct a pairing on $\mathcal{Y}$ with a non-trivial auxiliary component. However, there is a problem. The pairing of Definition~\ref{def:pairing-mu} extends to $\mathcal{Y}$ only as
$\langle\!\langle Y,Y'\rangle\!\rangle=\langle y,y'\rangle$. Indeed, the pairing is integrated against $\Omega$, so it is non-zero only when the anti-holomorphic form degrees of its two arguments add up to three. By \eqref{eq:Ydef}, for $Y\in\mathcal{Y}_n$ and $Y'\in\mathcal{Y}_{n'}$ the $Q$-components have form degrees $n$ and $n'$, and the auxiliary components $n+1$ and $n'+1$. The three possible blocks are therefore non-zero only at
\begin{equation}
  Q\times Q:\ n+n'=3\:,\qquad b\times Q:\ n+n'=2\:,\qquad b\times b:\ n+n'=1\:.
\end{equation}
A cyclic pairing has a single degree, so neither auxiliary block is compatible with the $Q\times Q$ block of degree three. 
The extension is therefore degenerate along the auxiliary direction, and
$(\mathcal{Y},\ell_1,\ell_2,\ell_3)$ is not a cyclic $L_3$ algebra. This is the
situation of \cite{Ashmore:2018ybe} as well, where it is noted that no
action exists whose equations of motion are the $L_3$ field equation, and none
is expected. In \cite{Ashmore:2018ybe} the Maurer--Cartan field equation encodes the vanishing of the superpotential together with that of its first derivative, and these are two separate conditions. A similar, though slightly different, pattern happens here.

The action is instead the following. Following \cite{Ashmore:2018ybe}, we no longer impose the divergence-free condition \eqref{eq:divfree} on the
off-shell field space, but enforce it through a Lagrange multiplier. As remarked in \cite{Ashmore:2018ybe}, this condition is imposed from the outset in Kodaira--Spencer theory but is
better regarded as the requirement that the deformed holomorphic top-form remain closed. The extended action reads
\begin{equation}
    \label{eq:Sext}
    S_b=\int_X\Bigl(\tfrac12\langle y,\bar Dy\rangle
      +\tfrac16\langle y,[y,y]\rangle
      +\langle y,\del b\rangle\Bigr)\:,
\end{equation}
which agrees in form with equation $(4.23)$ of \cite{Ashmore:2018ybe}, modulo potential sign conventions and rescalings. Here
\begin{equation}
    \langle y,\del b\rangle=\mu^a\del_ab\wedge\Omega\: ,
    \qquad
    \int_X\langle y,\del b\rangle=-\int_X(\del_a\mu^a)\,b\wedge\Omega\: .    
\end{equation}
Varying $b$ now returns $\del\chi=0$, that is
\begin{equation}
    C\coloneqq\del_a\mu^a=0\:,
\end{equation}
i.e. the divergence-free condition.

Varying $y$, now unconstrained, needs one more step, because Lemma~\ref{lem:cyclic} was proved on the constraint surface where $\mu$ is divergence-free. The constraint enters that proof only in the family $(\mu,x,x)$, through the divergence of
$\mu$ in step (iii); the families $(\alpha,\alpha,\alpha)$ and
$(x,\alpha,\alpha)$ do not involve $\mu$ and are unaffected. Keeping the
divergence term in step (iii), the same computation gives
\begin{equation}
    \label{eq:cyclicdefect}
    \int_X\langle x_3,[\mu_1,x_2]\rangle
    =\int_X\langle\mu_1,[x_2,x_3]\rangle
     +\int_X x_{3a}\,(\del_b\mu_1^b)\,V_2^a\wedge\Omega\:,
\end{equation}
so that cyclicity fails off the constraint surface by a term proportional to
$C$. The general-degree form of this defect is Lemma~\ref{lem:defect}.

Rather than permute the arguments of the cubic term of \eqref{eq:Sext} cyclically, which off the constraint surface produces the defect \eqref{eq:cyclicdefect}, 
we expand it. By \eqref{eq:pairing-mu}, \eqref{eq:bracket_cotangent} and
\eqref{eq:cyclicdefect}, the dependence of $S_b$ on $\mu$ is
\begin{equation}
    \label{eq:Sexpanded}
    S_b=\tfrac12\int_X\langle y,\bar Dy\rangle
     +\tfrac12\int_X\mu^a\Omega_{abc}V^bV^c\wedge\Omega
     -\int_X C\,\bigl(b+\tfrac13\,x_cV^c\bigr)\wedge\Omega
     +\tfrac16\int_X\langle y,[y,y]\rangle\Big|_{\mu=0}\:,
\end{equation}
the last term collecting the cubic couplings of $\alpha$ and $x$ alone. Note that the Lagrange multiplier is defined only up to the shift $b\mapsto b+\lambda\,x_a(\del x)^a$,
which changes \eqref{eq:Sext} by a multiple of
$\int_X(\del_a\mu^a)\,x_c(\del x)^c\wedge\Omega$. We fix $\lambda=-\tfrac13$, i.e.\ we replace $b$ by
$b-\tfrac13x_cV^c$, which removes the $x_cV^c$ term from \eqref{eq:Sexpanded}.\footnote{This is the
counterpart of the redefinition $b\mapsto b-\mu^ax_a$ made in \cite{Ashmore:2018ybe}, equation~(4.11), under the exchange of $\mu$ and
$(\del x)$.} From now on $S_b$ denotes the shifted action
\begin{equation}
    \label{eq:Sfinal}
    S_b=\int_X\Bigl(\tfrac12\langle y,\bar Dy\rangle
      +\tfrac16\langle y,[y,y]\rangle
      +\langle y,\del b\rangle\Bigr)
      +\tfrac13\int_X C\,x_cV^c\wedge\Omega\:,
\end{equation}
whose cubic coupling with $\mu$ is the Yukawa vertex \eqref{eq:yukawa}.

We now vary the action. The
quadratic term gives $\int_X\langle\delta y,\bar Dy\rangle$ by Lemma~\ref{lem:IBP}. The $\mu$-independent cubic terms give their share of
$\tfrac12\int_X\langle\delta y,[y,y]\rangle$ by the families (i) and (ii) of Lemma~\ref{lem:cyclic}, which hold off the constraint surface. Then \eqref{eq:cyclicdefect} with $(\mu_1,x_2,x_3)=(\mu,x,\delta x)$ gives the rest, up to one term proportional to $C$. Altogether, for arbitrary $\delta y$ and $\delta b$, we find
\begin{equation}
    \label{eq:varSext}
    \delta S_b=\int_X\bigl\langle\delta y,\ \bar Dy+\tfrac12[y,y]+\del b\bigr\rangle
    -\int_X C\wedge\bigl(\delta b-\delta x_a V^a\bigr)\wedge\Omega\: .
\end{equation}
The $b$-equation is $C=0$, i.e.\ the divergence-free constraint on $\mu$ is imposed. On this constraint surface, the $y$-equation of motion becomes
\begin{equation}
    \label{eq:MCext}
     E\coloneqq\bar Dy+\tfrac12[y,y]+\del b=0\: ,
\end{equation}
which is \eqref{eq:MC} with the $\del$-exact remainder identified. The quotient noted in footnote \ref{foot:quotient} is thereby avoided.

Note that the equation of motion \eqref{eq:MCext} forms the $y$-part of the Maurer--Cartan equation of the $L_3$-algebra of Definition \ref{def:products}, now restricted to the divergence-free constraint surface.\footnote{It is straightforward to check that the $L_3$-products of Definition \ref{def:products} preserve the constraint surface.} The equation of motion is thus covariant under $L_3$ gauge transformations. Indeed, the extended $L_3$-structure now generates the gauge transformations of the theory. For $Y=(y,b)$ and $\Lambda=(\xi,\beta)\in\mathcal{Y}_0$ with $\xi=(\xi_\mu,\xi_\alpha,\xi_x)\in\Omega^{0,0}(Q)$ and $\xi_\mu$ divergence-free, the products of Definition~\ref{def:products} give
\begin{equation}
    \label{eq:L3gauge}
    \delta Y=\ell_1(\Lambda)+\ell_2(Y,\Lambda)-\tfrac12\,\ell_3(Y,Y,\Lambda)\:,
\end{equation}
the counterpart of (5.10) of \cite{Ashmore:2018ybe}. For $\xi=0$ this is $\delta y=\del\beta$, $\delta b=\delb\beta$, whose $y$-component is the transformation \eqref{eq:gaugeMC}, now produced by the structure rather than added to it. One might wonder about the auxiliary component of the Maurer--Cartan equation. Below, we show that this equation actually also follows from the equation of motion, assuming $y$ sufficiently small.\footnote{This is in contrast to the situation in \cite{Ashmore:2018ybe}, where the auxiliary equation came from the requirement that the action vanished on-shell.}

Before we consider the full Maurer--Cartan equation, we make some comments on the (non-)invariance of the action under $L_3$-gauge transformations. Indeed, the action is invariant under \eqref{eq:L3gauge} only with the same restriction as in \cite{Ashmore:2018ybe}. For $\beta$, for $\xi_\alpha$, and for $\xi_\mu$ with $\del_a\xi_\mu^a=0$, the variation \eqref{eq:varSext} vanishes identically. For the cotangent parameter $\xi_x$, with $V_\xi^a\coloneqq(\del\xi_x)^a$, and for $\xi_\mu$ with $\del_a\xi_\mu^a\neq0$, it equals respectively
\begin{align}
    \label{eq:VarS1}
    \delta_{\xi_x}S_b&=\int_X C\wedge V_\xi^aE_a\wedge\Omega\:,\\
    \delta_{\xi_\mu}S_b&=\int_X(\del_a\xi_\mu^a)\bigl(\delb b+\tfrac16W-V^aE_a\bigr)\wedge\Omega\:.
    \label{eq:VarS2}
\end{align}
The first is a product of the $b$- and $x$-equations, and can be removed by the trivial transformation $\delta b\mapsto\delta b+V_\xi^aE_a$.\footnote{Whether this extra transformation of $b$ can be incorporated into the $L_3$-algebra above, which would bring it closer to the structure of \cite{Ashmore:2018ybe}, is left to future work.} The second is also linear in the equations of motion, since $\del_a(\delb b+\tfrac16W)$ is the cotangent component of $\bar DE+[y,E]$, but the transformation needed to remove it does not vanish on shell, so it cannot be compensated by a trivial symmetry. Such parameters do not preserve the constraint surface. The computations leading to \eqref{eq:VarS1}--\eqref{eq:VarS2} are given in Appendix~\ref{app:gauge}.

Collecting these results, we have the following theorem.
\begin{theorem}[Cyclicity and $L_3$ gauge invariance on the constraint surface]\label{thm:constraint}
On the constraint surface $C=\del_a\mu^a=0$,
\begin{enumerate}[label=(\roman*)]
\item The pairing is cyclic with respect to the bracket in all degrees, and $\bar D$, and equivalently $\ell_1$, integrates by parts, so that the cubic action is cyclic.
\item The extended action \eqref{eq:Sfinal} is invariant, off shell, under all $L_3$ gauge transformations \eqref{eq:L3gauge} that preserve the constraint surface, that is, those with $\del_a\xi_\mu^a=0$.
\end{enumerate}
\end{theorem}
\begin{proof}
Part (i) is Lemma~\ref{lem:cyclic} together with Lemma~\ref{lem:IBP}. By Lemma~\ref{lem:defect}, the only obstruction to cyclicity off the constraint surface is $\int_X w_a\,C\,V^a\wedge\Omega$, which vanishes for $C=0$. Part (ii) follows from Proposition~\ref{prop:varS}. The $\beta$ and $\xi_\alpha$ variations vanish identically, the $\xi_\mu$ variation \eqref{eq:VarS2} is proportional to $\del_a\xi_\mu^a$, and the $\xi_x$ variation \eqref{eq:VarS1} is proportional to $C$.
\end{proof}
\begin{remark}
    On the constraint surface the extended action reduces to the cubic action \eqref{eq:cubicAction}, since $\int_X\langle y,\del b\rangle=-\int_X C\,b\wedge\Omega$ and the last term of \eqref{eq:Sfinal} are both proportional to $C$. The multiplier then decouples, and since $\langle y,(0,0,\del\beta)\rangle=-\int_X C\,\beta\wedge\Omega$ also vanishes, the theory descends to the complex $K$ of \eqref{eq:Kcomplex}, where $\ell_3$ and the Jacobiator drop out and $(\bar D,[\,,\,])$ is a differential graded Lie algebra with an invariant pairing. This parallels \cite[Section~6.1]{Ashmore:2018ybe}, where the same reduction arises by integrating out the multiplier. The difference is that in \cite{Ashmore:2018ybe} invariance under the vector-type parameters is shown only for the reduced action, whereas by Proposition~\ref{prop:varS} the extended action is invariant under divergence-free parameters also off the constraint surface, with the multiplier present, up to the on-shell-trivial transformation removing \eqref{eq:VarS1}. 
\end{remark}

We now derive the auxiliary component of the Maurer--Cartan equation from
\eqref{eq:MCext}, following the argument of \cite{Ashmore:2018ybe},
Section~4.4. Apply $\bar D$ to \eqref{eq:MCext}. By Proposition~\ref{prop:nilp}
the first term drops, by Proposition~\ref{prop:leibniz} and Lemma~\ref{lem:brasym}
the second gives $[\bar Dy,y]$, and substituting \eqref{eq:MCext} back,
\begin{equation}
    -\tfrac12\,[[y,y],y]-[\del b,\,y]+\bar D(\del b)=0 .
\end{equation}
Here $[\del b,y]=0$ by Lemma~\ref{lem:central}, and $\bar D(\del b)=(0,0,\delb\del b)$
for the same reason. The Jacobiator at three equal arguments of degree one is
$J(y,y,y)=3[[y,y],y]$, and by Proposition~\ref{prop:jacobi} its only nonzero
component is $J_a=-\del_aW$. Hence
\begin{equation}
    \label{eq:Wconst}
    \del_a\Bigl(\tfrac16W+\delb b\Bigr)=0,
    \qquad\text{so that}\qquad
    \tfrac16W+\delb b=k\,\bar\Omega
\end{equation}
for a constant $k$, the form $\tfrac16W+\delb b\in\Omega^{0,3}(X)$ being
annihilated by $\del$ and $X$ compact. This is the counterpart of equation
(4.29) of \cite{Ashmore:2018ybe}, with $[\del b,y]$ absent and the Yukawa
cubic $W$ in place of the cubic vertex. The auxiliary component of the
Maurer--Cartan equation is the statement $k=0$,
\begin{equation}
    \label{eq:auxMC}
    \delb b=-\tfrac16\,W\: ,
\end{equation}
which is the auxiliary component of the full Maurer--Cartan equation
\begin{equation}
    \label{eq:MCL3}
    \ell_1(Y)+\tfrac12\ell_2(Y,Y)+\tfrac16\ell_3(Y,Y,Y)=0
\end{equation}
for the $L_3$-algebra of
Definition~\ref{def:products}: the Yukawa cubic is $\delb$-exact, with
potential the same $b$ that appears in \eqref{eq:MCext}. 

If we integrate \eqref{eq:Wconst} against $\Omega$, the $\delb b$ term drops by compactness
and $k=0$ is then equivalent to
\begin{equation}
    \label{eq:Wzero}
    \int_XW\wedge\Omega=0\: .
\end{equation}
The Maurer--Cartan equation \eqref{eq:MCL3} of the extended structure is
therefore the equation of motion \eqref{eq:MCext} together with \eqref{eq:auxMC}, equivalently with
the vanishing of the integrated Yukawa cubic. We now show that on a compact Calabi--Yau three-fold the second condition, when $\del x$ is sufficiently small, is in fact implied by the first.

\subsection{The auxiliary equation and the deformed three-form}
\label{sec:torelli}
Note that the cotangent component of \eqref{eq:MCext} involves $x$ and $b$ only. The
cotangent slot of $\bar D$ is $\delb$, the extension maps of \eqref{eq:barD}
act out of that slot and not into it, and the cotangent component
\eqref{eq:bracket_cotangent} of the bracket is built from $(\del x)$ alone.
Writing $V^a=(\del x)^a$ as before, the component reads
\begin{equation}
    \label{eq:Ecot}
    E_c\coloneqq \delb x_c+\tfrac12\,\Omega_{cab}V^aV^b+\del_cb=0\: ,
\end{equation}
and \eqref{eq:Wconst} was obtained by applying $\delb$ to it. It is
convenient to package $x$, $b$ and $W$ into the deformed three-form
\begin{equation}
    \label{eq:OmV}
    \Omega_V\coloneqq e^{\iota_V}\Omega
    =\Omega+\iota_V\Omega+\tfrac12\iota_V^2\Omega+\tfrac16\iota_V^3\Omega ,
\end{equation}
where $\iota_V=V^a\wedge\iota_{\del_a}$ is the contraction with the
vector-valued $(0,1)$-form $V$. Since $\iota_V$ is a derivation of degree
zero, $e^{\iota_V}$ is an algebra automorphism and
$\Omega_V=\bigwedge_a(dz^a+V^a)$ in a local frame; in particular $\Omega_V$
is decomposable. We then have the following proposition:

\begin{proposition}
\label{prop:OmV}
In the conventions of Section~\ref{sec:normalisation}, for arbitrary
$x\in\Omega^{0,1}(T^{*1,0}X)$ and $b\in\Omega^{0,2}(X)$,
\begin{equation}
    \label{eq:OmVidentity}
    \Omega_V-\Omega=-\,d(x+b)+E_c\wedge dz^c+\bigl(\tfrac16W+\delb b\bigr),
\end{equation}
where $x=x_c\wedge dz^c$ is regarded as a $(1,1)$-form and
$E_c$ is the cotangent component \eqref{eq:Ecot}.
\end{proposition}
\begin{proof}
The three terms on the right are the $(2,1)$, $(1,2)$ and $(0,3)$ components
of the left. In degree $(2,1)$, $\iota_V\Omega=\tfrac12V^a\Omega_{abc}\,dz^bdz^c$
and \eqref{eq:dxdef} with \eqref{eq:Omnorm} give $\iota_V\Omega=-\del x$. In
degree $(1,2)$, $\tfrac12\iota_V^2\Omega=\tfrac12\Omega_{cab}V^aV^b\wedge dz^c$,
which is $E_c\wedge dz^c-\delb x-\del b$ by \eqref{eq:Ecot}. In degree
$(0,3)$, $\tfrac16\iota_V^3\Omega=\tfrac16W$, and $-\delb b+(\tfrac16W+\delb b)$
is the same. Signs follow Convention~\ref{conv:koszul}, with $\del$ and $\delb$
acting from the left.
\end{proof}

\begin{corollary}
\label{cor:OmV}
Let $(x,b)$ satisfy the cotangent equation \eqref{eq:Ecot}. Then
\begin{equation}
    \Omega_V=\Omega-d(x+b)+k\bar\Omega    
\end{equation}
with $k$ the constant of \eqref{eq:Wconst}. In particular, $\Omega_V$ is $d$-closed and
$[\Omega_V]=[\Omega]+k[\bar\Omega]$. Consequently the auxiliary
equation \eqref{eq:auxMC}, which is the statement $k=0$, is equivalent to
\begin{equation}
    \label{eq:OmVclass}
    [\Omega_V]=[\Omega]\in H^3(X,\mathbb C),
\end{equation}
and the integrated form \eqref{eq:Wzero} is
$\int_X\Omega\wedge\Omega_V=0$.
\end{corollary}
\begin{proof}
Set $E_c=0$ in \eqref{eq:OmVidentity} and use \eqref{eq:Wconst}, which
follows from \eqref{eq:Ecot} alone. The last
statement is $\int_X\Omega\wedge\Omega_V=\tfrac16\int_X\Omega\wedge W$, the
lower components of $\Omega_V$ having positive holomorphic degree.
\end{proof}

The combination $x+b$ is the one left invariant, up to $d\beta$, by the
transformation $\delta x=\del\beta$, $\delta b=\delb\beta$ generated by
$\ell_1$, which is why the Lagrange multiplier enters \eqref{eq:OmVidentity} at all.
Applying $\bar\Omega^{abc}\del_b$ to \eqref{eq:Ecot} removes $b$ and gives
\begin{equation}
    \label{eq:KSeq}
    \delb V^a-V^d\del_dV^a=(\del E)^a\:,
\end{equation}
so on shell, $V$ satisfies the Kodaira--Spencer equation and the closedness of $\Omega_V$ in Corollary~\ref{cor:OmV} is the Tian--Todorov lemma \cite{todorov_weil-petersson_1989} for the divergence-free vector $V$. The auxiliary equation adds to this the cohomological statement
\eqref{eq:OmVclass}, and on a compact Calabi--Yau three-fold (at least for small enough but finite $\del x$) that statement is automatic.
This is easiest to see at linear order in $\del x$. There \eqref{eq:KSeq} reduces to $\delb V^a=0$, so $V$ defines a class in $H^{0,1}(T^{1,0}X)$, that is, an infinitesimal deformation of the complex structure. However, $\iota_V\Omega=-\del x$ is $\del$-exact and $\delb$-closed, and hence $\delb$-exact by the $\del\delb$-lemma. Equivalently, the harmonic representative of $V^a=(\del x)^a$ is zero. So $V$ is not a true deformation of the complex structure, and \eqref{eq:OmVclass} holds to this order. We now present a formal proof, valid at finite order.

\begin{proposition}
\label{prop:torelli}
Let $X$ be a compact Calabi--Yau three-fold and let $(x,b)$ satisfy the
cotangent equation \eqref{eq:Ecot}, with $V=(\del x)$ small enough that the
almost complex structure $J_V$ it defines lies in the neighbourhood of $J$ on
which the Kuranishi family of $X$ is complete. Then $k=0$ in \eqref{eq:Wconst}; that is, the auxiliary equation
\eqref{eq:auxMC} holds. Moreover, the complex structure defined by $\Omega_V$
is isomorphic to that of $X$.
\end{proposition}
\begin{proof}
By Corollary~\ref{cor:OmV}, $\Omega_V$ is a $d$-closed decomposable complex
three-form close to $\Omega$, so it is the holomorphic volume form of an
integrable complex structure $J_V$ on $X$ close to $J$, namely the one
obtained from $J$ by the Kodaira--Spencer solution $V$. By completeness of the
Kuranishi family \cite{kuranishi_new_1965}, $J_V=\phi^*J_t$ for some $t$ in
the Kuranishi space $\mathcal K$ near $0$ and some diffeomorphism $\phi$
isotopic to the identity, and then $\Omega_V=c\,\phi^*\Omega_t$ for a
constant $c$, where $\Omega_t$ is the holomorphic volume form of $J_t$. Hence
$[\Omega_V]=c\,[\Omega_t]$ in $H^3(X,\mathbb C)$.

Let $P\colon\mathcal K\to\mathbb P(H^3(X,\mathbb C))$, $t\mapsto[\Omega_t]$, be
the period map. For Calabi--Yau three-folds $P$ is an immersion near $0$ with
tangent space $H^{2,1}(X)$ at $[\Omega]$, by Griffiths transversality and
the local Torelli theorem for Calabi--Yau manifolds
\cite{tian_smoothness_1987, todorov_weil-petersson_1989, griffiths_principles_1994, huybrechts_complex_2005}.
In the affine chart of $\mathbb P(H^3)$ consisting of classes whose
$H^{3,0}$-component is $[\Omega]$, the image of $P$ near $[\Omega]$ is
therefore the graph of a holomorphic map $g$ from a neighbourhood of $0$ in
$H^{2,1}(X)$ to $H^{1,2}(X)\oplus H^{0,3}(X)$ with $g(0)=0$. By the first
statement of Corollary~\ref{cor:OmV}, which uses only \eqref{eq:Ecot} and
leaves $k$ free, $[\Omega_V]=[\Omega]+k[\bar\Omega]$ lies in this
chart with $H^{2,1}$-component zero. It lies on the graph only if
$k[\bar\Omega]=g(0)=0$, so $k=0$ and $t=0$; the latter gives
$J_V=\phi^*J$.
\end{proof}

It follows that on a compact Calabi--Yau three-fold, and for $V=(\del x)$ small enough that Proposition~\ref{prop:torelli} applies, the equations of motion of the action \eqref{eq:Sfinal} are equivalent to the Maurer--Cartan equation of the $L_3$ algebra of Definition~\ref{def:products}, together with the constraint \eqref{eq:constraint}. This differs from \cite{Ashmore:2018ybe}, where the auxiliary component of the Maurer--Cartan equation is instead equivalent, given the equations of motion, to the vanishing of the action. Since the argument rests on the local Torelli theorem, it need not persist on non-K\"ahler backgrounds, for instance once gravitational anomalies are included and the Hull--Strominger system is considered. We conclude this section with the following remark, comparing with \cite{Ashmore:2018ybe}.

\begin{remark}
\label{rem:admss}
Note that in \cite{Ashmore:2018ybe} the constant in the counterpart of \eqref{eq:Wconst} is fixed by the vanishing of the superpotential, 
because the potential of their Jacobiator 
is the cubic $\langle y,[y,y]\rangle$ to which the superpotential reduces on shell. Here the potential is $W$ while our dual action $S_b$ reduces on shell to $-\tfrac1{12}\int_X\langle y,[y,y]\rangle$. The two cubics differ. In particular, the vanishing of our dual action does not fix $k$. For Calabi--Yau backgrounds, it is fixed instead by Proposition~\ref{prop:torelli}, and the quantity whose vanishing is equivalent to \eqref{eq:auxMC} is not the action $S_b$, but the period pairing $\int_X\Omega\wedge\Omega_V$ of Corollary~\ref{cor:OmV}. 
Both differences trace back to the exchange of Remark~\ref{rem:nomumu}: our Jacobiator is built from $(\del x)$ alone, and $[\del b,y]=0$ by Lemma~\ref{lem:central}, whereas their bracket forces an auxiliary component in $\ell_2$, their (4.28). Whether the ``duality'', exchanging $\mu$ and $(\del x)$, extends to a map of $L_3$ algebras carrying their Jacobiator
to ours is left open.
\end{remark}

\section{Local trivialisation and \v{C}ech cohomology}
\label{sec:local-triv}

As noted, the kinetic operator $\bar{D}$ of our dual Kodaira--Spencer theory is not a standard Dolbeault operator: its off-diagonal entries carry explicit holomorphic derivatives, so $\bar{D}$ does not define a holomorphic bundle in the usual sense. Such operators have appeared in the context of heterotic moduli and elsewhere \cite{McOrist:2021dnd, deLazari:2024zkg, Chisamanga:2024xbm, McOrist:2024ivz, Kjelsnes:2026oox}, and appear to be somewhat ubiquitous in higher-order supergravity. 
The classical Dolbeault theorem therefore does not immediately apply for $\bar D$, and the cohomology $H^{0,q}_{\bar{D}}(Q)$ cannot immediately be identified with any sheaf cohomology group. Generalising the results of \cite{MurgasIbarra:2025ylk} to our particular example,\footnote{Our operator $\bar D$ differs from the type of operator studied in \cite{MurgasIbarra:2025ylk}, in that the extension map $T$ is second order in holomorphic derivatives rather than first order.} we remedy this by constructing, on each
ball of a suitable cover of $X$, an explicit local gauge transformation that puts $\bar{D}$ into standard $\bar{\partial}$ form. The (non-tensorial) overlap maps between patches turn out to be holomorphic in the appropriate sense, so they glue into a
well-defined ``quasi-holomorphic sheaf'' $\widetilde{\mathcal{Q}}$. We then establish a Dolbeault isomorphism $H^{0,q}_{\bar{D}}(Q) \cong \check{H}^q(\widetilde{\mathcal{Q}})$ in Theorem \ref{thm:main}.

\subsection{Local trivialisations}
We begin by considering the geometric structure on a local patch. Let $\{U_i\}_{i\in I}$ be a cover of the Calabi--Yau three-fold $X$ by open balls in holomorphic coordinate charts, chosen small enough that $V$ is holomorphically trivial on each $U_i$ and that $\Omega$ has constant coefficients as in Convention~\ref{conv:frame}; any open cover can be refined to one of this kind. Such balls are convex, hence Stein, and finite intersections of Stein open subsets of $X$ are again Stein, so the $\bar\partial$-Poincar\'e lemma holds on each $U_i$ and on every finite intersection. On each patch $U_i$ the holomorphicity condition $F^{0,2}=0$ allows us to pass to a local holomorphic frame, in which the connection has a $(1,0)$-form potential $A_i=A^i_a\,\dd z^a\in\Omega^{1,0}(U_i,\mathfrak{g})$ with $F=\bar\partial A_i$. Since
\[
  \bar\partial A_i
  =\bar\partial A^i_a\wedge \dd z^a
  =-\,\dd z^a\wedge\bar\partial A^i_a\: ,
\]
the convention $F=\dd z^a\wedge F_a$ of Section~\ref{sec:quad} 
gives
\begin{equation}
  \label{eq:Fa-local}
  F_a = -\,\bar\partial A^i_a\:,
\end{equation}
the sign arising from transposing the $(0,1)$-form
$\bar\partial A^i_a$ past the $(1,0)$-form $\dd z^a$.
The off-diagonal entries of
$\bar{D}$ acting on a section $y=(\mu,\alpha,x)$ are:
\begin{itemize}
  \item \emph{gauge-to-complex}: $\mathrm{tr}(F\wedge\partial_A\alpha)
    = (-1)^p\,\mathrm{tr}(F_a\wedge\partial_A\alpha)\wedge\dd z^a$ for $\alpha\in\Omega^{0,p}(\mathfrak g)$, 
    mapping $\alpha$ to
    $\Omega^{0,\bullet+1}(T^{1,0})$,
  \item \emph{metric-to-gauge}: $F_a\,\partial_b x_c\,\bar\Omega^{abc}$,
    mapping $x$ to $\Omega^{0,\bullet+1}(\mathfrak{g})$,
  \item \emph{metric-to-complex}: $T(x)
    = -\tfrac{1}{2}\,\partial\!\bigl(\omega^{\mathrm{CS}}_a\,
      \partial_b x_c\,\bar\Omega^{abc}\bigr)$,
    mapping $x$ to $\Omega^{2,\bullet+1}(X)$,
\end{itemize}
where $\omega^{\mathrm{CS}}_a = \iota_{\partial_a}\omega^{2,1}$ are the
components of the global Chern--Simons form \eqref{eq:dCS} along
$\partial_a$, satisfying the Chern--Weil identity \eqref{eq:CW}; they carry
no patch index.

Note that on $U_i$ both $\omega^{2,1}$ and $\mathrm{tr}(A_i\wedge F)$
satisfy $\bar\partial(\,\cdot\,)=\mathrm{tr}(F\wedge F)$, so they differ by a
$\bar\partial$-closed $(2,1)$-form; since $U_i$ is a ball, hence Stein, that difference
is $\bar\partial$-exact: 
\begin{equation}\label{eq:CSlocal}
  \omega^{2,1}=\mathrm{tr}(A_i\wedge F)+\bar\partial\eta_i\:,
  \qquad \eta_i\in\Omega^{2,0}(U_i)\:.
\end{equation}
Neither term on the right is global, and neither is $\partial$-closed.
Indeed, in the holomorphic frame
\begin{equation}
  \partial\,\mathrm{tr}(A_i\wedge F)=\mathrm{tr}(A_i\wedge A_i\wedge F)\:,
\end{equation}
which is cancelled by $\partial\bar\partial\eta_i$. With
$\iota_{\partial_a}F=F_a$, one then finds
\begin{equation}
    \iota_{\partial_a}\mathrm{tr}(A_i\wedge F)=\mathrm{tr}(A^i_a F)-\mathrm{tr}(A_i\wedge F_a)\: ,
  \qquad
  \bar\partial\,\mathrm{tr}(A^i_a A_i)
  =\mathrm{tr}(A^i_a F)+\mathrm{tr}(A_i\wedge F_a)\:,
\end{equation}
the second identity following from \eqref{eq:Fa-local}. Subtracting, and using
$\iota_{\partial_a}\bar\partial=-\bar\partial\iota_{\partial_a}$ on the last term of
\eqref{eq:CSlocal}, we find that
\begin{equation}\label{eq:CSrimitive}
  \omega^{\mathrm{CS}}_a
  = \bar\partial\bigl(\mathrm{tr}(A^i_a A_i)-\iota_{\partial_a}\eta_i\bigr)
    - 2\,\mathrm{tr}\bigl(A_i\wedge F_a\bigr)
  \qquad\text{on } U_i\: .
\end{equation}
Applying $\bar\partial$ then recovers \eqref{eq:CW}.

We now let
\begin{equation}\label{eq:Qai}
  Q_a^i \coloneqq  \mathrm{tr}\bigl(A^i_a A_i\bigr)-\iota_{\partial_a}\eta_i\;\in\;\Omega^{1,0}(U_i)\:,
\end{equation}
so that by \eqref{eq:CSrimitive}
\begin{equation}\label{eq:dbarQ}
  \bar\partial Q_a^i
  = \omega^{\mathrm{CS}}_a + 2\,\mathrm{tr}\bigl(A_i\wedge F_a\bigr)\:.
\end{equation}
Next, we define the following local operators on sections over $U_i$:
\begin{align}
  N_i(x) &\coloneqq  -\,A_a^i\,\partial_b x_c\,\bar\Omega^{abc}\:,
    \label{eq:Ni}\\
  P_i(\alpha) &\coloneqq  \mathrm{tr}(A_i\wedge\partial\alpha)
    + \mathrm{tr}(A_i\wedge A_i\wedge\alpha)\:,
    \label{eq:Pi}\\
  \lambda_i(x) &\coloneqq  \tfrac{1}{2}\,\partial\!\bigl(Q_a^i\,
    \partial_b x_c\,\bar\Omega^{abc}\bigr)\:.
    \label{eq:lambdai}
\end{align}
Note that $N_i$ uses the local connection $A_a^i$, whereas
$\bar{D}$ uses the curvature $F_a=-\bar\partial A_a^i$ in the metric-to-gauge block. The sign in \eqref{eq:Ni} is that of \eqref{eq:Fa-local}. We then have the following proposition:

\begin{proposition}[Local trivialisations]\label{prop:triv}
  Define the local automorphisms
  \[
    G_i \;\coloneqq \;
    \begin{pmatrix}
      1 & P_i & \lambda_i \\
      0 & 1   & N_i \\
      0 & 0   & 1
    \end{pmatrix},
    \qquad
    G_i^{-1} =
    \begin{pmatrix}
      1 & -P_i & P_i N_i - \lambda_i \\
      0 & 1    & -N_i \\
      0 & 0    & 1
    \end{pmatrix}.
  \]
  Then on each $U_i$,
  \begin{equation}\label{eq:triv}
    \bar{D} \;=\; G_i^{-1} \circ
    \begin{pmatrix}
      \bar\partial & 0 & 0 \\
      0 & \bar\partial & 0 \\
      0 & 0 & \bar\partial
    \end{pmatrix}
    \circ\, G_i\:.
  \end{equation}
\end{proposition}

\begin{proof}
The identity \eqref{eq:triv} is equivalent to $\bar\partial G_i = G_i\bar{D}$. Both sides are upper-triangular, so we can match the entries block by block. Comparing the $(1,2)$, $(2,3)$ and $(1,3)$ entries,  we find that the three required commutator identities to be checked are:
\begin{align}
  [\bar\partial, N_i](x)
    &= -\,(\bar\partial A_a^i)\,\partial_b x_c\,\bar\Omega^{abc}
     \;=\; F_a\,\partial_b x_c\,\bar\Omega^{abc}\:,
     \label{eq:commN}\\
  [\bar\partial, P_i](\alpha)
    &= \mathrm{tr}(F\wedge\partial\alpha)
     + \mathrm{tr}\bigl([F,A_i]\wedge\alpha\bigr)
     \;=\; \mathrm{tr}(F\wedge\partial_{A_i}\alpha)\:,
     \label{eq:commP}\\
  [\bar\partial, \lambda_i](x)
    &= T(x) + P_i\bigl(F_a\,\partial_b x_c\,\bar\Omega^{abc}\bigr)\:,
     \label{eq:commL}
\end{align}
matching, respectively, the metric-to-gauge, gauge-to-complex and metric-to-complex couplings of $\bar D$.

Equation \eqref{eq:commN} is immediate from \eqref{eq:Fa-local}. For \eqref{eq:commP}, the Leibniz rule and
$\bar\partial\partial=-\partial\bar\partial$ give
\begin{equation}
    [\bar\partial,\mathrm{tr}(A_i\wedge\partial\,\cdot\,)](\alpha)
  =\mathrm{tr}(F\wedge\partial\alpha) \:,
\end{equation}
while
\begin{equation}
      [\bar\partial,\mathrm{tr}(A_i^2\,\cdot\,)](\alpha)
  =\mathrm{tr}\bigl(\bar\partial(A_i\wedge A_i)\wedge\alpha\bigr)
  =\mathrm{tr}([F,A_i]\wedge\alpha)\: .
\end{equation}
The second equality in \eqref{eq:commP} is
\begin{equation}
    \mathrm{tr}(F\wedge[A_i,\alpha])
  =\mathrm{tr}([F,A_i]\wedge\alpha)
\end{equation}
by cyclicity of the trace.

For \eqref{eq:commL}, the Leibniz rule and
$\bar\partial\partial=-\partial\bar\partial$ give, as in the proof of Proposition~\ref{prop:nilp},
\begin{align}
[\bar\partial,\lambda_i](x)
&= -\tfrac12\,\partial\bigl(\bar\partial Q_a^i\wedge(\partial x)^a\bigr)\notag\\
&\overset{\eqref{eq:dbarQ}}{=}
 -\tfrac12\,\partial\bigl(\omega^{\mathrm{CS}}_a(\partial x)^a\bigr)
 - \partial\bigl(\mathrm{tr}(A_i\wedge F_a)\wedge(\partial x)^a\bigr)\notag\\
&= T(x)
 - \partial\bigl(\mathrm{tr}(A_i\wedge F_a)\wedge(\partial x)^a\bigr)\:.
\end{align}
Expanding the last term by the Leibniz rule and using
$\partial A_i=-A_i\wedge A_i$ (the frame form of $F^{2,0}=0$), we get
\begin{align}
    -\,\partial\bigl(\mathrm{tr}(A_i\wedge F_a)\wedge(\partial x)^a\bigr)
    &= \bigl[\mathrm{tr}(A_i\wedge\partial F_a)
       +\mathrm{tr}(A_i\wedge A_i\wedge F_a)\bigr](\partial x)^a\notag\\
    &- \mathrm{tr}(A_i\wedge F_a)\wedge\partial(\partial x)^a\: ,
\end{align}
which is precisely $P_i\bigl(F_a(\partial x)^a\bigr)$ computed from
\eqref{eq:Pi}. With \eqref{eq:commN}--\eqref{eq:commL} established,
both sides of $\bar\partial G_i = G_i\bar{D}$ agree entry by entry.
\end{proof}

The operators $N_i$, $P_i$, $\lambda_i$ are thus the local
$\bar\partial$-primitives of the three off-diagonal entries of $\bar D$:
the metric-to-gauge entry $F_a\,\partial_b x_c\,\bar\Omega^{abc}$, the
gauge-to-complex entry $\mathrm{tr}(F\wedge\partial_{A}\alpha)$, and the
metric-to-complex entry $T(x)$ up to $P_i\circ(F\!\cdot\!\nabla)$. They
exist on every ball $U_i$ of the cover precisely because all
three entries are locally $\bar\partial$-exact.

\subsection{The quasi-holomorphic sheaf
  \texorpdfstring{$\widetilde{\mathcal{Q}}$}{Qt}}

Using the local data, we can now define the sheaves $\Omega^{0,q}(\widetilde{\mathcal{Q}})$, and the quasi-holomorphic sheaf $\widetilde{\mathcal{Q}}$. On each overlap $U_{ij} \coloneqq  U_i \cap U_j$, define the
\emph{transition operator}
\begin{equation}
  \psi_{ij} \;\coloneqq \; G_i \circ G_j^{-1}.
\end{equation}

\begin{proposition}\label{prop:trans}
  The transition operators take the explicit form
  \begin{equation}\label{eq:psi-explicit}
    \psi_{ij} =
    \begin{pmatrix}
      1 & P_i - P_j &
        \lambda_i - \lambda_j + (P_j - P_i)N_j \\
      0 & 1 & N_i - N_j \\
      0 & 0 & 1
    \end{pmatrix},
  \end{equation}
  and satisfy:
  \begin{enumerate}[label=(\roman*)]
    \item \emph{Cocycle condition:}
      $\psi_{ij}\circ\psi_{jk}\circ\psi_{ki} = \mathrm{Id}$
      on $U_i\cap U_j\cap U_k$.
    \item \emph{Holomorphicity:}
      $[\bar\partial, \psi_{ij}] = 0$ on $U_{ij}$.
  \end{enumerate}
\end{proposition}

\begin{proof}
The formula \eqref{eq:psi-explicit} follows from direct matrix
multiplication. Part~(i) holds tautologically:
$\psi_{ij}\psi_{jk}\psi_{ki} = G_i G_j^{-1}G_j G_k^{-1}G_k G_i^{-1}
= \mathrm{Id}$.
For~(ii), on $U_{ij}$ both trivialisations hold simultaneously:
\begin{equation}
    \bar{D} = G_i^{-1}\bar\partial G_i = G_j^{-1}\bar\partial G_j\:.
\end{equation}
This gives
\begin{equation}
    \bar\partial G_i = \psi_{ij}\bar\partial G_j\:,
\end{equation}
from which $[\bar\partial,\psi_{ij}] = 0$ follows immediately.
\end{proof}

Since $\{\psi_{ij}\}$ satisfies the cocycle condition, the standard
sheaf-gluing construction~\cite{hartshorne_algebraic_1977} produces a
globally well-defined smooth sheaf $\Omega^{0,q}(\widetilde{\mathcal{Q}})$
on $X$, whose local trivialisations over $U_i$ are
$\Omega^{0,q}(U_i, T^{1,0}\oplus\mathfrak{g}\oplus T^{*1,0})$.
Since $\bar\partial$ commutes with each $\psi_{ij}$ by
Proposition~\ref{prop:trans}(ii), it descends to a well-defined
operator on $\Omega^{0,q}(\widetilde{\mathcal{Q}})$.
\begin{definition}
  The \emph{quasi-holomorphic extension sheaf} is

  \[
    \widetilde{\mathcal{Q}} \;\coloneqq \;
    \ker\!\bigl(\bar\partial :
    \Omega^{0,0}(\widetilde{\mathcal{Q}}) \to
    \Omega^{0,1}(\widetilde{\mathcal{Q}})\bigr).
  \]
\end{definition}

The glued sheaf is in fact isomorphic to the sheaf of smooth sections of the
fixed bundle $Q=T^{1,0}\oplus\mathfrak{g}\oplus T^{*1,0}$, the whole content of
the gluing being the operator that $\bar\partial$ becomes under the
isomorphism.

\begin{proposition}[Sheaf isomorphism]\label{prop:iso}
For each $s\geq0$ the assignment
\begin{equation}\label{eq:Phi}
  \Phi:\ \Omega^{0,s}(\widetilde{\mathcal{Q}})\longrightarrow\Omega^{0,s}(Q),
  \qquad (\sigma_i)_i\longmapsto\gamma,\quad
  \gamma|_{U\cap U_i}\coloneqq G_i^{-1}\sigma_i ,
\end{equation}
is an isomorphism of sheaves of $\mathbb{C}$-vector spaces, with inverse
$\gamma\mapsto(G_i\gamma|_{U\cap U_i})_i$, and it intertwines $\bar\partial$
on $\Omega^{0,\bullet}(\widetilde{\mathcal{Q}})$ with $\bar D$ on
$\Omega^{0,\bullet}(Q)$: $\Phi\circ\bar\partial=\bar D\circ\Phi$.
\end{proposition}

\begin{proof}
On $U\cap U_{ij}$ one has $G_i^{-1}\sigma_i=G_i^{-1}\psi_{ij}\sigma_j
=G_j^{-1}\sigma_j$, so the local expressions agree and $\gamma$ is a section
of $\Omega^{0,s}(Q)$ over $U$. Conversely, for $\gamma\in\Omega^{0,s}(U,Q)$ the
family $\sigma_i\coloneqq G_i\gamma$ satisfies $\psi_{ij}\sigma_j=\sigma_i$. The two
maps are mutually inverse since each $G_i$ is invertible, and both are
compatible with restriction because $G_i^{\pm1}$ are differential operators.
Finally $\bar\partial\sigma_i=\bar\partial(G_i\gamma)=G_i\bar D\gamma$ by
\eqref{eq:triv}.
\end{proof}

\begin{corollary}[$\widetilde{\mathcal{Q}}$ is canonical]\label{cor:kerD}
$\Phi$ restricts to an isomorphism of sheaves of $\mathbb{C}$-vector spaces
\begin{equation}\label{eq:kerD}
  \widetilde{\mathcal{Q}}\;\cong\;
  \ker\bigl(\bar D:\Omega^{0,0}(Q)\to\Omega^{0,1}(Q)\bigr),
\end{equation}
the sheaf of $\bar D$-closed smooth sections of $Q$. In particular
$\widetilde{\mathcal{Q}}$ does not depend on the cover, on the holomorphic
frames in which the $A_i$ are written, or on the $\bar\partial$-exact
ambiguity absorbed into $Q^i_a$.
\end{corollary}

Concretely, a local section of $\widetilde{\mathcal{Q}}$ is a triple
$(\mu,\alpha,x)$ of smooth sections with $\bar\partial x=0$,
$\bar\partial\alpha=-F_a(\del x)^a$ and
$\bar\partial\mu=-\tr(F\wedge\del_A\alpha)-T(x)$, and similarly for $\delb$-closed local sections of $\Omega^{0,q}(\widetilde{\mathcal{Q}})$. The local trivialisations
enter only in showing that such triples exist locally in sufficient number,
which is the content of the following lemma.

\begin{lemma}[$\bar{D}$-Poincar\'e lemma]\label{lem:poincare}
  Every $\bar{D}$-closed form in $\Omega^{0,s}(Q)$, for $s\ge 1$, is locally $\bar{D}$-exact.
\end{lemma}

\begin{proof}
On $U_i$, $\bar{D}\phi = 0$ iff $\bar\partial(G_i\phi) = 0$. The
ordinary $\bar\partial$-Poincar\'e lemma on the ball
$U_i$ gives $\eta$ with $G_i\phi = \bar\partial\eta$. Then
$\bar{D}(G_i^{-1}\eta) = G_i^{-1}\bar\partial\eta = \phi$.
\end{proof}

\begin{corollary}\label{cor:res}
  The sequence of sheaves
  \[
    0 \;\longrightarrow\; \widetilde{\mathcal{Q}}
      \;\longrightarrow\; \Omega^{0,0}(\widetilde{\mathcal{Q}})
      \;\xrightarrow{\;\bar\partial\;}
      \Omega^{0,1}(\widetilde{\mathcal{Q}})
      \;\xrightarrow{\;\bar\partial\;} \cdots
  \]
  is an exact fine resolution of $\widetilde{\mathcal{Q}}$.
\end{corollary}

\begin{proof}
Local exactness at each $\Omega^{0,s}(\widetilde{\mathcal{Q}})$
for $s\geq 1$ is Lemma~\ref{lem:poincare}. For fineness, transport the
$C^\infty_X$-module structure of $\Omega^{0,s}(Q)$ along
Proposition~\ref{prop:iso}: a smooth function $f$ acts on a family
$(\sigma_i)$ by
\begin{equation}\label{eq:twisted}
  f\cdot(\sigma_i)\coloneqq \bigl(G_i\,f\,G_i^{-1}\sigma_i\bigr)_i ,
\end{equation}
which is compatible with the gluing since
$\psi_{ij}G_jfG_j^{-1}=G_ifG_i^{-1}\psi_{ij}$. A sheaf of
$C^\infty_X$-modules on a paracompact Hausdorff space is fine. Note that
\eqref{eq:twisted} is not multiplication of the local components by $f$. That
operation does not commute with the $\psi_{ij}$ and does not descend to
$\Omega^{0,s}(\widetilde{\mathcal{Q}})$.
\end{proof}

Fine sheaves have vanishing \v{C}ech cohomology in positive degree
\cite{griffiths_principles_1994}, so Corollary~\ref{cor:res} gives the
acyclicity needed for the abstract de~Rham, or Dolbeault theorem, which is Theorem \ref{thm:main} below.

\subsection{The Dolbeault theorem}

The proof of the Dolbeault theorem relating $\bar D$-cohomology to \v{C}ech cohomology now follows in the standard way. First,  we define the subsheaves of $\bar\partial$-closed forms:
\begin{equation}
    Z^{0,q} \coloneqq  \ker\!\bigl(\bar\partial :
  \Omega^{0,q}(\widetilde{\mathcal{Q}}) \to
  \Omega^{0,q+1}(\widetilde{\mathcal{Q}})\bigr)\,, \qquad q \geq 0\:.
\end{equation}
The fine resolution of Corollary~\ref{cor:res} then breaks into short
exact sequences
\begin{equation}
    0 \to \widetilde{\mathcal{Q}} \to \Omega^{0,0}(\widetilde{\mathcal{Q}})
  \xrightarrow{\bar\partial} Z^{0,1} \to 0\,,
  \qquad
  0 \to Z^{0,q} \to \Omega^{0,q}(\widetilde{\mathcal{Q}})
  \xrightarrow{\bar\partial} Z^{0,q+1} \to 0\;\;(q\geq 1)\:.
  \label{eq:sesZ}    
\end{equation}
By Corollary~\ref{cor:res}, $Z^{0,0}=\widetilde{\mathcal{Q}}$, so the first sequence in \eqref{eq:sesZ} is the $q=0$ case of the second sequence. Each of these short exact sequences gives a long exact sequence in \v{C}ech cohomology. By Corollary~\ref{cor:res} the sheaves $\Omega^{0,q}(\widetilde{\mathcal{Q}})$ are fine, so their \v{C}ech cohomology vanishes in positive degree, and for $p\geq1$ the long exact sequence contains the segment
\begin{equation}
    0=\check{H}^p\bigl(\Omega^{0,q}(\widetilde{\mathcal{Q}})\bigr)\to\check{H}^p(Z^{0,q+1})\to\check{H}^{p+1}(Z^{0,q})\to\check{H}^{p+1}\bigl(\Omega^{0,q}(\widetilde{\mathcal{Q}})\bigr)=0\:.
\end{equation}
Hence $\check{H}^{p+1}(Z^{0,q})\cong\check{H}^p(Z^{0,q+1})$ for $p\geq1$, which lowers the \v{C}ech degree by one and raises the form degree by one. Starting from $Z^{0,0}=\widetilde{\mathcal{Q}}$ and iterating, we find for $q\geq2$
\begin{equation}
    \check{H}^q(X,\widetilde{\mathcal{Q}})\cong\check{H}^{q-1}(Z^{0,1})\cong\cdots\cong\check{H}^1(Z^{0,q-1})\:.
\end{equation}
For the last step we use the beginning of the long exact sequence of \eqref{eq:sesZ} for $Z^{0,q-1}$,
\begin{equation}
    0\to\check{H}^0(Z^{0,q-1})\to\check{H}^0\bigl(\Omega^{0,q-1}(\widetilde{\mathcal{Q}})\bigr)\xrightarrow{\;\bar\partial\;}\check{H}^0(Z^{0,q})\to\check{H}^1(Z^{0,q-1})\to0\:,
\end{equation}
where the last zero is again $\check{H}^1(\Omega^{0,q-1}(\widetilde{\mathcal{Q}}))=0$. Since $\check{H}^0$ is the space of global sections, this gives
\begin{equation}
    \check{H}^1(Z^{0,q-1})\cong\frac{\{\sigma\in\Omega^{0,q}(X,\widetilde{\mathcal{Q}}):\bar\partial\sigma=0\}}{\bar\partial\,\Omega^{0,q-1}(X,\widetilde{\mathcal{Q}})}=H^{0,q}_{\bar\partial}(X,\widetilde{\mathcal{Q}})\:.
\end{equation}
For $q=1$ only this last step is needed, and for $q=0$ one has directly $\check{H}^0(\widetilde{\mathcal{Q}})=H^{0,0}_{\bar\partial}(X,\widetilde{\mathcal{Q}})$. Altogether,
\begin{equation}\label{eq:chain}
  \check{H}^q(X,\widetilde{\mathcal{Q}})\;\cong\;H^{0,q}_{\bar\partial}(X,\widetilde{\mathcal{Q}})\:.
\end{equation}

Finally, we identify the right-hand side with the $\bar D$-cohomology. Applied to global sections, Proposition~\ref{prop:iso} identifies $\bar\partial$-closed, respectively $\bar\partial$-exact, families $(\sigma_i)$ with $\bar D$-closed, respectively $\bar D$-exact, sections $\gamma=G_i^{-1}\sigma_i$ of $Q$, through the mutually inverse maps
\begin{equation}
    \label{eq:Gmap}
    [\gamma]\;\mapsto\;[(G_i\gamma)_{i\in I}]\:,\qquad
    [(\sigma_i)]\;\mapsto\;[G_i^{-1}\sigma_i]\:.
\end{equation}
By Corollary \ref{cor:kerD} and Proposition \ref{prop:iso}, this does not depend on the choices made. For two sets of choices $G_i$ and $\hat G_i$, the map $\sigma_i\mapsto\hat G_iG_i^{-1}\sigma_i$ identifies the two complexes $\Omega^{0,\bullet}(X,\widetilde{\mathcal Q})$ and corresponds to the identity on $\Omega^{0,\bullet}(X,Q)$. 
This gives a canonical isomorphism
\begin{equation}\label{eq:id}
  H^{0,q}_{\bar\partial}(X,\widetilde{\mathcal{Q}})
  \;\cong\;
  H^{0,q}_{\bar{D}}\!\bigl(X,\,
  T^{1,0}\oplus\mathfrak{g}\oplus T^{*1,0}\bigr).
\end{equation}
Finally, combining \eqref{eq:chain} and \eqref{eq:id} gives the Dolbeault theorem:

\begin{theorem}[Dolbeault theorem for $\bar{D}$]\label{thm:main}
  Let $X$ be a compact Calabi--Yau three-fold and $\bar{D}$ the
  heterotic deformation operator. For each $q \geq 0$ there is a
  canonical isomorphism
  \begin{equation}\label{eq:dolbeault}
    \boxed{H^{0,q}_{\bar{D}}\!\bigl(X,\,
    T^{1,0}\oplus\mathfrak{g}\oplus T^{*1,0}\bigr)
    \;\cong\;
    \check{H}^q(X,\,\widetilde{\mathcal{Q}}),}
  \end{equation}
    the composite of \eqref{eq:chain} with \eqref{eq:id}. The latter is
  given explicitly by \eqref{eq:Gmap}.
\end{theorem}

We note that when $F=0$ and $\omega^{\mathrm{CS}}_a=0$, in flat holomorphic frames ($A_i=0$, $\eta_i=0$) the operators $N_i$, $P_i$, $\lambda_i$ vanish, and $G_i=\mathrm{Id}$. Then $\widetilde{\mathcal{Q}}$ reduces to the direct sum $T^{1,0}\oplus\mathfrak{g}\oplus T^{*1,0}$ of holomorphic bundles, and Theorem~\ref{thm:main} reduces to the usual Dolbeault theorem.

It is worth being explicit about what Theorem~\ref{thm:main} does and does not give on the backgrounds considered here. Proposition~\ref{prop:kahler}
already computes $H^{0,q}_{\bar D}(Q)$ on K\"ahler Calabi--Yau backgrounds in terms of standard cohomology groups, so on such a background the theorem
identifies two groups both of which are known independently. The content of Theorem~\ref{thm:main} is therefore not computational. Rather, it is the statement that a sheaf-theoretic realisation of the $\bar D$-cohomology exists at all, which is not obvious for an operator with derivative-dependent extension maps. 

Note also that the arguments of this section do not use the $\del\delb$-lemma or Hodge theory. Indeed, the local trivialisations, the transition operators and the resulting Dolbeault isomorphism use only $\bar D^2=0$, the ordinary $\delb$-Poincar\'e lemma on balls, and partitions of unity. They therefore
apply unchanged on a compact complex three-fold $X$ that is not K\"ahler, provided $\bar D$ can be defined there, that is, provided $X$ carries
\begin{equation}\label{eq:nonKahler}
\begin{gathered}
  \Omega\in\Omega^{3,0}(X)\ \text{holomorphic and nowhere vanishing}\:,\\
  \omega^{2,1}\in\Omega^{2,1}(X)\ \text{with}\ \delb\omega^{2,1}=\tr(F\wedge F)\:.
\end{gathered}
\end{equation}
The cubic theory of Section~\ref{sec:cubic} needs in addition
$\del\omega^{2,1}=0$. On K\"ahler backgrounds the $\del\delb$-lemma supplies $\omega^{2,1}$, as in \eqref{eq:dCScomp}. On non-K\"ahler backgrounds its
existence is an assumption. Such backgrounds lie outside the reach of Proposition~\ref{prop:kahler}, and it is there that $\widetilde{\mathcal Q}$ could carry extra information beyond the split complex. In particular, the Hull--Strominger system is not covered here. There $\tr(F\wedge F)$ is balanced against $\tr(R\wedge R)$ in the heterotic Bianchi identity (cf.\ footnote~\ref{foot:YMflat}). Including these non-Kähler backgrounds requires the inclusion of the gravitational anomaly, which we leave to future work.

\section{Conclusions and outlook}
\label{sec:concl}

In this paper we have constructed a dual version of the Kodaira--Spencer theory of \cite{Ashmore:2018ybe}, incorporating the Green--Schwarz mechanism of Costello--Li \cite{Costello:2015xsa,Costello:2019jsy, Li:2026eyh}. We work on Kähler Calabi--Yau backgrounds and ignore the gravitational anomaly throughout, leaving the study of torsional backgrounds and the Hull--Strominger system to future work. At quadratic order this produces the nilpotent operator $\bar D$ of \eqref{eq:barD}, whose extension maps carry explicit holomorphic derivatives, so that $(Q,\bar D)$ is not a holomorphic vector bundle, but a ``quasi-holomorphic sheaf'' \cite{MurgasIbarra:2025ylk}. At cubic order the same structure admits a compatible bracket and cyclic pairing, and the failure of the Jacobi identity is controlled by the Yukawa cubic. Enlarging the complex by a scalar $(0,\bullet)$-form as in \cite{Ashmore:2018ybe} turns this into an extended 
$L_3$ algebra whose products are completely determined. Finally, we have shown that $\bar D$ trivialises locally to $\delb$, and the resulting transition operators glue into the quasi-holomorphic extension sheaf $\widetilde{\mathcal Q}$ of Theorem~\ref{thm:main}.

The nature of $\widetilde{\mathcal Q}$ deserves a caveat. The transition operators $\psi_{ij}$ contain holomorphic derivatives and are therefore not $\mathcal{O}_X$-linear: for holomorphic $f$ one has $N_i(fx)\neq fN_i(x)$. The object glued in Section~\ref{sec:local-triv} is consequently a sheaf of $\mathbb{C}$-vector spaces and not a coherent analytic sheaf. The sheaves $\Omega^{0,s}(\widetilde{\mathcal Q})$ are nonetheless fine, but with respect to the module structure transported from $\Omega^{0,s}(Q)$ by Proposition~\ref{prop:iso} and Corollary \ref{cor:res}, rather than the naive multiplication of local components. What survives is what we have used: \v{C}ech cohomology, the fine resolution, the identification of $\widetilde{\mathcal Q}$ with the sheaf of $\bar D$-closed sections of $Q$, and the Dolbeault theorem $H^{0,q}_{\bar D} (Q)\cong \check H^q(\widetilde{\mathcal{Q}})$. The heavier machinery of coherent sheaves (Serre duality, Riemann--Roch, and with them the determinant-line constructions underlying holomorphic Ray--Singer torsion) does not transfer for free, and extending the present description far enough to reach those invariants remains to be done. Note that an object glued by differential operators and filtered by their order is naturally $\mathcal{D}$-module-shaped, and that is perhaps the natural setting in which to look.

The scope of the construction is set by the assumption that $\tr(F^2)$ be trivial in cohomology, which on a compact K\"ahler manifold is the statement that $[\mathrm{ch}_2(V)]$ vanish in $H^4(X,\mathbb{R})$. We emphasise that the Hermitian Yang--Mills condition is nowhere imposed: the connection on $V$ is only assumed to be the Chern connection of some Hermitian metric, so the
rigidity result of \cite{Garcia-Fernandez:2023vah} invoked in footnote \ref{foot:YMflat} in Section~\ref{sec:quad} does not apply. Restoring the gravitational anomaly would replace $\tr(F^2)$ by $\tr(F^2)-\tr(R^2)$, which is the heterotic Bianchi identity, and would therefore admit torsional Hull--Strominger backgrounds that the present assumption excludes. That is the most immediate extension of this work, and the one which would connect it to the compactifications that motivate the problem.

Several other questions are left open. The present analysis of anomaly cancellation is carried out at one loop, where the Green--Schwarz mechanism of Costello--Li cancels the ${\rm ch}_4$ anomaly of the quadratic theory. The mixed term $c_2(X)\,{\rm ch}_2({\rm End}\,V)$ remains, and should be treated together with the gravitational anomalies, possibly by a counterterm $\int_Xc_2(X)\wedge x$. At higher loop order, new counterterms may be required, and the structure of the corrected differential $\bar{D}$ may itself be modified. A systematic treatment of higher-loop anomalies in the present framework -- in particular the question of whether the quasi-holomorphic structure of $(Q, \bar{D})$ is stable under such corrections -- remains to be carried out. Furthermore, a recurring difficulty in the present paper is that $\bar{D}$ is not the $(0,1)$-part of a connection, owing to the holomorphic-derivative-dependent extension maps. The standard derivation of the one-loop anomaly -- which proceeds by transforming the path-integral measure against a background gauge rotation and uses the curvature of the gauge-fixed kinetic operator -- does not apply directly to such operators. Proposition~\ref{prop:gaugeFix} shows that the standard diagonal $\bar\partial^\dagger$-closed gauge condition, supplemented by $\del^\dagger x=0$, selects exactly one representative of each gauge-equivalence class, and we have argued somewhat heuristically that the anomaly then reduces to that of the flux-free theory. A more systematic treatment, establishing how the anomaly formulas are corrected for operators with non-tensorial extensions, is left for future work.

The one-loop partition function of the theory might also suffer from geometric anomalies, that is, 
a dependence on the background geometric structures, such as a metric and complex structure dependence, leading to a potential holomorphic anomaly equation \cite{Bershadsky:1993ta, Bershadsky:1993cx}. The one-loop partition function is governed by the determinant of the gauge-fixed kinetic operator, and on compact backgrounds this is related to the holomorphic Ray--Singer torsion of the complex $(Q, \bar{D})$. However, as discussed, the quasi-holomorphic sheaf $\tilde{\mathcal{Q}}$ is a sheaf of $\mathbb{C}$-vector spaces rather than a coherent analytic sheaf, so the determinant-line constructions underlying holomorphic Ray--Singer torsion do not transfer immediately. Extending the present description far enough to define and compute such torsion invariants, and to understand what replaces the standard Quillen metric in this non-tensorial setting, is an interesting open problem.

The one-loop partition function also provides a starting point for a study of enumerative geometry and topological invariants in this setting. In the context of derived geometry, such enumerative problems are best behaved for operators of vanishing index, or Euler characteristic, which has been established to be the case for the heterotic moduli problem \cite{deLazari:2024zkg, Chisamanga:2024xbm, Ibarra:2024ntg}. This leads to a virtual dimension zero moduli space, where solutions correspond to a set of points which may in principle be counted. For an ordinary holomorphic bundle $V$ on a Calabi--Yau manifold, Serre duality relates $H^{0,q}_{\bar\partial}(V)$ to $H^{0,n-q}_{\bar\partial}(V)$ of the dual bundle, and the Hirzebruch--Riemann--Roch theorem expresses the Euler characteristic in terms of characteristic classes. For the quasi-holomorphic complex $(Q, \bar{D})$ the analogous statements are not immediate, since the transition operators are not $\mathcal{O}_X$-linear and the standard coherent-sheaf machinery does not apply. Establishing Serre duality and an index theorem for $\bar{D}$, and identifying the correct replacement for the Atiyah class, which in the ordinary setting measures the obstruction to holomorphicity of the connection, are natural problems. As noted above, a $\mathcal{D}$-module perspective may be the appropriate one here.

Concerning the $L_3$ structure of Section~\ref{sec:extended}, we note that the structure is not cyclic: the extended pairing
is degenerate along the auxiliary direction, for the degree reason given in Section~\ref{sec:MCext}, and no known action produces the $L_3$ Maurer--Cartan equation directly as its equations of motion. This matches the situation of \cite{Ashmore:2018ybe}, where the same is observed, and reflects the fact that the Maurer--Cartan equation encodes two conditions -- the equation of motion and the auxiliary equation \eqref{eq:auxMC} -- rather than one. What distinguishes the two cases is that in \cite{Ashmore:2018ybe} the auxiliary equation is equivalent, given the equation of motion, to
the vanishing of the action, whereas here it is the statement that the deformed three-form $e^{\iota_V}\Omega$ is cohomologous to $\Omega$, and Proposition~\ref{prop:torelli} shows that on a compact Calabi--Yau three-fold this is implied by the equation of motion for small fields. The Maurer--Cartan equation then carries no information beyond the equation of motion, which is
a stronger statement than the one available in \cite{Ashmore:2018ybe}, though here we work on Kähler backgrounds, whereas the backgrounds of \cite{Ashmore:2018ybe} were torsional in general. Whether an action, for instance with a further ``dilaton'' multiplier coupling to $\delb b$, can be made to reproduce the Maurer--Cartan equation directly is left open, though see \cite{Kupka:2026usx, Jonker:2026cok} for the case of the action of \cite{Ashmore:2018ybe}. The truncation $\ell_{k\ge4}=0$ was shown in Proposition~\ref{prop:l4} to follow from the Jacobiator identity contracted with a fourth vector, using the complex dimension ${\rm dim}_{\mathbb{C}}(X)=3$. Whether the deformation problem in other dimensions is governed by a genuinely non-truncating $L_\infty$ structure, and what the higher brackets are, is left to future work.

The quasi-holomorphic extension sheaf $\tilde{\mathcal{Q}}$ constructed in Section~\ref{sec:local-triv} is an instance of a more general class of sheaves whose transition operators are differential rather than algebraic. The present paper handles extension maps that are at most second order in holomorphic derivatives, as required by the Hermitian-to-Beltrami entry $T$. A systematic theory of such objects, including a classification order by order of the differential operators involved, functorial constructions such as tensor products and duals, and the analogue of the Atiyah class measuring the obstruction to a holomorphic connection, would provide the right framework for the heterotic moduli problem and its higher-derivative (torsional) corrections, and is also of independent mathematical interest.


\section*{Acknowledgements}
We thank Logan Arnold-Lee, Alex S. Arvanitakis, Anthony Ashmore,  Xenia de la Ossa, Fabian Hahner, Eirik Høgmoe Kjelsnes, Julian Kupka, Jock McOrist, Ruben Minasian, Javier Murgas Ibarra, Sébastien Picard, Surya Raghavendran, Ingmar Saberi, Charles Strickland-Constable, Markus Upmeier and Fridrich Valach for enlightening conversations surrounding the work presented here. In some places Claude Code has been used for verifying the rigor of the claims. All AI-assisted output was reviewed and independently verified by the authors, and the authors take full responsibility for the mathematical content and exposition of this paper.

\appendix

\section{Proofs of the bracket identities}
\label{app:cubic}

This appendix supplies the computations behind
Proposition~\ref{prop:leibniz}, Lemma~\ref{lem:cyclic},
Proposition~\ref{prop:jacobi} and Proposition~\ref{prop:l4}. The conventions of
Section~\ref{sec:normalisation} are in force throughout.

\subsection{The graded Leibniz rule}
\label{app:leibniz}

\begin{proof}[Proof of Proposition~\ref{prop:leibniz}]
Throughout we abbreviate $V^a=(\partial x)^a$, $V'^a=(\partial x')^a$, and
write $[\,\cdot,\cdot\,]$ for the graded commutator on
$\mathrm{End}(V)$-valued forms. We shall use the following four
identities, each of which follows from the conventions of
Section~\ref{sec:quad}.

\emph{(a) Commutator of $\bar\partial$ with the covariant derivative.}
For $\alpha\in\Omega^{0,p}(\mathfrak g)$,
\begin{equation}
    \label{eq:dbnabla}
    [\bar\partial,\nabla_a]\,\alpha=-\,[F_a,\alpha]\:,
\end{equation}
which is the component form of $[\bar\partial,\del_A]=-[F,\,\cdot\,]$
together with $F=\dd z^a\wedge F_a$.

\emph{(b) Bianchi identity in components.} The $(2,1)$-part of
$\del_AF=0$ reads
\begin{equation}
    \label{eq:bianchicomp}
    \nabla_aF_b=\nabla_bF_a\:.
\end{equation}

\emph{(c) The cotangent component and the Lie bracket.} Applying the map
\eqref{eq:dxdef} to \eqref{eq:bracket_cotangent} gives
\eqref{eq:cot_char}, that is
\begin{equation}
    \label{eq:cotlie}
    \bigl(\partial[y,y']_\bullet\bigr)^a
    =-\bigl(V^b\partial_bV'^a-(-1)^{pq}\,V'^b\partial_bV^a\bigr)\:.
\end{equation}

\emph{(d) Contraction identity.} For every $e$,
\begin{equation}
    \label{eq:contract}
    F\wedge\iota_{\del_e}\Omega=\Omega\wedge F_e\:.
\end{equation}

Since $\bar D$ is block upper-triangular, \eqref{eq:leibniz} splits into
one identity per component of the bracket. We treat them in turn,
and within each we group terms by which of $\mu,\alpha,x$ occur.

\medskip\noindent\textbf{Cotangent component.} Only $\bar\partial$ acts
on this slot, and \eqref{eq:bracket_cotangent} is bilinear in $x$, so
the identity to verify is
\[
\bar\partial\bigl(\Omega_{abc}V^b\wedge V'^c\bigr)
=\Omega_{abc}\,(\partial\bar\partial x)^b\wedge V'^c
 +(-1)^p\,\Omega_{abc}\,V^b\wedge(\partial\bar\partial x')^c\:.
\]
This is the Leibniz rule for $\bar\partial$, using that $\Omega_{abc}$
is holomorphic, hence $\bar\partial$-closed, and that $\bar\partial$
commutes with the holomorphic derivatives in \eqref{eq:dxdef}.

\medskip\noindent\textbf{Gauge component.} Write
$G=[y,y']_{\mathfrak g}$ and $X=[y,y']_\bullet$. The gauge row
of $\bar D$ contributes $\bar\partial G+F_a(\partial X)^a$, so we must
show
\begin{equation}
    \label{eq:gaugeblock}
    \bar\partial G+F_a(\partial X)^a
    =[\bar Dy,y']_{\mathfrak g}
     +(-1)^p[y,\bar Dy']_{\mathfrak g}\:.
\end{equation}
The terms bilinear in $\alpha$ are
$\bar\partial[\alpha,\alpha']_{\mathfrak g}
=[\bar\partial\alpha,\alpha']_{\mathfrak g}
+(-1)^p[\alpha,\bar\partial\alpha']_{\mathfrak g}$, the Leibniz rule
for the pointwise bracket.

For the terms with one $\alpha$ and one $x$, apply $\bar\partial$ to
$-V^a\nabla_a\alpha'$ and use \eqref{eq:dbnabla}:
\[
\bar\partial\bigl(-V^a\nabla_a\alpha'\bigr)
=-(\partial\bar\partial x)^a\nabla_a\alpha'
 -(-1)^pV^a\nabla_a\bar\partial\alpha'
 +(-1)^pV^a[F_a,\alpha']\:.
\]
The first two terms are exactly the corresponding terms of the
right-hand side of \eqref{eq:gaugeblock} obtained by inserting
$\bar\partial x$ and $\bar\partial\alpha'$. The third equals
the contribution of the metric-to-gauge entry of $\bar Dy$ to the
pointwise bracket,
$[F_aV^a,\alpha']_{\mathfrak g}=(-1)^pV^a[F_a,\alpha']$, on the
right-hand side of \eqref{eq:gaugeblock}. The terms with
$\alpha$ and $x'$ cancel in the same way.

For the terms bilinear in $x$, the left-hand side of
\eqref{eq:gaugeblock} contributes $F_a(\partial X)^a$, which by
\eqref{eq:cotlie} equals
$-F_a\bigl(V^b\partial_bV'^a+V'^b\partial_bV^a\bigr)$ at $p=q=1$. The right-hand
side contributes the two terms in which the metric-to-gauge entry of
$\bar D$ has acted on one argument and the resulting gauge element is
paired against the cotangent component of the other, namely
$(-1)^{(p+1)q}V'^a\nabla_a(F_bV^b)-(-1)^pV^a\nabla_a(F_bV'^b)$, the
two prefactors being those of \eqref{eq:bracket_gauge} and of
\eqref{eq:leibniz} respectively; at $p=q=1$ both equal $+1$. Expanding
by the Leibniz rule and transposing each $\nabla_aF_b$ to the right
past one form factor, the terms free of derivatives of $V$, $V'$
assemble into
\[
V^aV'^b\bigl(\nabla_bF_a-\nabla_aF_b\bigr)\:,
\]
which vanishes by the Bianchi identity \eqref{eq:bianchicomp}. Note
that the anti-symmetry which effects this cancellation is carried by
the derivative indices of $\nabla_aF_b$, and not by the pair
$V^aV'^b$: the latter is graded symmetric in $(a,b)$ at
$p=q=1$, since transposing the two $(0,1)$-form factors costs
$(-1)^{pq}=-1$. The remaining terms, after the same transpositions
and a relabelling of dummy indices, equal
$-F_a\bigl(V^b\partial_bV'^a+V'^b\partial_bV^a\bigr)
=F_a(\partial X)^a$ by \eqref{eq:cotlie}, reproducing the left-hand
side and establishing \eqref{eq:gaugeblock}.

\medskip\noindent\textbf{Tangent component.} Write
$M=[y,y']^a$. The tangent row of $\bar D$ contributes
$\bar\partial M+\tr(F\wedge\del_AG)+T(X)$, the last two terms being
converted to a vector index through $\Omega$.

The terms bilinear in $(\mu,x)$ involve only $\bar\partial$, and close
by the Leibniz rule together with $[\bar\partial,\del_b]=0$; no
curvature enters.

For the terms bilinear in $\alpha$, apply $\bar\partial$ to the second
line of \eqref{eq:bracket_tangent}. Since the trace is a scalar,
$\nabla_c$ acts on it as $\del_c$, which commutes with $\bar\partial$,
so it suffices to expand $\bar\partial\tr(\alpha\nabla_b\alpha')$.
Using \eqref{eq:dbnabla},
\begin{equation}
    \label{eq:trexp}
    \bar\partial\tr(\alpha\nabla_b\alpha')
    =\tr(\bar\partial\alpha\,\nabla_b\alpha')
     +(-1)^p\tr(\alpha\,\nabla_b\bar\partial\alpha')
     -(-1)^p\tr\bigl(\alpha\,[F_b,\alpha']\bigr)\:.
\end{equation}
The first two terms on the right of \eqref{eq:trexp} reproduce the
second line of \eqref{eq:bracket_tangent} with $\bar\partial\alpha$
and $\bar\partial\alpha'$ inserted, that is, the corresponding terms
of $[\bar Dy,y']$ and $(-1)^p[y,\bar Dy']$. The residue is the
last term of \eqref{eq:trexp} together with its
$\alpha\leftrightarrow\alpha'$ partner, and after contraction with
$\bar\Omega^{cba}$ and use of the cyclicity of the trace this residue
is exactly $-\tr(F\wedge\del_A[\alpha,\alpha']_{\mathfrak g})$,
which is the contribution of the gauge-to-complex entry of $\bar D$
acting on the gauge component of the bracket. Moving $\nabla_c$ onto the
commutator alone also generates
$\bar\Omega^{cba}\tr\bigl((\nabla_cF_b)\,[\alpha,\alpha']_{\mathfrak g}\bigr)$,
in which the derivative falls on the curvature; this vanishes because
$\nabla_cF_b$ is symmetric in $c$ and $b$ by the Bianchi identity
\eqref{eq:bianchicomp} while $\bar\Omega^{cba}$ is antisymmetric. The two
cancel.

For the remaining two families we work at $p=q=1$ and use two facts. First,
a $\del$-exact two-form in the tangent slot corresponds to a vector
$(\del\theta)^a=\bar\Omega^{abc}\del_b\theta_c$. The gauge-to-complex entry
acting on $\beta$ is $(\del\vartheta(\beta))^a$ with
$\vartheta(\beta)_c\coloneqq\tr(F_c\beta)$, and, writing
$\omega^{2,1}=\tfrac12w_{bc}\,\dd z^{bc}$, the metric-to-complex entry acting on
a cotangent element with vector $U$ is $-\tfrac12(\del\theta(U))^a$ with
$\theta(U)_c\coloneqq w_{bc}U^b$. Second, for an even one-form $\theta$ and an odd
divergence-free vector $B$, \eqref{eq:inv} and \eqref{eq:OO} give
\begin{equation}
  \label{eq:Liecurl}
  [(\del\theta),B]=-\bigl(\del(L_B\theta)\bigr)\:,\qquad
  (L_B\theta)_c\coloneqq B^d\del_d\theta_c+(\del_cB^d)\,\theta_d\:.
\end{equation}
Each identity below then has the form $(\del\,\cdot\,)^a=(\del\,\cdot\,)^a$, and
it suffices that the two arguments differ by a gradient.

For the terms with $\alpha$ in $y$ and $x'$ in $y'$, the bracket has only the
gauge component $G=-V'^e\nabla_e\alpha$, so the left-hand side is
$(\del\vartheta(G))^a$ with $\vartheta(G)_c=V'^e\tr(F_c\nabla_e\alpha)$. On the
right, $[\bar Dy,y']$ contributes $-[(\del\vartheta(\alpha)),V']
=(\del L_{V'}\vartheta(\alpha))$ by \eqref{eq:Liecurl}, and
$-[y,\bar Dy']$ contributes the gauge bilinear of \eqref{eq:bracket_tangent} on
$(\alpha,F_dV'^d)$, which is $(\del K)^a$ with
\[
  K_c=\tfrac12\Bigl(\tr\bigl(\alpha\nabla_c(F_dV'^d)\bigr)
  -\tr\bigl(F_dV'^d\nabla_c\alpha\bigr)\Bigr)\:.
\]
Expanding with $\del_d\tr(F_c\alpha)=\tr(\nabla_dF_c\,\alpha)+\tr(F_c\nabla_d\alpha)$,
the Bianchi identity \eqref{eq:bianchicomp} and graded cyclicity of the trace,
\[
  L_{V'}\vartheta(\alpha)_c+K_c-\vartheta(G)_c
  =\tfrac12\,\del_c\bigl(V'^d\tr(F_d\alpha)\bigr)\:.
\]
The terms with $x$ in $y$ and $\alpha'$ in $y'$ follow by
Lemma~\ref{lem:brasym}.

For the terms bilinear in $x$, the bracket has only the cotangent component
$X$, with $(\del X)=-[V,V']$ by \eqref{eq:cotlie}, so the left-hand side is
$T(X)=\tfrac12(\del\theta([V,V']))^a$. On the right, $T$ acts on one argument
and the result enters the first line of \eqref{eq:bracket_tangent}, giving
$-[\tau,V']-[\tau',V]$ with $\tau=-\tfrac12(\del\theta(V))$ and
$\tau'=-\tfrac12(\del\theta(V'))$. By \eqref{eq:Liecurl} the identity reduces
to
\[
  \theta([V,V'])_c+\bigl(L_{V'}\theta(V)\bigr)_c+\bigl(L_V\theta(V')\bigr)_c
  =-\del_c\bigl(V'^dw_{db}V^b\bigr)\:,
\]
which follows on expanding the left-hand side: the terms containing
$\del V$ or $\del V'$ cancel in pairs, and the remaining two combine by
$\del\omega^{2,1}=0$, that is $\del_{[d}w_{bc]}=0$. This is the second
condition in \eqref{eq:dCScomp}; the Chern--Weil identity \eqref{eq:CW} plays
no role, and this sector does not constrain the normalisation $c\kappa$, which
is fixed by Proposition~\ref{prop:nilp} alone.
Collecting the three components completes the proof.
\end{proof}

\subsection{Cyclicity of the pairing}
\label{app:cyclic}

\begin{proof}[Proof of Lemma~\ref{lem:cyclic}]
Both sides are trilinear, so it suffices to check the identity on
arguments each of which lies in a single summand of $Q$. The pairing
is anti-diagonal in the $(\mu,x)$ slots and diagonal in the gauge
slot, while the components of the bracket have the following field
content: the cotangent component \eqref{eq:bracket_cotangent} needs
$(x,x)$; the gauge component \eqref{eq:bracket_gauge} needs
$(\alpha,\alpha)$ or $(x,\alpha)$; the tangent component
\eqref{eq:bracket_tangent} needs $(\mu,x)$ or $(\alpha,\alpha)$.
Matching this content against the pairing, the nonvanishing
configurations fall into exactly three families of unordered content:
$(\alpha,\alpha,\alpha)$, $(x,\alpha,\alpha)$ and $(\mu,x,x)$.
Configurations with two or more $\mu$-arguments vanish, because $\mu$ enters the bracket only linearly. Configurations with three cotangent arguments vanish on both sides, since the cotangent component of the bracket pairs only against the $\mu$-slot, which is absent.
By Lemma~\ref{lem:brasym} and
trilinearity it suffices to verify one generating identity per
family; we work at degrees $(1,1,1)$, where the prefactor in
\eqref{eq:cyclic} equals $+1$, the general case following by carrying the
Koszul factors through the transpositions below.

\medskip\noindent\textbf{(i) Three gauge arguments.} Only
$[\alpha_2,\alpha_3]_{\mathfrak g}$ contributes, and the statement is
\[
\int_X\tr\bigl(\alpha_1\wedge[\alpha_2,\alpha_3]\bigr)\wedge\Omega
=\int_X\tr\bigl(\alpha_3\wedge[\alpha_1,\alpha_2]\bigr)\wedge\Omega\:,
\]
which is the cyclicity of the trace applied pointwise, no integration
by parts being required; both sides carry the same factor from the
gauge term of \eqref{eq:pairing-mu}, which therefore drops out.

\medskip\noindent\textbf{(ii) The family $(x,\alpha,\alpha)$.} Take $y_1=x_1$,
$y_2=\alpha_2$, $y_3=\alpha_3$. On the left, $x_1$ pairs against the
gauge bilinear of the tangent component, the second line of
\eqref{eq:bracket_tangent}; converting the tangent component to a
two-form through $\Omega$, the integrand is $x_{1a}M^a\wedge\Omega$
with
\[
M^a=-\tfrac12\,\bar\Omega^{cba}\Bigl(\nabla_c\tr(\alpha_2\nabla_b\alpha_3)
+\nabla_c\tr(\alpha_3\nabla_b\alpha_2)\Bigr)\:,
\]
the relative sign being the graded antisymmetrisation of
\eqref{eq:bracket_tangent} at $p_2=p_3=1$. Since the trace is a
scalar, $\nabla_c$ acts on it as $\partial_c$; integrating
$\partial_c$ by parts against $\Omega$ and using
$\bar\Omega^{cba}\partial_cx_{1a}=-V_1^b$ gives
\[
\int_X x_{1a}M^a\wedge\Omega
=-\tfrac12\int_X V_1^b\Bigl(\tr(\alpha_2\nabla_b\alpha_3)
+\tr(\alpha_3\nabla_b\alpha_2)\Bigr)\wedge\Omega\:.
\]
In the first summand, integrating $\nabla_b$ off $\alpha_3$, the
divergence term vanishes by \eqref{eq:autodiv} and one graded
transposition inside the trace makes the summand equal to the second.
Hence
\begin{align*}
\int_X\langle x_1,[\alpha_2,\alpha_3]\rangle
&=-\int_X V_1^b\,\tr(\alpha_3\nabla_b\alpha_2)\wedge\Omega\\
&=-\int_X\tr\bigl(\alpha_3\wedge(-V_1^b\nabla_b\alpha_2)\bigr)\wedge\Omega\\
&=\int_X\langle\alpha_3,[x_1,\alpha_2]\rangle\:,
\end{align*}
the second equality moving the $(0,1)$-form $V_1^b$ past the
$(0,1)$-form $\alpha_3$, and the last being the pairing of $\alpha_3$
against the transport term of \eqref{eq:bracket_gauge}, whose gauge
term carries the sign of \eqref{eq:pairing-mu}. This computation is what
fixes the coefficient $-\tfrac12$ of the gauge bilinear in
\eqref{eq:bracket_tangent}, consistently with
Proposition~\ref{prop:leibniz}.

\medskip\noindent\textbf{(iii) The family $(\mu,x,x)$.} Take $y_1=\mu_1$, $y_2=x_2$,
$y_3=x_3$. On the left,
$\chi_1=\tfrac12\mu_1^d\Omega_{dbc}\dd z^{bc}$ pairs against the
cotangent component $\Omega_{aef}V_2^eV_3^f\,\dd z^a$, and the
contraction identity \eqref{eq:contrOm} gives \eqref{eq:yukawa}.
On the right, $x_3$ pairs against the tangent component of
$[\mu_1,x_2]$, whose only contribution is the first line of
\eqref{eq:bracket_tangent},
$[\mu_1,x_2]^a=-[\mu_1,V_2]^a
=-(\mu_1^b\partial_bV_2^a+V_2^b\partial_b\mu_1^a)$ at odd degrees.
Setting $I\coloneqq \int_X x_{3a}\,[\mu_1,V_2]^a\wedge\Omega$ and integrating
$\partial_b$ by parts in each term, the divergence terms vanish by
$\partial_b\mu_1^b=0$ and \eqref{eq:autodiv}, leaving
\[
I=\int_X\partial_bx_{3a}\bigl(\mu_1^aV_2^b-\mu_1^bV_2^a\bigr)\wedge\Omega
=\tfrac12\int_X\bigl(\partial_bx_{3a}-\partial_ax_{3b}\bigr)
\bigl(\mu_1^aV_2^b-\mu_1^bV_2^a\bigr)\wedge\Omega\:,
\]
since only the part of $\partial x_3$ antisymmetric in $(a,b)$
survives the contraction. By the inversion identity \eqref{eq:inv},
$\partial_bx_{3a}-\partial_ax_{3b}=\Omega_{cba}V_3^c$, and after
relabelling and one even transposition,
\[
I=-\int_X\Omega_{aef}\,\mu_1^a\,V_2^e\,V_3^f\wedge\Omega\:.
\]
By \eqref{eq:pairing-mu} the left-hand side is
$\int_Xx_{3a}\,[\mu_1,x_2]^a\wedge\Omega=-I$, with no additional sign,
and this reproduces \eqref{eq:yukawa}. This computation fixes the
relative normalisation between \eqref{eq:bracket_cotangent} and the
first line of \eqref{eq:bracket_tangent}.

Collecting (i)--(iii) gives \eqref{eq:cyclic}.
\end{proof}

\subsection{The Jacobiator}
\label{app:jacobi}

\begin{proof}[Proof of Proposition~\ref{prop:jacobi}]
We work at degrees $p_1=p_2=p_3=1$, the case relevant for the
physical fields, so that all Koszul prefactors in \eqref{eq:Sdef}
equal $+1$; the general case follows by carrying the corresponding
factor $(-1)^{p_ip_j}$ through each transposition below. The general-degree
statement is the one used by Proposition~\ref{prop:relations} and
Appendix~\ref{app:gauge}; Lemma~\ref{lem:jac110} there is its $(1,1,0)$ case,
written out. Write
$V_i\coloneqq (\partial x_i)^a$; each $V_i$ is odd, divergence free by
\eqref{eq:autodiv}, and the graded Lie bracket reads
$[V_i,V_j]^a=V_i^b\partial_bV_j^a+V_j^b\partial_bV_i^a$.

\medskip\noindent\textbf{Cotangent component.} The cotangent component
of a bracket requires two cotangent arguments, and the cotangent slot
of an inner bracket feeds the outer one through
$V_{[ij]}\coloneqq \bigl(\partial\,[y_i,y_j]_\bullet\bigr)^a=-[V_i,V_j]^a$ by
\eqref{eq:cot_char}. Hence
\[
J_a=-\,\Omega_{abc}\Bigl([V_1,V_2]^bV_3^c+[V_3,V_1]^bV_2^c
+[V_2,V_3]^bV_1^c\Bigr)\:.
\]
We evaluate this by Cartan calculus, using the identities
$[\partial,\iota_V]=L_V$, $[L_{V_i},\iota_{V_j}]=-\iota_{[V_i,V_j]}$ and
\eqref{eq:LVOmega} recorded in Section~\ref{sec:normalisation}. Now
$W=\iota_{V_1}\iota_{V_2}\iota_{V_3}\Omega$, the contractions
commuting pairwise for odd arguments, and since $\partial\Omega=0$,
\[
\partial W
=L_{V_1}\iota_{V_2}\iota_{V_3}\Omega
+\iota_{V_1}L_{V_2}\iota_{V_3}\Omega
+\iota_{V_1}\iota_{V_2}L_{V_3}\Omega
=-\bigl(\iota_{[V_1,V_2]}\iota_{V_3}\Omega
+\iota_{V_2}\iota_{[V_1,V_3]}\Omega
+\iota_{V_1}\iota_{[V_2,V_3]}\Omega\bigr)\:,
\]
where each $L_{V_i}$ has been commuted through the remaining
contractions and \eqref{eq:LVOmega} kills the terms in which it
reaches $\Omega$. Reading off components with $(\partial W)_a=-\partial_aW$, $W$ having
degree three, and
$(\iota_U\iota_V\Omega)_a=(-1)^{|V|+1}\,\Omega_{abc}U^bV^c$, and comparing
with the display above, gives $J_a=-\partial_aW$. Divergence-freeness enters
only through \eqref{eq:LVOmega}: for general vector-valued forms the
right-hand side would acquire divergence terms, and it is exactly
$\partial_aV^a=0$ that reduces the obstruction to an exact gradient.

\medskip\noindent\textbf{Gauge component.} By field content the
possible configurations are three $\alpha$'s, two $\alpha$'s and one
$x$, and one $\alpha$ and two $x$'s. For three $\alpha$'s the cyclic
sum vanishes by the pointwise graded Jacobi identity of
${\mathfrak g}$. For two $\alpha$'s and one $x$, writing
$L_V\alpha\coloneqq V^a\nabla_a\alpha$, the relevant combination is
$L_{V_1}[\alpha_2,\alpha_3]_{\mathfrak g}
-[L_{V_1}\alpha_2,\alpha_3]_{\mathfrak g}
-[\alpha_2,L_{V_1}\alpha_3]_{\mathfrak g}$ with graded signs, which
vanishes because $\nabla_a$ is a derivation of the pointwise bracket.
For one $\alpha$ and two $x$'s, the nested transport terms give the
graded commutator $[L_{V_1},L_{V_2}]\alpha_3$, while the cotangent
slot of the inner bracket feeds the transport term of the outer one,
contributing
$-V_{[12]}^a\nabla_a\alpha_3=[V_1,V_2]^a\nabla_a\alpha_3
=L_{[V_1,V_2]}\alpha_3$. Now
\[
[L_{V_1},L_{V_2}]\alpha_3
=L_{[V_1,V_2]}\alpha_3
+V_1^aV_2^b\,\bigl[F_{ab},\alpha_3\bigr]_{\rm sym}\:,
\]
where the curvature term involves
$[\nabla_a,\nabla_b]=[F_{ab},\,\cdot\,]$ on the index pair
symmetrised by the odd $V$'s. Since the connection is the Chern
connection, $F^{2,0}=0$ and the curvature term vanishes; the two
contributions then cancel. This is the one place the gauge component
uses $F^{2,0}=0$.

\medskip\noindent\textbf{Tangent component.} By field content the
configurations are $(\mu,x,x)$, $(\alpha,\alpha,\alpha)$ and
$(\alpha,\alpha,x)$; at most one argument is a $\mu$, since $\mu$
enters the bracket linearly.

For $(\mu,x,x)$, the first line of \eqref{eq:bracket_tangent} is the
graded Lie bracket $-[\mu,V']-[V,\mu']$ on
$\Omega^{0,\bullet}(T^{1,0}X)$. Nesting, the inner tangent output
enters the outer first line as a $\mu$-argument, and the inner
cotangent output enters through $V_{[ij]}=-[V_i,V_j]$; the cyclic sum
is then literally the graded Jacobi identity for the Lie bracket on
vector-valued forms, applied to $(\mu_1,V_2,V_3)$, and vanishes
identically, with no curvature or $\Omega$-identity required.

For $(\alpha,\alpha,\alpha)$, the only contributions are the gauge
bilinear of \eqref{eq:bracket_tangent}, call it $B(\alpha,\alpha')$,
evaluated on $([\alpha_i,\alpha_j]_{\mathfrak g},\alpha_k)$ and
summed cyclically. Expanding
$\nabla_b[\alpha_i,\alpha_j]=[\nabla_b\alpha_i,\alpha_j]
+[\alpha_i,\nabla_b\alpha_j]$ and using invariance and graded cyclicity
of the trace, the cyclic sum of the one-forms inside the outer derivative
of $B$ is
\[
-\,\partial_b\tr\bigl(\alpha_1[\alpha_2,\alpha_3]_{\mathfrak g}\bigr)\:,
\]
the gradient of a function. It is therefore annihilated by
$\partial_c(\cdot)\,\bar\Omega^{cba}$, and the cyclic sum of $B$ vanishes.

For $(\alpha,\alpha,x)$, the required identity is the equivariance of
the gauge bilinear under transport,
\begin{equation}
\label{eq:equiv}
L_{V}\bigl(B(\alpha,\alpha')\bigr)
=B\bigl(L_V\alpha,\alpha'\bigr)-B\bigl(\alpha,L_V\alpha'\bigr)\:,
\end{equation}
the relative sign being the Koszul factor $(-1)^{|L_V||\alpha|}=-1$
incurred in moving the odd operator $L_V$ past the odd argument
$\alpha$, where on the tangent output $L_V\mu=[V,\mu]$ is
the graded Lie derivative; given \eqref{eq:equiv}, the cyclic terms
in this family cancel pairwise exactly as in the gauge component. To
verify \eqref{eq:equiv}, expand both sides in the frames of
Convention~\ref{conv:frame}. Commuting $V^d\nabla_d$ past the two
covariant derivatives inside $B$ produces curvature commutators with
two holomorphic indices, which vanish by $F^{2,0}=0$, together with
derivative-of-$V$ terms $\partial_cV^d$, $\partial_bV^d$; these
reassemble against the term $-\partial_dV^a\,B^d$ produced by $L_V$
acting on the tangent output, by the Schouten identity
\eqref{eq:schouten} contracted with $\bar\Omega^{cba}$ together with
$\partial_dV^d=0$, the same mechanism by which \eqref{eq:cot_char}
was verified, applied one tensor slot deeper. Collecting the three
components completes the proof.
\end{proof}

\subsection{The \texorpdfstring{$L_\infty$}{L-infinity} relations with \texorpdfstring{$n\geq4$}{n>=4}}
\label{app:l4}

\begin{proof}[Proof of Proposition~\ref{prop:l4}]
For $Y_k\in\mathcal{Y}_{p_k}$, the $n$-th $L_\infty$ relation is \cite{Lada:1994mn}
\begin{equation}
    \label{eq:Linfn}
    \sum_{i+j=n+1}(-1)^{i(j-1)}\sum_{\sigma\in\mathrm{Sh}(i,n-i)}\chi(\sigma)\,
    \ell_j\bigl(\ell_i(Y_{\sigma1},\dots,Y_{\sigma i}),Y_{\sigma(i+1)},\dots,Y_{\sigma n}\bigr)=0\:.
\end{equation}
Here $\mathrm{Sh}(i,n-i)$ is the set of $(i,n-i)$-unshuffles, the permutations $\sigma$ of $\{1,\dots,n\}$ with $\sigma1<\dots<\sigma i$ and $\sigma(i+1)<\dots<\sigma n$, and $\chi(\sigma)=\mathrm{sgn}(\sigma)\,\varepsilon(\sigma)$, where $\varepsilon(\sigma)$ is the Koszul sign of reordering $Y_1,\dots,Y_n$ into $Y_{\sigma1},\dots,Y_{\sigma n}$, with $Y_k$ of degree $p_k$. For $n=3$ this is \eqref{eq:Linf3}, after the substitution described in the footnote there. That substitution multiplies $\ell_3$ by a sign which depends only on the total degree of its arguments; this is the same for every term below, so it plays no role for $n\geq4$.

We first determine which compositions survive. As noted in the sketch of the
proof in Section~\ref{sec:extended}, $\ell_2$ depends on its arguments only
through their $Q$-components, $\ell_3$ only through their cotangent
components, and $\ell_3$ takes values in the auxiliary slot, so that the
compositions \eqref{eq:l3compose} vanish.
Since $\ell_k=0$ for $k\geq4$, a term of \eqref{eq:Linfn} can be non-zero only if $i,j\leq3$, that is only if $n=i+j-1\leq5$. Every relation with $n\geq6$ therefore holds term by term. For $n=5$ the only remaining pair is $(i,j)=(3,3)$, which vanishes by \eqref{eq:l3compose}. For $n=4$ the remaining pairs are $(2,3)$ and $(3,2)$. The latter vanishes by \eqref{eq:l3compose}, and $(-1)^{i(j-1)}=1$ for the former, which leaves
\begin{equation}
    \label{eq:n4rel}
    \sum_{\sigma\in\mathrm{Sh}(2,2)}\chi(\sigma)\,
    \ell_3\bigl(\ell_2(Y_{\sigma1},Y_{\sigma2}),Y_{\sigma3},Y_{\sigma4}\bigr)=0\:,
\end{equation}
a sum over the six unshuffles
\begin{equation}
    \sigma=(12|34),\ (13|24),\ (14|23),\ (23|14),\ (24|13),\ (34|12)\:.
\end{equation}
The $Q$-component of \eqref{eq:n4rel} vanishes identically. In the auxiliary component, the cotangent component of $\ell_2(Y_i,Y_j)$ is \eqref{eq:bracket_cotangent}, and by \eqref{eq:cot_char} the associated vector is $-[V_i,V_j]$. Up to this overall sign, \eqref{eq:n4rel} is \eqref{eq:n4}.

Next we reduce to degree zero. By multilinearity we may take $V_i=v_i\,\omega_i$, with $\omega_i$ a constant basis $(0,p_i)$-form and $v_i$ a vector field with function coefficients. Each $v_i$ is again of the form $(\del x)$, and hence divergence-free. Since component forms stand to the left of the basis forms, $[V_i,V_j]=[v_i,v_j]\,\omega_i\omega_j$, where $[v_i,v_j]$ is the ordinary Lie bracket, and the term of \eqref{eq:n4} labelled by $\sigma$ is
\begin{multline}
    \chi(\sigma)\,W\bigl([v_{\sigma1},v_{\sigma2}],v_{\sigma3},v_{\sigma4}\bigr)\,
    \omega_{\sigma1}\omega_{\sigma2}\omega_{\sigma3}\omega_{\sigma4}\\
    =\mathrm{sgn}(\sigma)\,W\bigl([v_{\sigma1},v_{\sigma2}],v_{\sigma3},v_{\sigma4}\bigr)\,
    \omega_1\omega_2\omega_3\omega_4\:,
\end{multline}
because $\omega_{\sigma1}\cdots\omega_{\sigma4}=\varepsilon(\sigma)\,\omega_1\cdots\omega_4$ and $\chi(\sigma)\varepsilon(\sigma)=\mathrm{sgn}(\sigma)$. It therefore suffices to prove, for divergence-free vector fields $v_i=(\del x_i)$ of degree zero,
\begin{multline}
    \label{eq:n4even}
    W([v_1,v_2],v_3,v_4)-W([v_1,v_3],v_2,v_4)+W([v_1,v_4],v_2,v_3)\\
    +W([v_2,v_3],v_1,v_4)-W([v_2,v_4],v_1,v_3)+W([v_3,v_4],v_1,v_2)=0\:.
\end{multline}

To prove \eqref{eq:n4even}, write $W_{234}\coloneqq W(v_2,v_3,v_4)$, which is a function. Contracting the Schouten identity \eqref{eq:schouten} with $u^av^bw^c$ and with a matrix $N^e{}_d$ gives, in complex dimension three,
\begin{equation}
    \label{eq:trace}
    W(Nu,v,w)+W(u,Nv,w)+W(u,v,Nw)=N^d{}_d\,W(u,v,w)\:.
\end{equation}
Since $\Omega_{abc}$ is constant (Convention~\ref{conv:frame}) and $v_1^d\partial_dv_j=[v_1,v_j]+v_j^d\partial_dv_1$, using \eqref{eq:trace} with $N^e{}_d=\partial_dv_1^e$ gives
\begin{multline}
    \label{eq:n4first}
    v_1^d\partial_dW_{234}
    =W([v_1,v_2],v_3,v_4)+W(v_2,[v_1,v_3],v_4)\\
    +W(v_2,v_3,[v_1,v_4])+(\partial_dv_1^d)\,W_{234}\:.
\end{multline}
The last term vanishes by \eqref{eq:autodiv}. By the antisymmetry of $W$, the remaining three are the first three terms of \eqref{eq:n4even}. On the other hand, at degree zero Proposition~\ref{prop:jacobi}, together with \eqref{eq:bracket_cotangent} and \eqref{eq:cot_char}, reads
\begin{equation}
    \partial_aW_{234}=\Omega_{abc}\bigl([v_2,v_3]^bv_4^c+[v_4,v_2]^bv_3^c+[v_3,v_4]^bv_2^c\bigr)\:.
\end{equation}
Contracting with $v_1^a$ gives
\begin{equation}
    \label{eq:n4second}
    v_1^a\partial_aW_{234}
    =W(v_1,[v_2,v_3],v_4)+W(v_1,[v_4,v_2],v_3)+W(v_1,[v_3,v_4],v_2)\:,
\end{equation}
which, by the antisymmetry of $W$ and of the Lie bracket, is minus the last three terms of \eqref{eq:n4even}. The first three terms of \eqref{eq:n4even} therefore sum to $v_1^a\partial_aW_{234}$ and the last three to $-v_1^a\partial_aW_{234}$, which proves \eqref{eq:n4even}.
\end{proof}

\section{Gauge invariance of the extended action}
\label{app:gauge}

This appendix supplies the computations behind the statements made after
\eqref{eq:L3gauge}: the covariance of the Maurer--Cartan equation under the
$L_3$ gauge transformations, and the variation of the extended action
$S_b$ under them. We use the conventions of Section~\ref{sec:normalisation}.
The structural argument in the proof of Proposition~\ref{prop:varS} holds
for the full $\bar D$ and for $\alpha\neq0$; the explicit component
computations, given as a check, set $\alpha=0$ and $F=T=0$, so that
$\bar D=\delb$ on $\mu$ and $x$, except for the parameter $\beta$, where they
too are done for general $\bar D$. Compactness of $X$ enters only through
Stokes' theorem, and nothing below uses the constraint $\del_a\mu^a=0$. We
write
\begin{equation}
    V^a\coloneqq (\del x)^a,\qquad C\coloneqq \del_a\mu^a\in\Omega^{0,1}(X),\qquad
    E\coloneqq \bar Dy+\tfrac12[y,y]+\del b,\qquad
    F_b\coloneqq \delb b+\tfrac16W ,
\end{equation}
with components $E^a=\delb\mu^a+\tfrac12[y,y]^a$ and
$E_a=\delb x_a+\tfrac12\Omega_{abc}V^bV^c+\del_ab$, cf.\ \eqref{eq:Ecot},
so that the Maurer--Cartan equation of Definition~\ref{def:products} is
$E=0$, $F_b=0$. At degree one the bracket of Definition~\ref{def:bracket}
reads
\begin{equation}
    \label{eq:yy-app}
    [y,y]^a=-2[\mu,V]^a,\qquad
    [\mu,V]^a\coloneqq \mu^b\del_bV^a-\del_b\mu^a\,V^b,\qquad
    [y,y]_a=\Omega_{abc}V^bV^c ,
\end{equation}
and for a $0$-form vector $A$ and a $(0,1)$-form vector $B$ we write
$[A,B]^a=A^b\del_bB^a-B^b\del_bA^a$. The tools are holomorphic integration
by parts, $\int_X(\del_bu^b)\wedge\Omega=0$ since
$(\del_bu^b)\Omega=\pm\del(u^b\iota_{\del_b}\Omega)$, anti-holomorphic
integration by parts, $\int_X\delb\theta\wedge\Omega=0$, the inversion
identity \eqref{eq:inv} and the Schouten identity \eqref{eq:schouten}.
Holomorphic derivatives are even and carry no Koszul signs; the
$(0,1)$-forms $\mu^a$, $x_a$, $V^a$, $C$ anticommute among themselves.

\subsection{The gauge transformation}

For $\Lambda=(\xi,\beta)\in\mathcal{Y}_0$ with $\xi=(\xi_\mu,\xi_\alpha,\xi_x)$,
and $\xi_\alpha=0$ in the explicit computations, write
$V_\xi\coloneqq (\del\xi_x)$, a divergence-free vector field, and
$C_\xi\coloneqq \del_a\xi_\mu^a$. From Definition~\ref{def:products},
\eqref{eq:L3gauge} reads $\delta y=\bar D\xi+\del\beta+[y,\xi]$,
$\delta b=\delb\beta-\tfrac12\Omega_{abc}V^aV^bV_\xi^c$, that is,
\begin{equation}
    \label{eq:gaugecomp}
    \delta\mu=\delb\xi_\mu+[\xi_\mu,V]+[V_\xi,\mu],\qquad
    \delta x_a=\delb\xi_{x,a}+(\del\beta)_a+\Omega_{abc}V^bV_\xi^c ,\qquad
    \delta V=\delb V_\xi+[V_\xi,V],
\end{equation}
the last by applying $\bar\Omega^{abc}\del_b$ to $\delta x_c$ and using
\eqref{eq:inv} with $\del_aV^a=\del_aV_\xi^a=0$. The cotangent sector thus
transforms as a Kodaira--Spencer gauge transformation of $V$ with parameter
$V_\xi$. The constraint transforms as
\begin{equation}
    \label{eq:deltaC}
    \delta C=\delb C_\xi-V^b\del_bC_\xi+V_\xi^b\del_bC ,
\end{equation}
so the constraint surface is preserved precisely by the parameters with
$C_\xi=0$.

\subsection{Covariance of the Maurer--Cartan equation}

\begin{lemma}
\label{lem:jac110}
At degrees $(1,1,0)$ the Jacobiator is purely cotangent and
\begin{equation}
    2[y,[y,\xi]]_a-[[y,y],\xi]_a=\del_aW(V,V,V_\xi),\qquad
    W(V,V,V_\xi)\coloneqq \Omega_{bcd}V^bV^cV_\xi^d .
\end{equation}
\end{lemma}
\begin{proof}
With $u_a\coloneqq [y,\xi]_a=\Omega_{abc}V^bV_\xi^c$ one has $(\del u)^a=-[V,V_\xi]^a$
and $(\del[y,y])^a=-2V^d\del_dV^a$, both by \eqref{eq:inv} and
divergence-freeness. Hence
\[
    2[y,[y,\xi]]_a-[[y,y],\xi]_a=T_1+T_2+T_3\:,
\]
with
\[
    T_1=-2\Omega_{abc}V^bV^d\del_dV_\xi^c\:,\qquad
    T_2=2\Omega_{abc}V^bV_\xi^d\del_dV^c\:,\qquad
    T_3=2\Omega_{abc}V^d(\del_dV^b)V_\xi^c\:.
\]
Contracting \eqref{eq:schouten} with $V^bV^e\del_dV_\xi^c$ in $T_1$ and with
$V^bV_\xi^e\del_dV^c$ in $T_2$, and using $\del_cV_\xi^c=\del_cV^c=0$,
\[
    T_1=2\,\Omega_{bcd}V^bV^c\del_aV_\xi^d-T_1\:,\qquad
    T_2=2\,\Omega_{bcd}(\del_aV^b)V^cV_\xi^d-T_3\:.
\]
Hence
\[
    T_1+T_2+T_3=\Omega_{bcd}V^bV^c\del_aV_\xi^d
    +2\,\Omega_{bcd}(\del_aV^b)V^cV_\xi^d=\del_aW(V,V,V_\xi)\:,
\]
the last step using $\Omega_{bcd}V^b\del_aV^c=\Omega_{bcd}(\del_aV^b)V^c$ for
odd $V$. The tangent component vanishes by the Jacobi identity
for the Lie bracket of vector-valued forms.
\end{proof}

\begin{proposition}[Covariance]
\label{prop:cov}
Under \eqref{eq:L3gauge}, for all $\Lambda\in\mathcal{Y}_0$,
\begin{equation}
    \delta E=[E,\xi],\qquad
    \delta F_b=-\,\Omega_{abc}\,(\del E_x)^aV^bV_\xi^c ,
\end{equation}
where $(\del E_x)^a=\bar\Omega^{abc}\del_bE_c$ is built from the cotangent
component of $E$. In particular the Maurer--Cartan equation is covariant.
\end{proposition}
\begin{proof}
$\delta E=\bar D\delta y+[y,\delta y]+\del\delta b$, using
$\delta(\tfrac12[y,y])=[y,\delta y]$ for $\delta y$ of degree one. Here
$\bar D\bar D\xi=0$ by Proposition~\ref{prop:nilp};
$\bar D[y,\xi]=[\bar Dy,\xi]-[y,\bar D\xi]$ by Proposition~\ref{prop:leibniz};
$[y,\del\beta]=0$ by Lemma~\ref{lem:central}; and
$\bar D(\del\beta)+\del(\delb\beta)=0$ by Proposition~\ref{prop:l1sq}. What
remains is
\[
    \delta E=[\bar Dy,\xi]+[y,[y,\xi]]-\tfrac12\,\del W(V,V,V_\xi)
    =[\bar Dy+\tfrac12[y,y],\xi]
\]
by Lemma~\ref{lem:jac110}; the coefficient $-\tfrac12$ of $\ell_3$ in
\eqref{eq:L3gauge} is fixed by this cancellation. Since $[\del b,\xi]=0$,
the bracket is $[E,\xi]$.

For $F_b$, $\delta F_b=\delb\delta b+\tfrac12\Omega_{abc}\delta V^aV^bV^c$.
By \eqref{eq:gaugecomp} the $\delb V_\xi$ terms cancel between the two
contributions. Substituting $\delb V^a=(\del E_x)^a+V^d\del_dV^a$, which
is $\bar\Omega^{abc}\del_b$ applied to the definition of $E_c$, leaves the
$\del E_x$ term and three cubic expressions in $V,V_\xi$,
\[
    -\Omega_{abc}(V^d\del_dV^a)V^bV_\xi^c
    -\tfrac12\Omega_{abc}(V^d\del_dV_\xi^a)V^bV^c
    +\tfrac12\Omega_{abc}(\del_dV^a)V_\xi^dV^bV^c .
\]
By \eqref{eq:schouten} the third equals minus the first, and the second
vanishes, each term produced by \eqref{eq:schouten} containing either
$\del_aV_\xi^a=0$ or the totally antisymmetric product $V^dV^bV^c$
contracted through $\Omega$ against $\del_bV_\xi^a$. This is the mechanism
of Proposition~\ref{prop:l4}.
\end{proof}

\subsection{Variation of the action}

We first record the general-degree form of the cyclicity defect
\eqref{eq:cyclicdefect}, which is what the structural argument below needs.

\begin{lemma}[Cyclicity off the constraint surface]
\label{lem:defect}
For $w\in\Omega^{0,r}(T^{*1,0}X)$, $\mu\in\Omega^{0,p}(T^{1,0}X)$ and
$x\in\Omega^{0,q}(T^{*1,0}X)$ with $p+q+r=3$, with $V=(\del x)$ and
$C=\del_b\mu^b$,
\begin{equation}
    \label{eq:defect}
    \int_X\langle w,[\mu,x]\rangle
    =(-1)^{r(p+q)}\int_X\langle\mu,[x,w]\rangle
     +\int_Xw_a\,C\,V^a\wedge\Omega .
\end{equation}
The families (i) and (ii) of Lemma~\ref{lem:cyclic} involve no divergence of
$\mu$ and hold exactly at all degrees, so \eqref{eq:defect} is the only source
of defects in any cyclic rearrangement of the pairing, and the defects
involve $\mu$, $x$, $w$ and $C$ only.
\end{lemma}
\begin{proof}
$\langle w,[\mu,x]\rangle=-w_a(\mu^b\del_bV^a-\del_b\mu^aV^b)\,\Omega$.
Integrating $\del_b$ by parts and using $\del_bV^b=0$,
\[
    \int w_a\mu^b\del_bV^a=-\int(\del_bw_a)\mu^bV^a-\int w_aCV^a ,
    \qquad
    -\int w_a\del_b\mu^aV^b=\int(\del_bw_a)\mu^aV^b .
\]
Relabelling
$a\leftrightarrow b$ in the first, the sum is
$\int(\del_bw_a-\del_aw_b)\mu^aV^b-\int w_aCV^a$, and by \eqref{eq:inv}
$\del_bw_a-\del_aw_b=\Omega_{cba}U^c$ with $U=(\del w)$. Moving $U^c$ past
$\mu^a$ costs $(-1)^{pr}$, and cyclic relabelling gives
$(-1)^{pr}\Omega_{abc}\mu^aU^bV^c$. On the other side
\[
    \langle\mu,[x,w]\rangle=\mu^a\Omega_{abc}V^bU^c\,\Omega
    =-(-1)^{qr}\Omega_{abc}\mu^aU^bV^c\,\Omega .
\]
Assembling the signs gives \eqref{eq:defect}; at $(p,q,r)=(1,1,1)$ it is \eqref{eq:cyclicdefect}.
\end{proof}

\begin{proposition}
\label{prop:varS}
Let $S_b$ be the action \eqref{eq:Sfinal}, with the
full operator $\bar D$ and all components of $y$ present, and let
$\delta y$, $\delta b$ be given by \eqref{eq:L3gauge}. Then, off shell
and off the constraint surface,
\begin{align}
    \delta_\beta S_b&=0 ,\qquad \delta_{\xi_\alpha}S_b=0 ,\\
    \label{eq:deltaximu}
    \delta_{\xi_\mu}S_b
    &=\int_X C_\xi\bigl(F_b-V^aE_a\bigr)\wedge\Omega ,\\
    \label{eq:deltaxix}
    \delta_{\xi_x}S_b
    &=\int_X C\wedge V_\xi^aE_a\wedge\Omega ,
\end{align}
where $E_a$ is the cotangent component of $E$, which does not involve
$\alpha$, $F$ or $T$.
\end{proposition}
\begin{proof}
\medskip\noindent\textbf{Structural argument.} Start from \eqref{eq:varSext},
which was derived for the full $\bar D$ and $\alpha\neq0$,
\[
    \delta S_b=\int\langle\delta y,E\rangle-\int C(\delta b-\delta x_aV^a) ,
\]
and insert $\delta y=\bar D\xi+[y,\xi]+\del\beta$. By Lemma~\ref{lem:IBP},
\[
    \int\langle\bar D\xi,E\rangle=-\int\langle\xi,\bar DE\rangle ,
\]
and by Lemma~\ref{lem:sym}, Lemma~\ref{lem:cyclic} at degrees $(2,1,0)$ and
Lemma~\ref{lem:defect},
\[
    \int\langle[y,\xi],E\rangle=\int\langle\xi,[E,y]\rangle+\Delta
    =-\int\langle\xi,[y,E]\rangle+\Delta ,
\]
where $\Delta$ collects the defect terms of \eqref{eq:defect}. Now
\[
    \bar DE+[y,E]
    =[\bar Dy,y]+\delb\del b+[y,\bar Dy]+\tfrac12[y,[y,y]]
    =\delb\del b-\tfrac16J(y,y,y) ,
\]
by Proposition~\ref{prop:nilp}, Proposition~\ref{prop:leibniz},
Lemma~\ref{lem:central} and Lemma~\ref{lem:brasym}, all of which hold for
the full $\bar D$ and all components. This is purely cotangent, with
component $\del_a(\tfrac16W+\delb b)=\del_aF_b$ by
Proposition~\ref{prop:jacobi}, so
\[
    -\int\langle\xi,\bar DE+[y,E]\rangle
    =-\int\xi_\mu^a\del_aF_b\wedge\Omega
    =\int C_\xi F_b\wedge\Omega .
\]
Hence
\begin{equation}
    \label{eq:structural}
    \delta S_b=\int_X C_\xi F_b\wedge\Omega+\Delta
    +\int_X\langle\del\beta,E\rangle
    -\int_X C\,(\delta b-\delta x_aV^a)\wedge\Omega .
\end{equation}
Every term on the right is a functional of $\mu$, $x$, $\xi$, $\beta$, $b$,
the cotangent component $E_a$ and the divergence
$\del_aE^a=\delb C+(\del_bC)V^b$; the last holds for the full $\bar D$
because the extension outputs of \eqref{eq:barD} in the tangent slot and the
$\alpha$-bilinear of \eqref{eq:bracket_tangent} are $\del$-exact, hence
divergence free. None of these depends on $\alpha$, $F$ or $T$. Consequently
\eqref{eq:structural} is the same expression as at $\alpha=0$, $F=T=0$, and
its evaluation is the explicit computation below. The parameter $\xi_\alpha$
enters no defect, no $\delta b$ and no $\delta x$, and
$\langle\xi_\alpha,\cdot\rangle$ does not see a cotangent element, so
$\delta_{\xi_\alpha}S_b=0$. The $C_\xi$ terms can be read off directly: the
defect for the sub-triple $(E_x,x,\xi_\mu)$ is, by \eqref{eq:defect} with
$[x,\xi_\mu]=-[\xi_\mu,x]$, $-\int C_\xi V^aE_a$, giving \eqref{eq:deltaximu};
the defect for $(E_x,\mu,\xi_x)$ is $+\int C\,V_\xi^aE_a$, and the remaining
$\xi_x$ terms cancel among themselves, as the explicit computation shows.

\medskip\noindent\textbf{Explicit check at $\alpha=0$, $F=T=0$.}
Insert \eqref{eq:gaugecomp} into \eqref{eq:varSext}, or equivalently vary
the expanded form of \eqref{eq:Sfinal},
\[
    S_b=\tfrac12\int\langle y,\bar Dy\rangle
      +\tfrac12\int\mu^a\Omega_{abc}V^bV^c\wedge\Omega
      -\int C\,b\wedge\Omega
\]
at $\alpha=0$. The $\delb\xi$ parts of $\delta y$ drop from the quadratic
term by anti-holomorphic integration by parts, and the remaining terms
combine using
\eqref{eq:cyclicdefect}, \eqref{eq:inv}, \eqref{eq:schouten} and
holomorphic integration by parts; every term that is not proportional to
$C$ or to $C_\xi$ cancels. We record the one identity in each sector that
is not a direct cancellation.

For $\beta$, with $\delta x_a=(\del\beta)_a$, $\delta b=\delb\beta$ and
general $\bar D$, the variation is
\[
    \int\delta x_aE^a-\int C(\delta b-\delta x_aV^a) ,
\]
and one needs the off-shell divergence of the tangent equation,
$\del_aE^a=\delb C+(\del_bC)V^b$: the extension outputs of \eqref{eq:barD}
in the tangent slot are $\del$-exact, hence divergence-free, and by
\eqref{eq:yy-app} $\del_a[y,y]^a=2(\del_bC)V^b$. Integrating by parts,
\[
    \int\delta x_aE^a=\int C\bigl(\delb\beta+(\del_a\beta)V^a\bigr) ,
\]
which cancels the multiplier term exactly.

For $\xi_\mu$, only $\mu$ and $C$ vary. The cubic term gives
\[
    \tfrac12\int[\xi_\mu,V]^a\Omega_{abc}V^bV^c=-\tfrac13\int C_\xi W ,
\]
each half of the bracket contributing $-\tfrac16\int C_\xi W$, one directly
and one by \eqref{eq:OO} and the total anti-symmetry of $V^dV^bV^c$. The
quadratic term gives
\[
    -\int C_\xi V^a\delb x_a-\int\xi_\mu^a\Omega_{abc}V^b\delb V^c ,
\]
the second piece cancelling against the $\delb\xi_\mu$ part of the cubic
term. Finally, the multiplier term gives
$-\int(\delta C)\,b=\int C_\xi(\delb b-V^b\del_bb)$ by
\eqref{eq:deltaC}. The sum is
\[
    \int C_\xi(\delb b+\tfrac16W-V^aE_a) .
\]

For $\xi_x$, writing $\eta\coloneqq V_\xi$, the terms bilinear in $\mu$ and
$\eta$ cancel among themselves, those involving $\delb\eta$ after one
anti-holomorphic integration by parts and those involving $[\eta,\mu]$ and
$[\eta,V]$ by \eqref{eq:inv} and \eqref{eq:schouten} as in
Lemma~\ref{lem:jac110}; what survives is $\int C\eta^a\delb x_a$ from the
quadratic term, $\int C\eta^b\del_bb$ from $\delta C=\eta^b\del_bC$, and
$\tfrac12\int C\eta^a\Omega_{abc}V^bV^c$ from $-\int C\,\delta b$, which
add up to $\int C\,\eta^aE_a$.
\end{proof}

\bibliographystyle{JHEP}
\bibliography{References}

\end{document}